\documentclass[aps,prb,twocolumn,nofootinbib,citeautoscript,10pt,longbibliography,notitlepage]{revtex4-2}
\pdfoutput=1

\usepackage{graphicx}% Include figure files
\usepackage{dcolumn}% Align table columns on decimal point
\usepackage{bm,amsfonts,amsmath}
\usepackage{color} 
\usepackage[tight]{subfigure}
\usepackage{ulem}

\usepackage[papersize={8.5in,11in}]{geometry}
\usepackage{xcolor}
\usepackage{tabularx}
\usepackage{comment}

\definecolor{darkblue}{rgb}{0.,0.,0.4}
\definecolor{darkred}{rgb}{0.5,0.,0.}
\definecolor{BlueViolet}{RGB}{138,43,226}
\definecolor{SkyBlue}{RGB}{30,144,255}
\definecolor{DarkGreen}{RGB}{0,100,0}
\usepackage[pdftex,colorlinks=true,linkcolor=darkblue,citecolor=blue,urlcolor=darkred]{hyperref}
\usepackage{float}
\usepackage{physics}

\def \nn{\nonumber \\}
\renewcommand{\vec}{\mathbf}

\begin{document}

\title{Thermoelectric response in nodal-point semimetals}

\author{Ipsita Mandal}
\affiliation{Department of Physics, Shiv Nadar Institution of Eminence (SNIoE), Gautam Buddha Nagar, Uttar Pradesh 201314, India}
\affiliation{Freiburg Institute for Advanced Studies (FRIAS), University of Freiburg, D-79104 Freiburg, Germany}

\author{Kush Saha}
\affiliation{National Institute of Science Education and Research, Jatni, Khurda 752050, Odisha, India\\
%%%%%%%%%%%%%%%%%
Homi Bhabha National Institute, Training School Complex, Anushakti Nagar, Mumbai 400094, India}

%%%%%%%%%%%%%%%%%%%%%%%%%
\begin{abstract}
In this review, the thermoelectric properties in nodal-point semimetals with two bands are discussed. For the two-dimensional (2D) cases, it is shown that the expressions of the thermoelectric coefficients take different values depending on the nature of the scattering mechanism responsible for transport, by considering examples of short-ranged disorder potential and screened charged impurities. An anisotropy in the energy dispersion spectrum invariably affects the thermopower quite significantly, as illustrated by the results for a node of semi-Dirac semimetal and a single valley of graphene. The scenario when a magnetic field of magnitude $B$ is applied perpendicular to the plane of the 2D semimetal is also considered. The computations for three-dimensional (3D) cases necessarily involve the inclusion of nontrivial Berry phase effects. In addition to demonstrating the expressions for the response tensors, the exotic behaviour observed in planar Hall and planar thermal Hall set-ups is also discussed.
\end{abstract}
%%%%%%%%%%%%%%%%%%%%%%
\maketitle

\tableofcontents

%%%%%%%%%%%%%%%%%%%%%%%%%%%%%%%%%%%%%%%%%

 \section{Introduction}
 
The measurement of the thermoelectric effects is one of the most widely used experimental probes for investigating transport mechanisms in metals, semimetals, and semiconductors. The behaviour of thermopower has proved to be a powerful and versatile tool in characterizing material properties as it provides information complementary to that obtained from electrical resistivity measurements, for example, by shedding light on both the conduction processes and thermodynamics. In this review, we will elucidate the derivation of the analytical expressions for the thermoelectric coefficients of the quasiparticles emerging in two-dimenstional (2D) and three-dimensional (3D) semimetals. In particular, we will focus on semimetals with a pair of bands. At a nodal point of such a semimetal, the two bands cross each other giving a zero density of states right at the band-crossing point.
We will consider the cases when an electric field $\mathbf E $ (or a temperature gradient $ {\nabla}_{\mathbf r} T $) is applied externally across a sample. The thermoelectric coefficients take different values depending on the relaxation processes involved, which include short-ranged disorder and scattering off charged impurities. The results obtained using the semiclassical Boltzmann equation approach turn out to be in good agreement with other theoretical approaches (like the Kubo formula and the quantum Boltzmann equations). We will also discuss the situation when a magnetic field $\mathbf B$ is applied in addition. For the 2D cases, the direction of $\mathbf B $ is perpendicular to the plane of the semimetal. For the 3D cases, if the magnetic field has a parallel component along $\mathbf E $  (or $ {\nabla}_{\mathbf r} T $), we observe the so-called planar Hall (or planar thermal Hall) phenomenon.
%%%%%%%%%%%%%5
For a weak magnetic field, when the formation of the Landau levels can be ignored, we will continue to use the semiclassical Boltzmann formalism. However, this fails for strong magnetic fields, leading to quantized energy eigenvalues in the form of Landau levels. We deal with this quantizing magnetic field regime by using entropy to derive the form of the transport coefficients \cite{Skinner2018,Bergman2010}.
In all our calculations, we consider a single node and focus on the cases when the relaxation time involves only the intranode scattering processes.

%For the zero and weak magnetic field cases, we show the derivations of the response tensors using the semiclassical Boltzmann equation framework \cite{mermin,tong,arovas,soto}. 

There is an extensive literature devoted to the study of the thermoelectric properties of 2D isotropic materials like graphene and related 2D Dirac materials \cite{Wei2009,zhu2010}. For 3D nodal-point semimetals, one needs to include  the effects of a nontrivial Berry curvature \cite{girish1,Girish2017,Gegory2014,GirishTiwari2017,Liang2017,Vozmediano2018,2021nag_nandy,ips-serena,amit-magnus,papaj_magnus,sajid_magnus}.
%%%%%%%%%%%%%%%%%%%%%%
The characterization of transport properties in isotropic Dirac/Weyl materials has spanned both 2D and 3D \cite{Wei2009,zhu2010,piet2014,Huang2013,mikito2014,hosur2012,karl2014,kamran2013,prl_niu,Liang2017,kamran2015,
Bardarson2017,gorbar2017,trivedi2017,emil-magneto,ips-magneto}. Subsequently, the task of computing the thermoelectric properties in 2D anisotropic Dirac/Weyl materials, such as VO$_2$/TiO$_3$ \cite{pardo,pardo2,banerjee}, organic salts \cite{kobayashi,suzumura}, and deformed graphene \cite{hasegawa,orignac,montambaux1,montambaux2}, has been taken up \cite{ips-kush}. Such a system is represented by a semi-Dirac semimetal featuring two bands, whose low-energy bandstructure harbours a linear dispersion in one direction and a quadratic dispersion along the direction perpendicular to it. 
Since transport coefficients are determined by the bandstructure and the relevant scattering processes for the emergent quasiparticles, the anisotropic dispersion of the 2D semi-Dirac materials invariably leads to unconventional electric and magnetic properties, as opposed to the isotropic Weyl/Dirac systems \cite{landau-level,moon}. There also have been studies incorporating the 3D anisotropic cases \cite{Gegory2016,2021nag_nandy,ips-serena}.
In particular, the behaviour of the transport coefficients reflects how anisotropy can give rise to interesting field-, temperature-, and doping-dependence.

The review is organized as follows. In Sec.~\ref{secboltz} and Sec.~\ref{secboltz2}, we outline the derivation of the semiclassical Boltzmann equations and, subsequently, the expressions for the transport coefficients under the relaxation time approximation. The second one expands on the first to include nontrivial Berry phase effects. Sec.~\ref{sec2d}---Sec.~\ref{secmag2} consider the 2D cases, providing comparisons for linear-in-momentum isotropic (by considering a single valley of graphene) and hybrid/anisotropic (by considering semi-Dirac semimetal) dispersions. 
%%%%%%%%%%%%%%%%%
In Sec.~\ref{sec2d}, we provide the analytical expressions for thermoelectric coefficients in zero magnetic field, assuming a constant (i.e., independent of energy or momentum) relaxation time. In Sec.~\ref{secdisorder}, we discuss the form of the transport coefficients by using a relaxation time resulting from the scatterings off short-ranged disorder. This is followed by Sec.~\ref{seccoulomb}, where we consider a relaxation time caused by the presence of a screened Coulomb potential for the carriers, resulting from charged impurities. In Sec.~\ref{secmag1} and Sec.~\ref{secmag2}, we add an external magnetic field directed along the line perpendicular to the plane of the 2D semimetals. These two sections elucidate the behaviour of the transport coefficients for the weak (non-quantizing) and strong regimes (i.e., when Landau levels emerge) of the strength of the external magnetic field, respectively. Finally, we conclude with a summary and outlook in Sec.~\ref{conclusion}.

%%%%%%%%%%%%%%%%%%%%%%%%%%%%%%%%
\section{Semiclassical Boltzmann equation for zero Berry curvature}
\label{secboltz}

Our fundamental understanding of the electronic properties of crystalline solids is primarily based on the Bloch theory for periodic systems. One of the most widely used descriptions is the semiclassical theory for quasiparticle dynamics within a band, supplemented by the simple and efficient framework of the Boltzmann transport equations. Hence, we review the Boltzmann’s transport theory, which allows us to deal with dissipation and momentum relaxation of non-stationary electronic states in metals and semimetals. 

The task of computing a finite conductivity can be accomplished by using a formalism based on the distribution function of quasiparticles. A system, isolated from any external influence, reaches equilibrium through relaxation after some characteristic time, accompanied by an increase of entropy (as can be explained using the tools of statistical physics). Considering a system in $d$ spatial dimensions, we define the distribution function (alternatively, the probability density function) $ f_n( \mathbf r , \mathbf k, t) $ for the Bloch band (labelled by the index $n$) with the crystal momentum $\mathbf k$ and dispersion $\epsilon_n(\mathbf k)$, such that
\begin{align}
\label{eqdist}
dN_n = g_n \,f_n( \mathbf r , \mathbf k, t) \,
\frac{ d^d \mathbf k}{(2\, \pi)^d } 
\,d^d \mathbf r
\end{align}
is the number of particles in an infinitesimal phase space volume $
dV_p = \frac{ d^d \mathbf k}{(2\, \pi)^d } 
\,d^d \mathbf r $ centered at $\left \lbrace \mathbf r , \mathbf k \right \rbrace $ at time $t$, and $g_n$ is the degeneracy
\footnote{The degeneracy may arise due to some extra quantum numbers present in the description of the system. One example is when we need to consider the spin degrees of freedom.} of the band. By definition, the distribution function $ f_n$ is dimensionless. 
By performing integrals over the momentum space involving $ f_n$ in the integrands, we can obtain various physical quantities.
%A simple example is the current density
%\begin{align}
%\mathbf J_n = g_n \,{\mathcal Q} \int_{BZ} \frac{ d^d \mathbf k}{(2\, \pi)^d } \,\,f_n( \mathbf r , \mathbf k, t)
%\, \boldsymbol{v}_n(\mathbf k)
%\end{align}
%at the position $\mathbf r$ and time $t$, where 
Let $\mathcal Q $ be the electric charge of a single quasiparticle, and 
\begin{align}
\boldsymbol{v}_n({\mathbf{k}})
= \frac{1}{\hbar} \, \nabla_{\vec k} \epsilon_n ({\mathbf k }) 
\end{align}
be the Bloch velocity (or group velocity), with $\hbar$ denoting the reduced Planck's constant.

In our set-up, we assume that the system is inhomogeneous on a large scale, and we are dividing it into subsystems
which are approximately homogeneous. Then, each of these subsystems can be characterized by the distribution function $ f_n( \mathbf r , \mathbf k, t) $, which depends on the position of the corresponding subsystem. The Liouville’s theorem, which describes the evolution of the distribution function in phase space for a Hamiltonian system, 
states that $\frac{df_n}{dt} = 0$. In other words, the distribution function is a probabilitiy distribution in the phase space and, because probability is locally conserved, it must obey a continuity equation just like an incompressible fluid. Consequently, in the course of the evolution of the probability distribution function, governed by the Hamilton's equations of motion, the probability does not change as we follow it along any trajectory in the phase space and, hence, represents an integral of motion. The Liouville's theorem thus implies that the distribution function remains constant.
%%%%%%%%
All of this follows from the facts that the phase space volume does not change, and the particle number is conserved in the phase space. If, however, collisions are taken into consideration, Liouville's theorem is violated and the distribution
function is no longer constant along the semiclassical phase space trajectories. Therefore, in order to explain dissipative transport phenomena resulting from scattering events, Ludwig Boltzmann modified the Liouville equation to
\begin{align}
\label{eqkin1}
& \frac{df_n}{dt} = \left[ \frac{\partial f_n}{\partial t} \right]_{\text{coll}} \nn
  \Rightarrow
 &  \left( \partial_t   
+ \dot{\mathbf r} \cdot \nabla_{\mathbf r} 
+ \dot{\mathbf k} \cdot \nabla_{\mathbf k} \right) f_n
= \left[ \frac{\partial f_n}{\partial t} \right]_{\text{coll}} \,,
\end{align}
where the the right-hand side contains the correction term $\left[ \frac{\partial f_n}{\partial t} \right]_{\text{coll}}$, which arises due to the collisions added as a perturbation. We have denoted total time derivatives by the widely used convention of overhead dots. The collision term must be such that the distribution function relaxes toward a thermal
equilibrium. Equations of this form generically represent kinetic equations.

The Hamilton's equations of motion for the Bloch electrons, under the influence of electromagnetic fields, are given by (cf. Chapter--12 of Ref.~\cite{mermin}):
\begin{align}
\label{eqkin2}
{\hbar} \, \dot{\mathbf r} = \partial_{\mathbf k} \epsilon_n ({\mathbf k})\,,
\quad 
\hbar\,\dot{\mathbf k}
= {\mathcal Q} \left( {\mathbf E} +  
\frac{ \dot{\mathbf r} \times {\mathbf B} } {c}
\right ) ,
\end{align}
where $\mathbf E $ and $\mathbf B $ are the externally applied electric and magnetic fields, respectively. Here, we have neglected the orbital magnetization of the Bloch wavepacket and the contributions from
the spin-orbit interactions. Furthermore, we have assumed that the Bloch bands are topologically trivial.
%%%%%%%%%%%%%%%%
Using Eqs.~\eqref{eqkin1} and \eqref{eqkin2}, this leads to the kinetic equation
\begin{align}
\label{eqkin3}
 \left [
 \partial_t  
+ {\boldsymbol v}_n  
 \cdot \nabla_{\mathbf r} 
+ \frac{\mathcal Q}  {\hbar } \left(
\mathbf E
+ \frac{ {\boldsymbol v}_n   \times {\mathbf B} } {c}
\right) 
\cdot \nabla_{\mathbf k} \right ] f_n
= \left[ \frac{\partial f_n}{\partial t} \right]_{\text{coll}} \,.
\end{align}
The terms on the left-hand side are often denoted as the \textit{drift terms}, constituting the ``co-moving'' total time derivative. They are also sometimes referred to as the ``streaming term'', because it tells us how the quasiparticles move in the absence of collisions. On the other hand, the right-hand side results from collisions, and it is also often referred to as the collision integral. Due to these two sets of terms, two effects cause $f_n$ to evolve with time $t$:
\\(1) There is a smooth evolution arising from the drift and acceleration
of the quasiparticles. Ignoring collisions, evolving from time $t$ to $t + \Delta t$, the new distribution
$f_n(\mathbf r , \mathbf k, t + \Delta t)$ will be the old distribution $ f_n(\mathbf r -\dot{\mathbf r} \,\Delta t, 
\mathbf k-\dot{\mathbf k} \,\Delta t, t)$. This part is the consequence of the Liouville's theorem.
%%%%%%%%%%%%%%%%%%
\\(2) Scattering processes cause discontinuous changes of the momentum $\mathbf  k$ at some
statistical rate.

One of the main assumptions of the Boltzmann transport equation is that the quasiparticles can be treated semiclassically, obeying Newton's laws of motion. Quantum mechanics enters into the equation only through the bandstructure and the description of the collision term. Needless to say, $ f_n( \mathbf r , \mathbf k, t) $ being a function of $\mathbf r $ and $ \mathbf k$ simultaneously will not violate the uncertainty principle, if the subsystems (into which the system is divided into) are large enough. In other words, the spatial variations and the temporal variations should occur at large distances (or long wavelengths denoted by $ \lambda $) and small frequencies (denoted by $\omega$), respectively. Mathematically, these imply the limits $ \lambda  \gg  2\,\pi / k_F $ and $\hbar \,\omega \ll E_F $, where $k_F$ and $E_F $ denote the Fermi momentum and the Fermi energy, respectively. This scenario is feasible if we demand that the external perturbations do not vary rapidly in space.
% https://github.molgen.mpg.de/pages/bs/pqm/pqm_boltzmann.html
Hence, the Boltzmann equation formalism is a semiclassical description that does not account for very fast processes in small areas (which are the restrictions arising due to the uncertainty principle).
The main idea behind the Boltzmann equation framework is that there
are two time scales in the problem \cite{tong} --- (1) the first is the time between two successive collisions ($\tau$), which is known as the scattering time (or the relaxation time); (2) the second is the collision time ($\tau_{\text{coll}}$), which
is roughly the time it takes for a collision between quasiparticles to take place. 
In the regime where the condition $\tau \gg \tau_{\text{coll}} $ holds, most of the time, $f_n$ simply follows its Hamiltonian evolution, with occasional perturbations caused by the collision events.

In any system, the quasiparticles transport thermal energy (i.e., heat) simultaneously with the electric charge. This is
why the transport of electric charge and heat are naturally interconnected. To demonstrate this connection, we now generalize the transport theory set up above. In order to derive the generalized Boltzmann equation,
we consider a metal with weakly space-dependent temperature $T (\mathbf r)$ and chemical potential $\mu(\mathbf r) $.
This necessitates the introduction of the electrochemical potential and the generalized (external) force field defined by
\begin{align}
\eta(\mathbf r) = \Phi(\mathbf r) 
- \frac{\mu (\mathbf r) } {\mathcal Q}
\text{ and }
\boldsymbol{\mathcal E} (\mathbf r) = 
-\nabla_{\mathbf r} \eta (\mathbf r) \,,
\end{align}
respectively, where $\Phi (\mathbf r) $ is the electrostatic potential such that $\mathbf E = -\nabla_{\mathbf r } \Phi $. Hence, Eqs.~\eqref{eqkin2} and \eqref{eqkin3} must be generalized to
\begin{align}
\label{eqkin21}
{\hbar} \, \dot{\mathbf r} = \partial_{\mathbf k} \epsilon_n ({\mathbf k})\,,
\quad 
\hbar\,\dot{\mathbf k}
= {\mathcal Q} \left( \boldsymbol{\mathcal E} +  
\frac{ \dot{\mathbf r} \times {\mathbf B} } {c}
\right )
\end{align}
and
\begin{align}
\label{eqkin32}
& \left [
 \partial_t  
+ {\boldsymbol v}_n  
 \cdot \nabla_{\mathbf r} 
+ \frac{\mathcal Q}  {\hbar } \left(  \boldsymbol{\mathcal E}
+ \frac{ {\boldsymbol v}_n   \times {\mathbf B} } {c}
\right) 
\cdot \nabla_{\mathbf k} \right ] f_n
% \nn & 
= \left[ \frac{\partial f_n}{\partial t} \right]_{\text{coll}} \,,
\end{align}
respectively.

In the following, we will consider a simple model of the collision integral, which is known as the
relaxation time approximation. The local value of the static distribution of the fermionic quasiparticles is given by
the function
\begin{align}
\label{eqfd}
 f^{(0)}_n (\mathbf r,\mathbf k) 
 = \frac{1}
 {e^{ \beta (\mathbf r )\,
 \left \lbrace \epsilon_n  (\mathbf k )
 -\mu (\mathbf r )
 \right \rbrace } + 1}\,,
\end{align} 
which describes a local equilibrium situation at the subsystem centred at position $\mathbf r$, at the local temperature $T(\mathbf r )$, and with local chemical potential $\mu (\mathbf r )$. We have used the symbol $\beta =1 /(k_B \, T)$ (where $k_B$ is the Boltzmann constant), which is sometimes referred to as the inverse temperature.
Now we make the ansatz
\begin{align}
\left[ \frac{\partial f_n}{\partial t} \right]_{\text{coll}}
= -
 \frac{ f_n(\mathbf r,\mathbf k, t)
 -f^{(0)}_n (\mathbf r,\mathbf k)
 }
 {\tau} \,,
\end{align}
where $\tau $ is called the relaxation time, which is generically $\mathbf k $-dependent.
The relaxation time quantifies the characteristic time scale, within which the
system relaxes to equilibrium, for the scattering processes relevant for the problem under consideration.
The mean free path of the quasiparticles can be defined in terms of $ \tau $ as
\begin{align}
\ell = \tau \, v_F \,,
\end{align}
where $ v_F$ is the Bloch velocity at the Fermi level.
Two assumptions are made while applying the relaxation time approximation:
\\(1) The distribution function of the quasiparticles, directly after a collision, does not
depend on their distribution function shortly before the collision. This assumption implies that the collisions destroy all information about the non-equilibrium distribution function of the quasiparticles (before the collision).
\\(2) The collisions do not change the shape of the equilibrium distribution function of the quasiparticles. This assumption actually says that the collisions themselves are shaping the distribution function or, in other words, stabilizing
the system as far as its thermodynamic equilibrium is concerned --- consequently, they will not change it.

This approximation holds when we study processes close to the equilibrium, i.e., when
\begin{align}
  |f_n(\mathbf r,\mathbf k, t)
 - f^{(0)}_n (\mathbf r,\mathbf k) | \ll f^{(0)}_n(\mathbf r,\mathbf k, t)\,.
\end{align}  
 Therefore, in order to obtain a solution to the full Boltzmann equation, we assume a slight deviation quantified by
\begin{align}
\label{eqpertf}
 f_n(\mathbf r,\mathbf k, t)
 =  f^{(0)}_n(\mathbf r,\mathbf k) +  \delta  f_n(\mathbf r,\mathbf k, t)\,,
\end{align} 
with $ f^{(0)}_n(\mathbf r,\mathbf k) $ given by Eq.~\eqref{eqfd}.
%%%%%%%%%%%%%%%%%%%%%%%%%%%%%%%%
Observing that
\begin{align}
\label{eqpertf1}
\nabla_{\mathbf r}  f^{(0)}_n (\mathbf r,\mathbf k) 
& = 
\left( \nabla_{\mathbf r} \mu + \frac{ \epsilon_n - \mu} {T} 
\, \nabla_{\mathbf r} T \right )
\left( - \frac{\partial  f^{(0)}_n } {\partial \epsilon_n } \right ),\nn
%%%%%%%%%%%%%%
\text{ and }
\nabla_{\mathbf k}  f^{(0)}_n (\mathbf r,\mathbf k) 
& = \hbar \, {\boldsymbol v}_n 
\, \frac{\partial  f^{(0)}_n } {\partial \epsilon_n } \,,
\end{align}
the Boltzmann equation in Eq.~\eqref{eqkin3} reduces to
\begin{widetext}
\begin{align}
\label{eqkin4}
\partial_t\delta f_n
+ {\boldsymbol v}_n \cdot 
{\nabla}_{\mathbf r} \,\delta f_n 
+
\frac{\mathcal Q}  {\hbar } \left(
\boldsymbol{\mathcal E}
+ 
\frac{ {\boldsymbol v}_n   \times {\mathbf B} } {c}
\right) 
\cdot \nabla_{\mathbf k}\, \delta f_n
+
{\boldsymbol v}_n \cdot
\left( -\,
\frac{ \epsilon_n - \mu} {T}  \, \nabla_{\mathbf r} T 
+ {\mathcal Q}\, \boldsymbol{\mathcal E}
\right )
 \frac{\partial  f^{(0)}_n } {\partial \epsilon_n } 
 = 
 - \frac{\delta f_n } {\tau} \,.
\end{align}
\end{widetext}
%%%%%%%%%%%
The form of the above equation reflects the following caveat of the relaxation time approximation of the Boltzmann equation \cite{arovas,soto}. In the absence of any external fields, or temperature and chemical potential gradients, Eq.~\eqref{eqkin4} reduces to $ \partial_t \delta f_n =  - \delta f_n / \tau $, giving the solution $ \delta f_n (t) = \delta f_n (t=0)  \, \exp (-t/\tau )$. This result is physically incorrect because the total particle number is a collisional invariant. Since this approximation lacks particle number conservation (or electric charge
conservation), this model cannot be used to determine the diffusion coefficient of the quasiparticles. However, this defect of the relaxation time approximation does not affect the validity of our conclusions regarding various transport coefficients, such as the electrical and the thermal conductivity tensors.

Let us consider a small region of the solid with a fixed volume $ d V $, centred around the position $\mathbf r $, where the temperature $T$ can be effectively taken to be constant. According to the first law of thermodynamics,
we have
\begin{align}
T \, d \mathcal S = d E -\mu \, d N\,,
\end{align}
where $ \mathcal S$ is the entropy, $E$ is the internal energy, and $N$ is the particle number.
We divide both the sides by $ d V $ in order to get the expressions in terms of the corresponding volume densities:
\begin{align}
\label{eqthermo}
T \,ds = d\epsilon  -\mu\, d{\mathcal N}\,,
\end{align}
where $T\, ds$ represents the change in the thermal energy (or heat) density.
The symbols $s$, $\epsilon$, and $ \mathcal N $ denote the entropy density, the internal energy density, and the particle number density, respectively. The rate at which the thermal energy appears in the region $dV$ is just equal to $T\, ds$, leading to the 
average thermal current density expression of
\begin{align}
\mathbf{J}^Q  = T\, \mathbf{J}^s\,,
\end{align}
where $ \mathbf{J}^s$ is the entropy current density. On the other hand, Eq.~\eqref{eqthermo} leads to the
relation
\begin{align}
\label{eqenpartcur}
\mathbf{J}^Q = \mathbf{J}^\epsilon  -\mu \, \mathbf{J}^{\mathcal N} \,,
\end{align}
when expressed in terms of the current densities, where $ \mathbf{J}^\epsilon $ is the energy current density and $ \mathbf{J}^{\mathcal N }$ is the particle number current density. If the quasiparticle number is not
conserved, ${\mathbf J}^{\mathcal N}$ is not well-defined, but $\mu = 0$.
For carriers with a conserved charge $ \mathcal Q $ (for example, electrons with charge $ - \,e$), the particle current implies that $ {\mathbf J}
 =  {\mathcal Q} \,{\mathbf J}^{\mathcal N}$.
%%%%%%%%%%%%%%%%%%%%%%%%%%%%%%%
For the conserved case, by definition, we have
\begin{align}
\mathbf{J}^{\mathcal N} & = \sum_n g_n  \int_{\text{ BZ} }
\frac{ d^d \mathbf k}{(2\, \pi)^d } \,
\delta f_n( \mathbf r , \mathbf k, t)
\, \boldsymbol{v}_n(\mathbf k) \text{ and} \nn
%%%%%%%%%%%%
\mathbf{J}^{\epsilon} & = \sum_n g_n  \int_{\text{ BZ} }
\frac{ d^d \mathbf k}{(2\, \pi)^d } \,\epsilon_n\,
\delta f_n( \mathbf r , \mathbf k, t)
\, \boldsymbol{v}_n(\mathbf k)\,,
\end{align}
where the subscript ``BZ'' in each integral sign indicates that the integral has to be performed over the first Brillouin zone.

From the above discussions, we find that the average electrical and thermal currents in the system are given by
\begin{align}
\label{eqcur}
{\mathbf J}
& =  {\mathcal Q} \,{\mathbf J}^{\mathcal N}
=  {\mathcal Q} \, \sum_n g_n  \int_{\text{ BZ} }
\frac{ d^d \mathbf k}{(2\, \pi)^d } \,
\delta f_n( \mathbf r , \mathbf k, t)
\, \boldsymbol{v}_n(\mathbf k)   \nonumber \\
%%%%%%%%%%%%%%%%
\text{and } \mathbf{J}^Q
& = \sum_n g_n  \int_{\text{BZ} }
\frac{ d^d \mathbf k}{(2\, \pi)^d } 
\left( \epsilon_n - \mu \right) \delta f_n( \mathbf r , \mathbf k, t)
\, \boldsymbol{v}_n(\mathbf k)\,,
\end{align}
%%%%%%%%%%%%%%%
respectively. The response matrix, which relates
the resulting generalized currents to the driving forces, can
be expressed as
\begin{align}
\label{eqcur1}
\begin{pmatrix}
 J_\alpha \vspace{0.2 cm} \\
{J}^Q_\alpha 
\end{pmatrix} & = 
 \sum \limits_\gamma
\begin{pmatrix}
L_{\alpha \gamma }^{11} & L_{\alpha \gamma }^{12} 
 \vspace{0.2 cm}  \\
L_{\alpha \gamma }^{21} & L_{\alpha \gamma }^{22}
\end{pmatrix}
%%%%%%%%%%%%%
\begin{pmatrix}
\mathcal{E}_\gamma 
 \vspace{0.2 cm}  \\
- { \partial_{r^\gamma} T } 
\end{pmatrix} ,
\end{align}
where the subscripts $ \lbrace \alpha, \gamma \rbrace  \in [1, d]$ indicate the Cartesian components of the current vectors and the transport tensors in $d$-dimensions. This gives us the \textit{Onsager matrix} of the transport coefficients.

For the case when $\mathbf E $, $\grad_{\mathbf r } \mu $, and $\grad_{\mathbf r } T$ are time-independent, with no magnetic field applied (i.e., $\mathbf B = 0$), Eq.~\eqref{eqkin4} reduces to
\begin{align}
\label{eqkin41}
& \partial_t\delta f_n
+ {\boldsymbol v}_n \cdot 
{\nabla}_{\mathbf r} \,\delta f_n 
+
{\boldsymbol v}_n \cdot
\left( 
\frac{ \mu - \epsilon_n} {T}  \, \nabla_{\mathbf r} T 
+ {\mathcal Q}\, \boldsymbol{\mathcal E}
\right )
 \frac{\partial  f^{(0)}_n } {\partial \epsilon_n } 
\nn &  = 
 - \frac{\delta f_n } {\tau} \,.
\end{align}
We can assume the solution $\delta f_n $ not to have any explicit time dependence, since the applied fields and gradients are time-independent, leading to $\partial_t\delta f_n = 0 $. Furthermore, we note that the inhomogeneous term in Eq.~\eqref{eqkin41} involves $\boldsymbol{\mathcal E} $ and $\nabla_{\mathbf r}  T $, implying that $\delta f_n $ is proportional to these quantities. Since we are forced to consider the external fields to be slowly varying in space, ${\nabla}_{\mathbf r} \,\delta f_n $ is second order in the smallness parameter which might be used to parametrize the smallness of $ |\boldsymbol{\mathcal E} |$ and $ |\nabla_{\mathbf r}  T | $. Therefore, to the leading lowest order in this smallness parameter,
the so-called \textit{linearized Boltzmann equation} is obtained as
%%%%%%%%%%%%%%%%%%
\begin{align}
\label{eqkin5}
\delta f_n =  \tau \, {\boldsymbol v}_n \cdot
\left( 
\frac{ \epsilon_n - \mu} {T}  \, \nabla_{\mathbf r} T 
- {\mathcal Q} \, \boldsymbol{\mathcal E}
\right )
 \frac{\partial  f^{(0)}_n } {\partial \epsilon_n } 
\end{align}
for Eq.~\eqref{eqkin4}.

Plugging in the results from Eq.~\eqref{eqkin5} in Eq.~\eqref{eqcur}, and setting
$\mathcal Q = -e $ (which is the charge of an electron), we get
\begin{align}
\label{eqresultsL}
& L _{\alpha \gamma }^{11} 
= - e^2 \, \sum_n g_n  \int_{\text{ BZ} }
\frac{ d^d \mathbf k}{(2\, \pi)^d } \,
\tau \, \frac{\partial  f^{(0)}_n } {\partial \epsilon_n } \,
\, { {v}_n }_{\alpha}  \, { {v}_n }_{\gamma} \,,\nn
%%%%%%%%%%%%%%%%%%%%%%%%%%%%%
& L _{\alpha \gamma }^{21}  =
T \, L _{\alpha \gamma }^{12}
\nn &
 =   e \, \sum_n g_n  \int_{\text{ BZ} }
\frac{ d^d \mathbf k}{(2\, \pi)^d } \,
\tau \, \frac{\partial  f^{(0)}_n } {\partial \epsilon_n } \,
\, { {v}_n }_{\alpha}  \, { {v}_n }_{\gamma} \left( \epsilon_n -\mu \right),\nn
%%%%%%%%%%%%%%%%%%
& L _{\alpha \gamma }^{22} 
 = - \sum_n g_n  \int_{\text{ BZ} }
\frac{ d^d \mathbf k}{(2\, \pi)^d } \,
\tau \, \frac{\partial  f^{(0)}_n } {\partial \epsilon_n } \,
\, { {v}_n }_{\alpha}  \, { {v}_n }_{\gamma} 
\frac{\left( \epsilon_n -\mu \right)^2}
{T}.
\end{align}
%%%%%%%%%%%%%%%%%%%%%%%%%%%%%%%%%
%%%%%%%%%%%%%%%%%%
We note that the relation $ L _{\alpha \gamma }^{12} = T \, L _{\alpha \gamma }^{21} $ (between the off-diagonal transport coefficient tensors) is a generic property that holds for any pair of cross transport coefficients and, as shown by Onsager, is a consequence of microscopic reversibility \cite{onsager}.

Using the condensed symbol
\begin{align}
\label{eq:transcoeff}
{\mathcal L}_{\alpha \gamma }^{ (\zeta) } & 
 =  - e^2 \sum_n g_n  \int_{\text{ BZ} }
\frac{ d^d \mathbf k}{(2\, \pi)^d } \,
\tau \, \frac{\partial  f^{(0)}_n } {\partial \epsilon_n } \,
\, { {v}_n }_{\alpha}  \, { {v}_n }_{\gamma} \left( \epsilon_n -\mu \right)^\zeta,
\end{align}
where $\zeta \in \lbrace 0, 1, 2 \rbrace $,
the thermoelectric transport coefficient tensors are given by
\begin{align}
& L_{\alpha \gamma }^{11} =  \mathcal{L}_{\alpha }^{(0)} \,, \quad
L_{\alpha \gamma }^{21} =  T\,  L_{\alpha \gamma }^{12}
= \frac{-\mathcal{L}_{\alpha \gamma}^{(1)}} {e}\,, 
\nn &
L_{\alpha \gamma }^{22} = \frac{\mathcal{L}_{\alpha \gamma}^{(2)}
}
{e^2 \, T }\,.
\end{align}
%%%%%%%%%%%%%%%%%%%%%%%%%%%%%%5
The thermal conductivity is measured under the conditions when there
is no electric current. To obtain this, it is convenient to transform
Eq.~\eqref{eqcur1} into a more convenient form as follows \cite{arovas,mermin}:
\begin{align}
\label{eqcur2}
\begin{pmatrix}
 \mathcal{E}_\alpha \vspace{0.2 cm} \\
{J}^Q_\alpha 
\end{pmatrix} & = \sum \limits_\gamma
\begin{pmatrix}
\rho_{\alpha \gamma } & S_{\alpha \gamma }
 \vspace{0.2 cm}  \\
\Pi_{\alpha \gamma } & - \kappa_{\alpha \gamma }
\end{pmatrix}
%%%%%%%%%%%%%
\begin{pmatrix}
J_\gamma 
 \vspace{0.2 cm}  \\
 \partial_{r^\gamma} T 
\end{pmatrix} \,,
\end{align}
where
\\(1) $\rho $ is the resistivity tensor, such that $  \mathcal{E}_\alpha = \rho_{\alpha \gamma } \, J_\gamma $ under the condition $\nabla_{\mathbf r} T = 0$;
\\(2) $S$ is the thermopower tensor (also known as the Seebeck coefficient), such that $  \mathcal{E}_\alpha = S_{\alpha \gamma } \, \partial_{r^\gamma} T $ under the condition $\mathbf J= 0$;
\\(3) $\Pi$ is the Peltier coefficient, such that $ {J}^Q_\alpha = \Pi_{\alpha \gamma } \, J_\gamma $ under the condition $\nabla_{\mathbf r} T = 0$;
\\(4) $\kappa$ is the thermal conductivity tensor, such that $ {J}^Q_\alpha = -\kappa_{\alpha \gamma } \, \partial_{r^\gamma} T$
under the condition $\mathbf J= 0$.
%%%%%%%%%%%%%%%%%%%%%%%%%%%%%%%%%%%%%%%
\\All the above ingredients allow us to formulate the final expressions for the electrical conductivity tensor $\sigma$, $S$, $\Pi$, and $\kappa$ as follows \cite{mermin,arovas}:
\begin{align}
\label{eq:kappa}
& \sigma_{\alpha \gamma}  =  L_{\alpha  \gamma}^{11} \,  ,
\quad
  S_{\alpha \gamma} = \sum \limits_{\alpha^\prime}
  \left(L^{11}\right)^{-1}_{\alpha  \alpha^\prime }
L_{\alpha^\prime \gamma}^{12} \, , \nn
%%%%%%%%%
& \Pi_{\alpha \gamma} = \sum \limits_{\alpha^\prime}
L_{\alpha  \alpha^\prime}^{21}   
\, \left(L^{11}\right)^{-1}_{\alpha^\prime \gamma} \,, \nn &
%%%%%%%%
 \kappa _{\alpha  \gamma} =
 L_{\alpha  \gamma}^{22}
- \sum \limits_{\alpha^\prime, \, \gamma^\prime }
L_{\alpha  \alpha^\prime }^{21}
\left(L^{11}\right)^{-1}_{ \alpha^\prime  \gamma^\prime }
L_{ \gamma^\prime \gamma }^{12}  \,.
\end{align}
For the time-dependent case with no magnetic field, we get the dc resistivity and the dc conductivity.

%%%%%%%%%%%%%%%%%%%%%%%%%%%%%%%%
\section{Semiclassical Boltzmann equation for 3D nodal-point semimetals with nonzero Berry curvature}
\label{secboltz2}

In the presence of a nontrivial topological charge in the bandstructure, the Boltzmann equation of Eq.~\eqref{eqkin3} gets modified. We specifically focus on 3D nodal-point semimetals with nonzero Chern numbers. The Nielson-Ninomiya theorem \cite{nielsen} imposes the condition that the nodes come in pairs, such that each pair carry Chern numbers which are equal in magnitude but opposite in signs. The sign of the monopole charge is often referred to as the chirality of the corresponding node.

Considering transport for a single node of chirality $\chi$ in a 3D nodal-point semimetal, Eq.~\eqref{eqkin21} has to be modified to \cite{Sundurum:1999,son13_chiral}
\begin{align}
\label{eqrkberry}
{\hbar}   \, \dot{\mathbf r} = 
\partial_{\mathbf k} \epsilon_n  (\mathbf k)
-
\hbar\, \dot{\mathbf k} \cross \mathbf{\Omega }_{\chi ,n}
\,,
\quad 
\hbar\,\dot{\mathbf k}
= {\mathcal Q} \left(\boldsymbol{\mathcal E} +  
\frac{ \dot{\mathbf r} \cross {\mathbf B} } {c}
\right ) ,
\end{align}
%%%%%%%%%%%%
where $\mathbf{\Omega }_{\chi ,n} (\mathbf k)$ is the Berry curvature of the node, which is a pseudovector
expressed by
\begin{align}
\mathbf{\Omega }_{\chi ,n} (\mathbf k) = 
i \left \langle  \nabla_{\mathbf k}
u^{\chi}_n  (\mathbf k) \right | \cross \left | \nabla_{\mathbf k}  u_n^{\chi}  (\mathbf k) \right \rangle .
\end{align} 
The Berry  curvature arises from the Berry phases generated by $|u_n^{\chi} (\mathbf k)\rangle$, where $\lbrace |u_n^{\chi} (\mathbf k)\rangle \rbrace $ denotes the set of orthonormal Bloch cell eigenstates for the one-particle Hamiltonian $H_\chi (\mathbf k)$. $H_\chi (\mathbf k)$ represents the low-energy effective description of the node with chirality $\chi$ and band energies $\lbrace \epsilon_n \rbrace $. It can be checked that $\mathbf{\Omega }_{\chi ,n} $  is proportional to $\chi$ and, hence, it has opposite signs for an energy band with index $n$ at the two nodes of opposite chiralities.
%%%%%%%%%%%%%%%%%%
Furthermore, in the presence of a nontrivial band topology, a weak nonquantizing magnetic field $\mathbf B$ necessitates the introduction of the shifted energy \cite{prl_niu,Sundurum:1999,xiao10_Berry}
 %%%%%%%%%%%%%%%%
 \begin{align}
 \tilde \epsilon_n(\mathbf k) = \epsilon_n (\mathbf k)
 - {\boldsymbol m}_{\chi, n}  (\mathbf k) \cdot {\mathbf B}(\mathbf r, t)\,, 
 \end{align}
 where
 \begin{align}
 {\boldsymbol m}_{\chi, n} (\mathbf k) 
 =  \frac{i\,\mathcal Q} {2\, \hbar }
 \left 
 \langle \nabla_{\mathbf k} u^{\chi}_n  (\mathbf k)  \right | \cross
 \left [ H_\chi(\mathbf k) - \epsilon_n (\mathbf k) \right ]
 \left | \nabla_{\mathbf k}  u_n^{\chi}  (\mathbf k)
 \right \rangle 
 %%%%%%%%%%%%%%%%%%%%%%%%
 \end{align}
is the orbital magnetic moment (OMM) generated by the Berry phase.
Similar to $\mathbf{\Omega }_{\chi ,n}$, the OMM is an intrinsic property of the band --- it depends only on the Bloch wavefunctions. In fact, under symmetry operations, ${\boldsymbol m}_{\chi, n}$ transforms exactly like the Berry curvature.
The OMM behaves exactly like the electron spin and, in the presence of a nonzero $\mathbf B$, it couples to the magnetic field through a Zeeman-like term --- this is the reason behind the shift in the energy eigenvalues.
The modifications necessitate redefining the Bloch velocity as
%UNITS of arovas lecture notes
 \begin{align}
 \tilde
 {\boldsymbol  v}_n =  
 \frac{1}{\hbar} \, \nabla_{\mathbf k} \tilde \epsilon_n ({\mathbf k })\,.
 \end{align}

 The two coupled equations in Eq.~\eqref{eqrkberry} can be solved to obtain
 \begin{align}
 \label{eqrkberry1}
 & \dot{\mathbf{r}} =  D_{\chi, n} 
 \left[ 
 {\tilde{\boldsymbol{v}}}_n
- \frac{\mathcal Q} {\hbar} \, \boldsymbol{\mathcal E} \cross \mathbf{\Omega}_{\chi , n}
-
 \frac{\mathcal Q}{c} \left(   \mathbf{\Omega }_{\chi ,n} \cdot 
 {\tilde{\boldsymbol{v}}}_n  \right)
 \mathbf{B}\right], \nn
 %%%%%%%%%%%%%%%%%%%%%%%%%%%%%%%%%%%
& \text{and } \hbar \, \mathbf{\dot{k}}  =  
 D_{\chi, n}  \,
 \mathcal Q 
 \left[  \boldsymbol{\mathcal E}
 +\frac{
 {\tilde{\boldsymbol{v}}}_n
 \cross \mathbf{B}}  {c} 
- \frac{\mathcal Q  } {c} \,
 \left ( \boldsymbol{\mathcal E}  \cdot {\mathbf B} \right )
 \mathbf{\Omega }_{\chi ,n} \right],
 \end{align}
%%%%%%%%%%%%%%%
where
\begin{align}
D_{\chi,n}^{-1} = 1- \frac {\mathcal Q} {\hbar\, c }
\, {\mathbf B} \cdot \mathbf{\Omega }_{\chi ,n}  \,.
\end{align} 
%%%%%%%%%%%%%%%
The physical significance of the function $D_{\chi,n} $ can be understood as follows.
Xiao \textit{et al.} \cite{prl_niu} observed that the Liouville’s theorem on the conservation of phase space volume element $ dV_p $ is violated by the Berry phase, when we consider the semiclassical dynamics of the Bloch electrons. This breakdown of the Liouville’s theorem is remedied by introducing a modified density of states in the phase space, such that the number of states in the volume element $D_{\chi,n}^{-1} \, dV_p$ remains conserved. In other words, based on the modifications in the presence of nonzero Berry phases, the classical phase-space probability density is now given by
\cite{son13_chiral,prl_niu,duval06_Berry,Son:2012}
\begin{align}
F_n (\mathbf r, \mathbf k, t) = D_{\chi,n}^{-1} (\mathbf r, \mathbf k) \, 
f_n( \mathbf r , \mathbf k, t) \,.
\end{align}
This implies that the probability conservation, in the absence of collisions, is equivalent to $F_n$ satisfying the continuity equation in the phase space, viz. $\frac{ d F_n} {dt} = 0 $.
% https://journals.aps.org/prb/pdf/10.1103/PhysRevB.90.165115

%%%%%%%%%%%%%%%%%%%%%%%%%%%%%
Incorporating all these ingredients, Eq.~\eqref{eqkin32} should be modified to
\cite{lundgren14_thermoelectric,das19_linear}
\begin{widetext}
\begin{align}
\label{eqkin33}
& D_{\chi, n}  \,
\left [ \partial_t  
+ \left \lbrace 
{\tilde{\boldsymbol{v}}}_n
-\frac{ \mathcal Q} {\hbar} \,\boldsymbol{\mathcal E} \cross \mathbf{\Omega}_{\chi, n}
-
\frac{\mathcal Q} {\hbar \, c} \left(   \mathbf{\Omega }_{\chi ,n} \cdot 
{\tilde{\boldsymbol{v}}}_n  \right)
\mathbf{B} \right \rbrace 
 \cdot \nabla_{\mathbf r} 
+ \frac{\mathcal Q} {\hbar}
\left(  \boldsymbol{\mathcal E}
+ \frac{ 
{\tilde{\boldsymbol{v}}}_n  \cross {\mathbf B} } {c}
\right) 
\cdot \nabla_{\mathbf k} 
%%%%%
- \frac{\mathcal Q^2 } {\hbar^2 \,c} \,
\left ( \boldsymbol{\mathcal E}  \cdot {\mathbf B} \right )
\mathbf{\Omega }_{\chi ,n}   \cdot \nabla_{\mathbf k} 
\right ] f_n
%%%%%%%%%%%%%%%%%%%%%%%%
\nn & \, =   
 \frac{ f^{(0)}_n (\mathbf r,\mathbf k)
 - f_n(\mathbf r,\mathbf k, t) }
 {\tau (\mathbf k)} \,.
\end{align}
\end{widetext}
We would like to point out that, here, $ f^{(0)}_n $ is a function of $ \tilde \epsilon_n $, i.e.,
$ f^{(0)}_n (\mathbf r,\mathbf k) 
 =  \left [ e^{ \beta (\mathbf r )
 \,\left
 \lbrace \tilde \epsilon_n (\mathbf k)
 -\mu (\mathbf r ) \right \rbrace  } + 1 \right]^{-1} $.
For the sake of simplicity, in the current analysis, we have have assumed that only the intranode scatterings are relevant, contributing to $\tau$, ignoring the internode processes. It is important to note that we have not assumed $|\mathbf B| $ to be very small.
To solve the above equation, we need to make an appropriate ansatz, as outlined in (a) Refs.~\cite{nandy_2017_chiral,das19_linear} for the planar Hall effect; and (b) Ref.~\cite{nandy_thermal_hall} for the planar thermal Hall effect. 

%%%%%%%%%%%%%%%%%%%%%%%%%%%%%%%%%%%%%%%%%%%%%%%%%%%%%%%%%
\begin{figure*}[t]
\includegraphics[width=0.5 \linewidth]{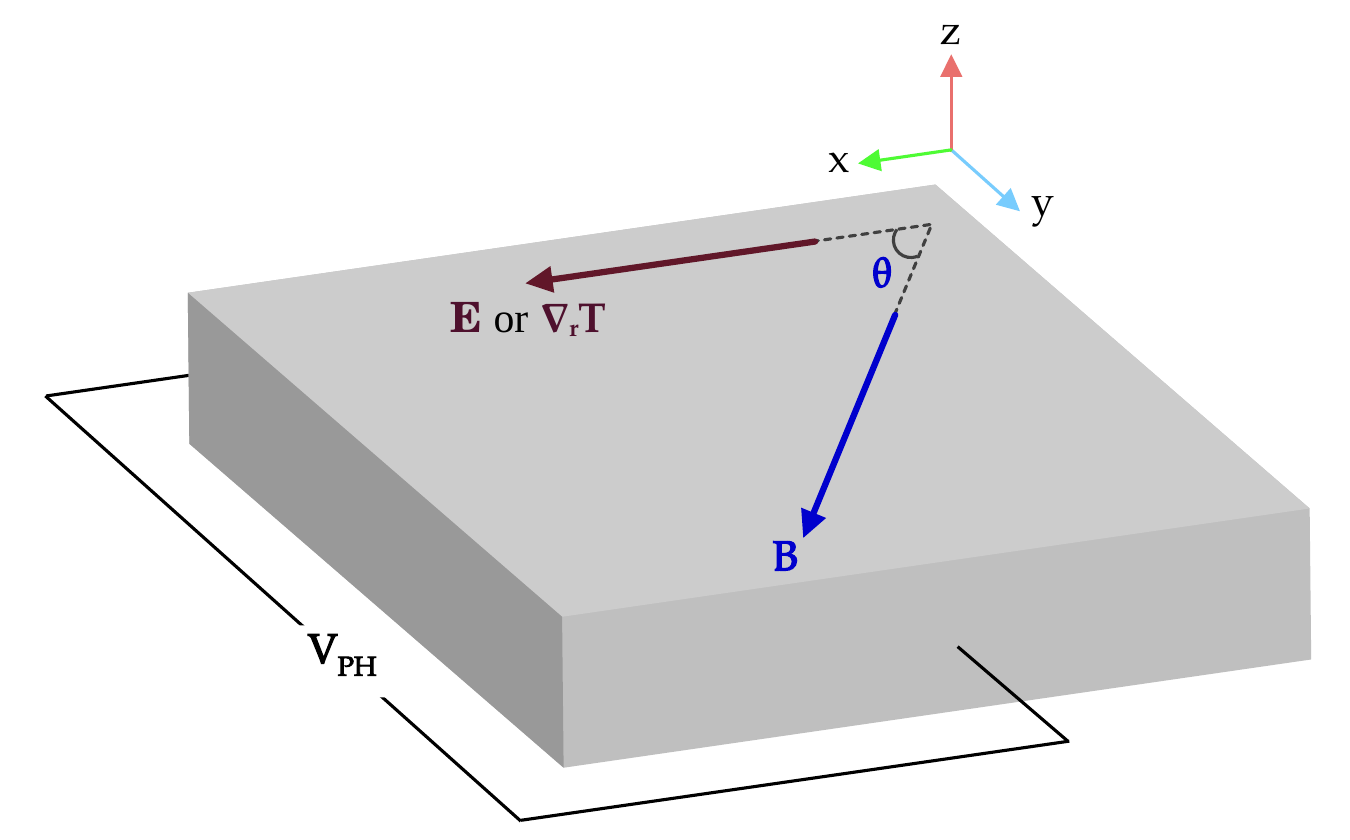}
\caption{\label{figsetup}
Schematics showing the planar Hall/thermal Hall experimental set-up, where the sample is subjected to an external electric field $ E\, {\mathbf{\hat x}} $/temperature gradient $\partial_x T\, {\mathbf{\hat x}}$. An external magnetic field $\mathbf B $ is applied such that it makes an angle $\theta $ with the existing electric field/temperature gradient. The resulting planar Hall/thermal Hall voltage, generated along the $y$-axis, is indicated by the symbol $V_{\rm PH}$.
}
\end{figure*}
%%%%%%%%%%%%%%%%%%%%

In order to obtain a solution to the full Boltzmann equation in- the time-independent scenarios, for small values of $\mathbf E$, $\nabla_{\mathbf r} \mu$, and $\nabla_{\mathbf r} T$, we assume a slight deviation $\delta  f_n(\mathbf r,\mathbf k)$ from the equilibrium distribution of the quasiparticles. It is assumed not to have any explicit time-dependence either, since the applied fields and gradients are time-independent. Hence, the non-equilibrium time-independent distribution function is assumed to be of the form shown in Eq.~\eqref{eqpertf}.
%%%%%%%%%%%%%%%%%%%%%%%%%%%%%%
The gradients of equilibrium distribution function $ f^{(0)}_n$ evaluate to
Eq.~\eqref{eqpertf1}.
%%%%%%%%%%%%%%%%%%%%%%%
In the following, we will consider a uniform chemical potential, such that $ \nabla_{\mathbf r} \mu =0 $, as we are mainly interested in the interplay of $\mathbf E $ (or $\nabla_{\mathbf r} T $) and $\mathbf B$.
We assume that all of the quantities (viz. $\mathbf E $, $\nabla_{\mathbf r} T$, and the resulting $\delta f_n$) are of the same order of smallness. The spatial gradient $\nabla_{\mathbf r}  f^{(0)}_n$ is parallel to
the thermal gradient $\nabla_{\mathbf r} T$, and we consider situations where $\mathbf E $ and  $\nabla_{\mathbf r} T$ are oriented along the same direction. Hence, the term $\mathcal Q \left (  {\mathbf E} \cross \mathbf{\Omega}^{\chi }_n \right) \cdot \nabla_{\mathbf r} f^{(0)}_n $ in Eq.~\eqref{eqkin33} gives zero.

The particle current density of Eq.~\eqref{eqenpartcur} is equal to  $ {\mathbf J}^\chi
 /  {\mathcal Q} $, when contributed by the single node, where $ {\mathbf J}^\chi $ has a circulating component in the form of the orbital magnetization current~\cite{prb101235430}
 \begin{align}
 \label{eqmagcur}
  {\mathbf J}^{\chi, \rm mag} 
 = 
 \sum_n g_n\, \nabla_{\mathbf r} \cross  \int
\frac{ d^3 \mathbf k}{(2\, \pi)^3 } \,
 {\boldsymbol m}_{\chi,n} \,
 f_n( \mathbf r , \mathbf k) \,.
 \end{align}
The overall contribution to the electrical and the thermal current densities from the concerned node are then captured by~\cite{prb101235430,nandy_2017_chiral,das19_linear}
% PHYSICAL REVIEW B 101, 235430 (2020)
% https://journals.aps.org/prb/pdf/10.1103/PhysRevB.101.235430 => Eq. 11
\begin{align}
\label{eqcur_ber}
 {\mathbf J}^\chi & =    
\sum_n g_n  \int
\frac{ d^3 \mathbf k}{(2\, \pi)^3 } 
{\mathcal Q} \, D_{\chi, n}^{-1}   \, \dot{\mathbf r}
\,  f_n( \mathbf r , \mathbf k)
 +  {\mathbf J}^{\chi, \rm mag} 
 \end{align}
and 
%%%%%%%%%%%%%%%555
\begin{align}
\mathbf{J}^{Q , \chi}
 & =    
\sum_n g_n  \int
\frac{ d^3 \mathbf k}{(2\, \pi)^3 } 
{\mathcal Q} \, D_{\chi, n}^{-1}   \, \dot{\mathbf r}
\left( \tilde \epsilon_n - \mu \right)  f_n( \mathbf r , \mathbf k)
\nn
& \quad + \sum_n g_n\, \nabla_{\mathbf r} \cross  \int
\frac{ d^3 \mathbf k}{(2\, \pi)^3 } \,
 {\boldsymbol m}_{\chi,n}  \left( \tilde \epsilon_n - \mu \right) 
\, f_n( \mathbf r , \mathbf k) 
\nn
& \quad + \sum_n g_n\, \nabla_{\mathbf r} \mu\cross  \int
\frac{ d^3 \mathbf k}{(2\, \pi)^3 } \,
   {\boldsymbol m}_{\chi,n}
\, f_n( \mathbf r , \mathbf k) 
 \,,
\end{align}
respectively. Note that the quantity $ {\boldsymbol m}_{\chi,n}^Q = \left( \tilde \epsilon_n - \mu \right)  {\boldsymbol m}_{\chi,n}$ is known as the thermal magnetic moment \cite{tao,Zhang_2016}.
%Clearly, the magnetic moment terms do not contribute for a uniform system. However, in presence of inhomogeneity, the local current densities are composed of the transport and magnetization parts.
The part of the response originating from the node with chirality $\chi$ be denoted by ${\mathbf J}^\chi$ and ${\mathbf J}^{Q,\chi}$. The response matrix, defined by the set $ [L_{\chi}] \equiv \lbrace L^{11}_{\alpha  \gamma}, \,  L^{12}_{\alpha  \gamma}, \, L^{21}_{\alpha  \gamma} , \, L^{22}_{\alpha  \gamma} \rbrace
\equiv  \lbrace \sigma^{\chi}_{\alpha  \gamma}, \,  \Upsilon^\chi_{\alpha  \gamma}, \, T\,\Upsilon^\chi_{\alpha  \gamma} , \, \ell^\chi_{\alpha  \gamma} \rbrace $, then represents the transport coefficients (analogous to the Berry-phase-independent case), relating the transport current parts to the driving electric potential gradient (cf. Refs.~\cite{tao, prb101235430}), defining the Onsager matrix [shown in Eq.~\eqref{eqcur1}] for inhomogeneous materials. It has been argued that the magnetization-sourced parts of the currents cannot be measured in conventional transport experiments \cite{cooper_omm}. Therefore, we will only be interested in the so-called transport currents, defined by
\begin{align}
& {\mathbf{\tilde J}}^\chi  =    
{\mathbf J}^\chi -   {\mathbf J}^{\chi, \rm mag} \,, \nn
& \mathbf{\tilde J}^{Q , \chi}
\nn & =    
\mathbf{J}^{Q , \chi}
 - \sum_n g_n\, \nabla_{\mathbf r} \cross  \int
\frac{ d^3 \mathbf k}{(2\, \pi)^3 } \,
 {\boldsymbol m}_{\chi,n}  \left( \tilde \epsilon_n - \mu \right) 
\, f_n( \mathbf r , \mathbf k) \,.
\end{align}
%%%%%%%%%%%%%%%%%%%%%%%

Henceforth, let us assume uniform materials (with a constant $\mu$) so that we do not have to worry about the modifications introduced by inhomogeneity. Furthermore, we consider only the case of a momentum-independent $\tau$, for the sake of simplicity. Let us investigate an experimental set-up with a semimetal subjected to a static external electric field $ \mathbf E $ (caused by an external potential gradient) along the $x$-axis, and a time-independent uniform magnetic field $ \mathbf B $ along the $y$-axis. Since $ \mathbf B $ is perpendicular to $ \mathbf E $, a potential difference (known as the Hall voltage) is generated along the $z$-axis. This phenomenon is the well-known Hall effect. 
However, if $ \mathbf B $ is applied such that it makes an angle $ \theta $ with $ \mathbf E $, where $  \theta  \neq \pi/2$, then, although the conventional Hall voltage induced from the Lorentz force is zero along the $y$-axis, transport involving a semimetal node with a nonzero topological charge gives rise to a voltage difference along this direction. This is known as the planar Hall effect (PHE), arising due to the chiral anomaly~\cite{son13_chiral,burkov17_giant,li_nmr17,nandy_2017_chiral,nandy18_Berry,2021nag_nandy,ips-serena}.
%%%%%%%
The associated transport coefficients, related to this voltage, are referred to as the planar Hall conductivity (PHC) and the longitudinal magnetoconductivity (LMC), which depend on the value of $ \theta $.
%%%%%%%%%%%%%
In an analogous set-up, we observe the planar thermal Hall effect (PTHE) [also referred to as the planar Nernst effect (PNE)], where, instead of an external electric field, a temperature gradient $ \mathbf \nabla_{\mathbf r} T$ is applied along the $x$-axis. The temperature gradient then induces a potential difference along the $y$-axis due to the chiral anomaly~\cite{girish1,ips-serena}. The associated transport coefficients are known as the longitudinal thermoelectric coefficient (LTEC) and the transverse thermoelectric coefficient (TTEC). The behaviour of these conductivity tensors has been extensively investigated in the literature~\cite{zhang16_linear,chen16_thermoelectric,das19_linear, das20_thermal,das22_nonlinear, pal22a_berry, pal22b_berry, fu22_thermoelectric, araki20_magnetic,mizuta14_contribution,onofre,ips-rahul-ph}.
To fix a coordinate system, we use the convention that the magnetic field is applied in the $xy$-plane, such that its components are given by $\mathbf B = B \left( \cos \theta \,{\mathbf{\hat x}} + \sin \theta \, {\mathbf{\hat y}}\right)$. The corresponding experimental set-up is schematically shown in Fig.~\ref{figsetup}.

To study the response in a PHE set-up [cf. Fig.~\ref{figsetup}], an electric field $\mathbf E = E \,  {\mathbf{\hat x}} $ is applied, making it coplanar with $\mathbf B$, and setting $\nabla_{\mathbf r} T  $ to zero. 
From the solutions obtained in Refs.~\cite{lundgren14_thermoelectric,nandy_2017_chiral,das19_linear}, and setting ${\mathcal Q} = - e $ (where $e$ is the magnitude of the charge of an electron) and $g_n= 1$ (ignoring the degeneracy due to electron's spin), we arrive at the following expression for the conductivity tensor:
\begin{widetext}
\begin{align}
\label{eqsigmatot}
& \sigma_{\alpha  \gamma}^\chi 
= \sigma_{\alpha  \gamma}^{\chi,\, \rm AHE}
+ \sigma_{\alpha  \gamma}^{\chi,\, \mathbf \Gamma}
+ \bar \sigma_{\alpha  \gamma}^\chi \,, \quad
%%%%%%%%
 \sigma_{\alpha  \gamma}^{\chi,\, \rm AHE}
 = -\frac{e^2} {\hbar } \,\epsilon_{ \alpha \gamma \lambda} 
 \int \frac{ d^3 \mathbf k}{(2\, \pi)^3 } \,
 \Omega_{\chi, n}^\lambda \,  f^{(0)}_n \,,\nn
%%%%%%%%%%%%%%%%
&
\bar \sigma_{\alpha  \gamma}^\chi 
=- e^2 \, \tau
\int \frac{ d^3 \mathbf k}{(2\, \pi)^3 } \, D_{\chi, n} 
\left [ {{\tilde{v}}}_{n, \alpha}  + \frac{e\, B_\alpha } {\hbar \, c} \left( 
{\tilde{\boldsymbol{v}}}_n \cdot \mathbf \Omega_{\chi, n} \right)
\right ]
\left [ {{\tilde{v}}}_{n, \gamma}  +\frac{e\,  B_\gamma  } {\hbar \, c} \left( 
{\tilde{\boldsymbol{v}}}_n \cdot \mathbf \Omega_{\chi, n} \right)
\right ]
\, \frac{\partial  f^{(0)}_n
} 
{\partial  \tilde \epsilon_n } \,,
\end{align}
%%%%%%%%%%%%%%%%%%%%%%%%
\end{widetext}
where $\sigma_{\alpha  \gamma}^{\chi,\, \rm AHE}$ represents the ``intrinsic anomalous'' Hall effect \cite{haldane,pallab_axionic,burkov_intrinsic_hall} (which is, evidently, completely independent of the scattering rate), $ \sigma_{\alpha  \gamma}^{\chi,\, \mathbf \Gamma}$ is the  Lorentz-force contribution to the conductivity, and $ \bar \sigma_{\alpha  \gamma}^\chi $ is the Berry-curvature-related coefficient.
%%%%%%%%%%%%%%%%%%%%%5
For a momentum-independent $\tau$, $ \sigma_{\alpha  \gamma}^{\chi,\, \mathbf \Gamma}$ is much smaller than the other terms~\cite{nandy_2017_chiral} and, hence, can be neglected. Furthermore, we will neglect $\sigma_{\alpha  \gamma}^{\chi,\, \rm AHE}$, as it leads to a zero contribution as far as the in-plane components of the response tensors are considered \cite{ips-ruiz}.

Investigating the response in a PTHE set-up  [cf. Fig.~\ref{figsetup} entails applying a temperature gradient $\nabla_{\mathbf r}  T = \partial_x T\, {\mathbf{\hat x}} $ coplanar with $\mathbf B $, with $\mathbf E$ set to zero. Using this set-up, we are interested in finding the form of the thermoelectric conductivity tensor ($ \Upsilon_{\alpha \gamma}^\chi$) for the same semimetallic node described above. This will allow us to determine the Peltier, Seebeck, and Nernst response coefficients. Using the solutions described in Refs.~\cite{lundgren14_thermoelectric,pal22b_berry,nandy_thermal_hall}, we get the expressions
%%%%%%%%%%%%%%%%%%%%%%%%%%%
\begin{widetext}
\begin{align}
\label{eqalphatot}
& \Upsilon^\chi_{\alpha  \gamma} 
=  \Upsilon_{\alpha  \gamma}^{\chi, \rm AHE}
+  \Upsilon_{\alpha  \gamma}^{\chi, \mathbf \Gamma}
+ {\bar \Upsilon}^\chi_{\alpha  \gamma} \,, \quad
\Upsilon_{\alpha  \gamma}^{\chi, \rm AHE}
 = 
 \frac{e} {\hbar } \,\epsilon_{ \alpha \gamma \lambda} 
 \int \frac{ d^3 \mathbf k}{(2\, \pi)^3 } \, \Omega_{\chi, n}^\lambda \, 
 \, \frac{\left( \tilde \epsilon_n - \mu \right)
 } {T} \, f^{(0)}_n \,,\nn
%%%%%%%%%%%%%%%%
&
{\bar \Upsilon}^\chi_{\alpha  \gamma}
= e \, \tau
\int \frac{ d^3 \mathbf k}{(2\, \pi)^3 } \, D_{\chi, n} 
\left [ 
{{\tilde{v}}}_{n, \alpha} +\frac{e\, B_\alpha}  {\hbar \, c} \left( 
{\tilde{\boldsymbol{v}}}_n \cdot \mathbf \Omega_{\chi, n} \right)
\right ]
\left [ {{\tilde{v}}}_{n, \gamma}  +\frac{e\,  B_\gamma } {\hbar \, c} \left( 
{\tilde{\boldsymbol{v}}}_n \cdot \mathbf \Omega_{\chi, n} \right)
\right ]
\, \frac{\left( \tilde \epsilon_s - \mu \right) } {T}
\, \frac{\partial  f^{(0)}_n } {\partial \tilde  \epsilon_n } \,.
\end{align}
\end{widetext}
%%%%%%%%%%%%%%%%%%%%%%%%
Analogous to the earlier case, $ \Upsilon_{\alpha  \gamma}^{\chi, \rm AHE}$ arises independent of an external magnetic field, $ \Upsilon_{\alpha  \gamma}^{\chi, \mathbf \Gamma}$ results from the Lorentz-force-like contributions, and $ \bar \Upsilon_{\alpha  \gamma}^{\chi}  $ is the Berry-curvature-related part. We ignore the first two contributions, because the sum of $  \Upsilon_{\alpha  \gamma}^{\chi, \rm AHE}$ from the two nodes of opposite chiralities gives zero, while $ \Upsilon_{\alpha  \gamma}^{\chi, \mathbf \Gamma}$ has a subleading contribution for a momentum-independent $\tau$.

Lastly, the dominant part for $ \ell^\chi_{\alpha  \gamma} $, after neglecting parts representing the intrinsic anomalous Hall and Lorentz-force-like contributions, is given by \cite{nandy_thermal_hall, pal22b_berry}
%%%%%%%%%%%%%%%%%%%%%%%%%%%
\begin{widetext}
\begin{align}
\bar \ell_{\alpha  \gamma}^\chi
=   - \, \tau 
\int \frac{ d^3 \mathbf k}{(2\, \pi)^3 } \, D_{\chi, n} 
\left [ 
{{\tilde{v}}}_{n, \alpha} +\frac{e\, B_\alpha}  {\hbar \, c} \left( 
{\tilde{\boldsymbol{v}}}_n \cdot \mathbf \Omega_{\chi, n} \right)
\right ]
\left [ {{\tilde{v}}}_{n, \gamma}  +\frac{e\,  B_\gamma } {\hbar \, c} \left( 
{\tilde{\boldsymbol{v}}}_n \cdot \mathbf \Omega_{\chi, n} \right)
\right ]
\, \frac{\left( \tilde \epsilon_s - \mu \right)^2 }
{T}
\, \frac{\partial  f^{(0)}_n } {\partial \tilde  \epsilon_n } \,.
\end{align}
\end{widetext}
%%%%%%%%%%%%%%%%%%%%%%%%

%%%%%%%%%%%%%%%%%%%%%%%%%%%%%%%%%%%%%%%%%%%%%%%%%%%%%%%%%
\begin{figure*}[t]
\subfigure[]{\includegraphics[width=0.22\linewidth]{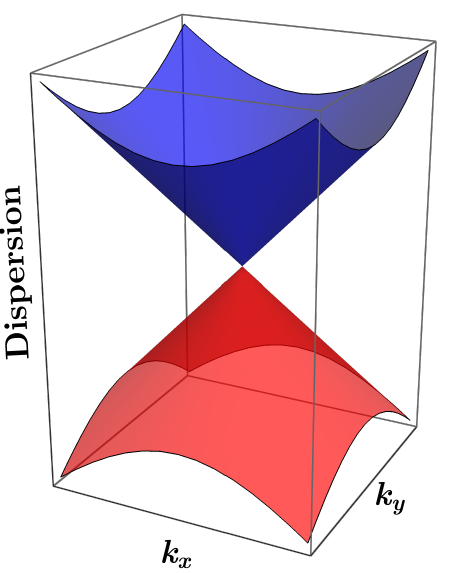}} \hspace{ 3 cm}
\subfigure[]{\includegraphics[width=0.22 \linewidth]{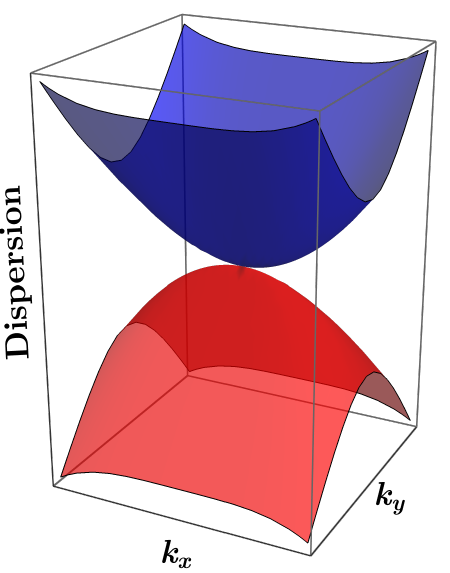}}
\caption{\label{figdis}
Schematic dispersion of (a) an isotropic Dirac semimetal (e.g., a single valley of graphene), and
(b) a semi-Dirac semimetal exhibiting anisotropy.}
\end{figure*}
%%%%%%%%%%%%%%%%%%%%

%%%%%%%%%%%%%%%%%%%%%%%%%%%%%%%%%%%%%%%%%%%%
\section{Thermoelectric response for 2D nodal phases assuming constant relaxation time}
\label{sec2d}

In this section, we will demonstrate the computation of the transport coefficients for some 2D semimetals having nodal points where two bands cross, and consider the scenario when the relaxation time $\tau $ can be approximated to be a constant. We therefore set $d=2$ in the integrals derived in Sec.~\ref{secboltz}, and label the 2D space by the Cartesian coordinates $x$ and $y$. In our notations, $1 \Leftrightarrow x $ and $ 2 \Leftrightarrow y$ for the Cartesian components of the vectors and the tensors. Here, we consider a single valley of graphene as an example of the isotropic dispersion case [cf. Fig.~\ref{figdis}(a)], and the anisotropic Weyl (also known as semi-Dirac) semimetal as an example of the anisotropic dispersion scenario [cf. Fig.~\ref{figdis}(b)].

%%%%%%%%%%%%%%%%%%%%%%%%%%%%%%%%%%%%%%%%%%%%%%%%%%%%%%%%%%%%%%%%%%%5
\subsection{Isotropic case: A single valley of graphene}
\label{secgraphene}

In this subsection, we compute the response matrix for graphene, which is a 2D isotropic Dirac  semimetal. The low-energy Hamiltonian describing a single valley is given by \cite{weiss}
\begin{align}
\label{diracham}
H_{D} (k_x, k_y)= \hbar \, v \left (  k_x  \, \sigma_x + k_y   \,\sigma_y \right ) ,
\end{align}
%%%%%%%%%%%%%%%
where $\lbrace \sigma_x, \sigma_y, \sigma_z \rbrace $ are the three Pauli matrices, $ \lbrace k_{x},k_{y} \rbrace $ are the momenta along the $x$ and $y$ directions, respectively, and $v$ is the Fermi velocity.

It is convenient to switch to the polar-coordinate parametrization $ k_x= k \,\cos \theta $ and $ k_y= k  \, \sin \theta$ (with $ k \equiv |\mathbf k| \, $), such that the energy eigenvalues are given by $ \epsilon_\pm ({\bf k} ) = \pm\,\hbar \, v\, k $.
The density of states (DOS) is given by 
\begin{align}
\label{eqdosg}
\rho_g (\epsilon) = \frac{ |\epsilon| } {2 \,\pi\,\hbar ^2 v^2} \,.
\end{align}
%%%%%%%
Using Eq.~\eqref{eqresultsL}, along with an energy and momentum independent scattering time $\tau$, we get
%%%%%%%%%%%%%%%%%%%%%%%
\begin{widetext}
\begin{align}
\label{free case}
\sigma_{\alpha \alpha }  
= \mathcal{L}_{ \alpha \alpha }^{(0)} 
 = 
\frac{\beta\,v^2\,e^2 \,\tau  }     { 8 \,\pi\, \hbar^2}
\int_0^\infty dk   \,k 
\left [
\text{sech}^2 \left(  \frac{\beta  \left( k + \mu \right ) } {2} \right )
+ \text{sech}^2 \left(  \frac{\beta  \left( k -\mu \right ) } {2} \right ) \right ]
%%%%%%%%%%%%%
= \frac{e^2\,v^2\, \tau \,
\ln \left( 2+ 2\,\cosh \left ( \beta \,\mu \right ) \right )}
{4\, \pi\, \hbar^2\,\beta }\,,
%%%%%%%%%%%%%%%%%%%%%%%%%
\end{align}
\begin{align}
& L_{\alpha \alpha }^{21}
  = \frac{-\mathcal{L}_{\alpha \alpha }^{(1)}} {e} 
\nn &= 
\frac{\beta\,v^2\,e  \,\tau}     {8\,\pi^2\,  \hbar^2}
\int_0^\infty dk \,k
\Big [
\mu \left \lbrace  \text{sech}^2 \left (  \frac{ \beta \left  ( k +\mu \right )  }  {2} \right )
+ \text{sech}^2 \left (  \frac{ \beta \left  ( k - \mu \right )  }  {2} \right )  \right \rbrace 
%%%%%
+  k \,  \left \lbrace  \text{sech}^2 \left (  \frac{ \beta \left  ( k + \mu \right )  }  {2} \right )
- \text{sech}^2 \left (  \frac{ \beta \left  ( k- \mu \right )  }  {2} \right )  \right \rbrace
\Big  ]\nn
%%%%%%%%%%%%%
& =  -\frac{v^2 \,e \,\tau}
{ \left(2\,\beta \, \pi \, \hbar  \right )^2 }
\left[ 
\beta \,\mu   \,\ln \left (  2+ 2\,\cosh \left ( \beta \,\mu \right ) \right  ) 
+ 2 \, {\text{Li}}_{2} (-e^{ \beta\,\mu} )  - 2 \, {\text{Li}}_{2}(-e^{ -\beta \,\mu} ) 
\right],
\end{align}
%%%%%%%%%%%%%%%%%%%%%%%%%%%%%%
\begin{align}
& T \, L_{\alpha \alpha}^{22 } 
= \frac{\mathcal{L}_{\alpha \alpha  }^{(2)}}    {e^2   } 
\nn &= 
\frac{\beta\,v^2\,\tau}     {8\,\pi\,  \hbar^2}
\int_0^\infty dk  \,k\,
\Big [
\text{sech}^2 \left (  \frac{ \beta \left  ( k +\mu \right )  }  {2} \right ) 
\left  ( k \,\epsilon_0 +\mu \right )^2
+ \text{sech}^2 \left (  \frac{ \beta \left  ( k- \mu \right )  }  {2} \right ) 
\left  (k-\mu \right )^2 
\Big  ]\nn
%%%%%%%%%%%%%
& =  \frac{v^2 \,\tau}
{4\,\pi\,\hbar^2}
\Big [  
\frac{4\,\mu 
	\left \lbrace {\text{Li}}_{2} (-e^{- \beta\,\mu} )  
	- {\text{Li}}_{2}(-e^{ \beta \,\mu} )  \right  \rbrace} {\beta} 
+ \frac{   6 \, {\text{Li}}_{3} (-e^{ \beta\,\mu} ) 
+  6 \, {\text{Li}}_{3} (-e^{ -\beta\,\mu} ) }   {\beta^2 }
- \mu^2  \,\ln \left ( 2+ 2\,\cosh \left ( \beta \,\mu \right ) \right )
\Big ]\,,
\end{align}
%%%%%%%%%%%%%%%
\end{widetext}
where
\begin{align}\text{Li}_\varsigma (z) =  \sum \limits_{ \ell = 1}^{\infty} \frac{z^\ell} { \ell^\varsigma } 
\end{align}
denotes the  polylogarithmic function of order $\varsigma$.
%%%%%%%%%%%%%%%%
At low temperatures [i.e., $\mu/(k_B T)\gg 1)$], the above results reduce to $ \sigma_{\alpha \alpha}= e^2 v^2 \tau\mu/(4\pi\hbar^2)$,
and
\begin{align}
\label{eqgrapheneL}
& 
L_{\alpha \alpha }^{21} = \frac{v^2\,e\,\tau}
{4\,\pi\,\hbar^2}\times \frac{\pi^2 \,(k_B \,T)^2}  {3}
\,, \nn
&  T\, L_{\alpha \alpha }^{22} =
\frac{v^2\,\tau}{4\,\pi\,\hbar^2}\times \frac{\mu\,\pi^2 \,(k_B \,T)^2 } {3}\,.
\end{align}
With these expressions, the Seebeck coefficients ($S$) at low temperatures are found to be 
\begin{align}
\label{eqthermozerob}
S_{xx}=\frac{L_{xx}^{21}}{T\sigma_{xx} }\simeq
-\frac{ k_B\,T} {3\,e\,\mu} \text{ and }
%%%%%%%%%%%
S_{yy}=\frac{L_{yy}^{21}} {T\sigma_{yy} }\simeq
-\frac{k_B\,T} {3\,e\,\mu} \,.
\end{align}
Similarly, we obtain the thermal conductivity to be
\begin{align}
& \kappa_{xx} = \kappa_{yy}
=\frac{v^2\,\tau}
{4\,\pi\,\hbar^2 \, T}
\left[\frac{\pi^2 \,\mu\, (k_B\,T)^2} {3}
-\frac{ \pi^4\,(k_B\,T)^4 } {9\,\mu}\right].
\end{align}
To the leading order in $T$, the longitudinal components of the thermal conductivity thus show a linear-in-temperature dependence.
%%%%%%%%%%%%%%%%%%%%%%%%%%%%%%%%%%%%%

\subsection{Anisotropic case: Semi-Dirac semimetal}

We consider a model of a 2D anisotropic semi-Dirac semimetal, captured by the Hamiltonian \cite{pardo,pardo2, banerjee, ips-kush, ips_cd} 
\begin{align}
\label{ham-continuum}
H_{\rm{semiD}}(k_x,k_y) =\frac{\hbar^2\, k_x^2}{2\,m}\,\sigma_x +\hbar \, v\, k_y \,\sigma_y ,
\end{align}
where $m$ is the effective mass along the $x$-axis and $v$ is the Fermi velocity along the $y$-axis. We will use $a= \frac{\hbar^2 }{2\,m}$ and
$b=\hbar \, v$ in the equations for simplifying the expressions. With these notations, the dispersion, resulting from Eq.~\eqref{ham-continuum}, is found to be $\epsilon_\pm ({\mathbf k}) = \pm\sqrt{a^2\,k_x^4+b^2\,k_y^2}$. This anisotropic nature of the spectrum invariably manifests itself in the transport coefficients for the system, which have been computed in Ref.~\cite{ips-kush}, by following the methods outlined in Ref.~\cite{park}. We review those calculations and results in the remainder of this subsection.

With the parametrization $ k_x= \text{sign}[\cos\theta] \left( \frac{   {\tilde r} \,|\cos\theta|} {a} \right )^{1/2}$ and $ k_y= \frac{   {\tilde r} \, \sin\theta}{b}$ for $  {\tilde r} \geq 0$, the energy eigenvalues take the simple form $\epsilon _\pm  ({\bf k})= \pm\,   {\tilde r}$. The Jacobian of this transformation is given by
\begin{align}
\label{eq:jacobian}
& \mathcal {J}(  {\tilde r},\theta) 
 = \left \vert \det
\begin{pmatrix}
\partial_  {\tilde r} k_x &  \partial_\theta k_x \vspace{0.2 cm}\\
\partial_  {\tilde r} k_y & \partial_\theta k_y
\end{pmatrix} \right \vert 
% %%%%%%%%%
\nn & =
\begin{vmatrix}
\frac{1} {2} \left( \frac{ |\cos\theta| } {a\,   {\tilde r}} \right )^{1/2}  \vspace{0.2 cm} & 
-\frac{\sin \theta} {2}   \left( \frac{  {\tilde r} } {a \, |\cos\theta|} \right )^{1/2} \\
\sin \theta &   {\tilde r} \cos \theta 
\end{vmatrix}
%%%%%%%%%%%
\nn & = \sqrt{\frac{  {\tilde r}} {4\,a \,b^2 \, |\cos\theta|}} \,.
\end{align} 
Let us apply this convenient parametrization for calculating the DOS at energy $\epsilon >0$, which is given by
\begin{align}
\rho_{sd} (\epsilon) & = \int \frac{d^2\vec k}{ \left( 2\, \pi \right)^2} \,
\delta \big( \epsilon -\epsilon_+({\bf k})  \big)
\nn & = \int_0^\infty d\tilde r 
\int_0^{2\,\pi}  \frac{ d\theta }
{ \left( 2\, \pi \right)^2} \, \mathcal {J}(  {\tilde r},\theta)\,\delta \big ( \epsilon -   {\tilde r} \big )
\nn & =  \int_0^{2\,\pi} \frac{ d\theta }{ \left( 2\, \pi \right)^2} 
\sqrt{\frac{\epsilon} 
{4\,a\,  b^2 \, |\cos\theta|}}
=\frac{ 10.4882} {8\,\pi^2 } \sqrt{\frac{\epsilon } {a\,  b^2}}\,.
\end{align}
Clearly, the DOS of the semi-Dirac semimetal differs from its isotropic counterpart, graphene, the latter featuring the DOS $\rho_ g (\epsilon)\sim |\epsilon|$. The effects of the anisotropic dispersion show up via the characteristic DOS, in addition to the different values of the components of the Fermi velocity along the two mutually perpendicular directions.

Using Eq.~\eqref{eqresultsL}, along with an energy and momentum independent scattering time $\tau$, we get
\begin{widetext}
\begin{align}\label{sigma}
\sigma _{x x} &=  \mathcal{L}_{x x }^{(0)}   
= 
	\frac{e^2\, \tau\,\sqrt{a}\,\beta} {8\,\pi^2\,\hbar^2\, b}
	\int_0^\infty d\tilde r \int_0^{2\,\pi} 
d\theta  \,  {\tilde r}^{3/2} \,|\cos \theta|^{5/2} 
	\left [\text{sech}^2 \left (  \frac{ \beta \left  (   {\tilde r}-\mu \right )  }  {2} \right )
	+ \text{sech}^2 \left (  \frac{ \beta \left  (   {\tilde r} + \mu \right )  }  {2} \right )  \right ]\nn
	%%%%%%%%%%%%%
	& = -\frac{2.16 \,e^2 \,\tau\sqrt{a}} {2\,\hbar^2\, b\,(\pi\,\beta)^{3/2}}
	\left[ \text{Li}_{3/2}(-e^{\beta\,\mu)} )  +  \text{Li}_{3/2}(-e^{-\beta\,\mu)} ) \right],
	%%%%%%%%%%%%%%%%%%%%%%%%%
\end{align}
\begin{align}
	\sigma _{ y y} 
&= \mathcal{L}_{y y}^{(0)} = 
	\frac{ e^2\,\tau\, b \,\beta}     { 32\,\pi^2\,\hbar^2\,\sqrt{a}}
\int_0^\infty  d {\tilde r} \int_0^{2\,\pi} d\theta  \,\sqrt{  {\tilde r} \, |\sec \theta| }\,\sin^2 \theta  
	\left [\text{sech}^2 \left (  \frac{ \beta \left  (   {\tilde r}-\mu \right )  }  {2} \right )
	+ \text{sech}^2 \left (  \frac{ \beta \left  (   {\tilde r} + \mu \right )  }  {2} \right )  \right ]\nn
%%%%%%%%%%%%%
& = -\frac{ 3.5\, e^2\,\tau\, b}   
{8\,\pi^{3/2} \, \hbar^2\,\sqrt{a\,\beta}}
	\left[ \text{Li}_{1/2}(-e^{\beta\,\mu} )  +  \text{Li}_{1/2}(-e^{-\beta\,\mu} ) \right].
\end{align}
%%%%%%%%%%%%%%%%%%%%%%%%%
\end{widetext} 
For $\mu/(k_B\,T)\gg1$, the above expressions reduce to
\begin{align}
& \sigma _{x x} = \frac{2.88\,e^2 \,\tau\,\sqrt{a}}
	 {2\,\pi^2\,\hbar^2 \, b }    \left [ \mu^{3/2}
+ \frac{\pi^2  \, (k_B\,T)^2 }   {8\,\sqrt{\mu}} \right ] ,\nn &
%%%%
\sigma _{y y} = \frac{7\,e^2\,\tau\, b}    
 { 8\,\pi^2\,\hbar^2\,\sqrt{a}}
 \left [ \sqrt{\mu}
 -\frac{\pi^2 \, (k_B\,T)^2} {24\,\mu^{3/2}}  \right ].
	\label{eq:sigmalowtemp}
	\end{align}
Evidently, the low-temperature longitudinal dc-conductivity components are direction-dependent, as expected from the fact that the components of the group velocity differ in the $x$- and $y$-directions. This is in contrast with the isotropic case (e.g., a single valley of graphene), where we have $\sigma_{xx}=\sigma_{yy}\sim \mu$.

The thermoelectric coefficients are obtained in a similar fashion, taking the forms as shown below:
\begin{widetext}
\begin{align}
\label{Lxx21}
L_{x x}^{21} &=
\frac{e \,\tau\,\sqrt{a}\,\beta} {8\,\pi^2\,\hbar^2\, b}
	\int_0^\infty  d {\tilde r} \int_0^{2\,\pi} 
	d\theta  \,  {\tilde r}^{3/2} \,|\cos \theta|^{5/2} 
	\Bigg [
	\mu \left \lbrace  \text{sech}^2 \left (  \frac{ \beta \left  (   {\tilde r} +\mu \right )  }  {2} \right )
	+ \text{sech}^2 \left (  \frac{ \beta \left  (   {\tilde r} - \mu \right )  }  {2} \right )  \right \rbrace \nn
	& \hspace{ 7 cm} 
+    {\tilde r}  \left \lbrace  \text{sech}^2 \left (  \frac{ \beta \left  (   {\tilde r} + \mu \right )  }  {2} \right )
	- \text{sech}^2 \left (  \frac{ \beta \left  (   {\tilde r} - \mu \right )  }  {2} \right )  \right \rbrace
	\Bigg  ]\nn
%%%%%%%%%%%%%
& =  -\frac{2.16\, e\, \tau\,\sqrt{a}} {2\hbar^2 \,b\,(\pi \, \beta)^{3/2}}
\left[ 
	\mu \left \lbrace \text{Li}_{3/2}(-e^{-\beta\,\mu)} )  +  \text{Li}_{3/2}(-e^{\beta\,\mu)} )
	\right \rbrace
	+ \frac{5}{2\,\beta} \left \lbrace \text{Li}_{5/2}(-e^{-\beta\,\mu)} )  -  \text{Li}_{5/2}(-e^{\beta\,\mu)} )
	\right \rbrace \right],
	\end{align}
%%%%%%%%%%%%%%%
\begin{align}
	L_{yy}^{21}&= \frac{ e\,\tau \,b \,\beta}     { 32\,\pi^2\,\hbar^2\,\sqrt{a}}
	\int_0^\infty  d {\tilde r} \int_0^{2\,\pi} d\theta  \,\sqrt{  {\tilde r} \, |\sec \theta| }\,\sin^2 \theta  
\Bigg [
	\mu \left \lbrace  \text{sech}^2 \left (  \frac{ \beta \left  (   {\tilde r} +\mu \right )  }  {2} \right )
	+ \text{sech}^2 \left (  \frac{ \beta \left  (   {\tilde r} - \mu \right )  }  {2} \right )  \right \rbrace \nn
	& \hspace{ 8.5 cm} +    {\tilde r}  \left \lbrace  \text{sech}^2 \left (  \frac{ \beta \left  (   {\tilde r} + \mu \right )  }  {2} \right )
	- \text{sech}^2 \left (  \frac{ \beta \left  (   {\tilde r} - \mu \right )  }  {2} \right )  \right \rbrace
\Bigg  ]\nn
%%%%%%%%%%%%%
	& = -\frac{ 3.5\, e\,\tau\, b}     { 8\,\pi^{3/2}\,\hbar^2\,\sqrt{ a   \,  \beta}}
	\left[ \mu \left \lbrace \text{Li}_{1/2}(-e^{-\beta\,\mu)} )  +  \text{Li}_{1/2}(-e^{\beta\,\mu)} )
	\right \rbrace
	+ \frac{3 }{2\,\beta} \left \lbrace \text{Li}_{3/2}(-e^{-\beta\,\mu)} )  -  \text{Li}_{3/2}(-e^{\beta\,\mu)} )
	\right \rbrace \right] ,
	\end{align}
\end{widetext}
%%%%%%%%%%%%%%%%%%%%%%%%%%%%%%%%%%%%%%%%%%%%%%%%%%%%%%%%%
At low temperatures [i.e., $\mu/(k_B\,T)\gg 1$], we obtain
\begin{align}
& L _{x x}^{21}= -\frac{2.88\,e \,\tau\,\sqrt{a}} {2\,\pi^2\,\hbar^2\, b} 
\left [ \frac{\pi^2\, \mu^{1/2}}{2}(k_B\,T)^2\right ],\nn
& ·L_{y y}^{21}= -\frac{7\,e\, \tau\, b} { 8\,\pi^2\,\hbar^2\,\sqrt{a}}
\left[ \frac{\pi^2}{6\,\mu^{1/2}}(k_B\,T)^2\right ].
\end{align}
As for the case of $\sigma$ analysed above, the low-temperature behavior of the longitudinal thermoelctric coefficients has a distinct dependence on the chemical potential, depending on the axis under consideration. In contrast, the behaviour $L_{xx}^{21 } =L_{yy}^{21} \sim  ( k_B \, T )^2 $ [cf. Eq.~\eqref{eqgrapheneL}] for graphene is independent of the chemical potential. Although the individual coefficients in the semi-Dirac semimetal differ from those in graphene, the Mott relation still prevails at low temperatures, as seen from
\begin{align}
\label{eqthermozerob2}
S_{xx}=\frac{L_{xx}^{21}}{T\sigma_{xx} }\simeq
-\frac{ k_B\,T} {2\,e\,\mu} \text{ and }
%%%%%%%%%%%
S_{yy}=\frac{L_{yy}^{21}} {T\sigma_{yy} }\simeq
-\frac{k_B\,T} {6\,e\,\mu} \,.
\end{align}
%%%%%%%%%%%%%%%%%%%%%%%%%%
These results show that, at low temeperatures and for momentum-independent relaxation time, there is no violation of the usual Mott relation. However, as will be evident in Sec.~\ref{secdisorder}, a momentum-dependent relaxation time may lead to deviations \cite{Sarma2009} from the the Mott relations.

To investigate the electronic contribution to the thermal conductivity  $\kappa$, we next compute
\begin{widetext}
%%%%%%%%%%%%%%%%%%%%%%
\begin{align}
T\, L_{x x}^{22 } & = \frac{\mathcal{L}_{xx}^{(2)}} {e^2  }\nonumber\\
& = 
\frac{\tau\sqrt{a} \,\beta} {8\, \pi^2 \,\hbar^2 \, b }
\int_0^\infty  d {\tilde r} \int_0^{2\,\pi} d\theta  \,  {\tilde r}^{3/2} \,|\cos \theta|^{5/2} 
	\Big [
	\text{sech}^2 \left (  \frac{ \beta \left  (   {\tilde r} +\mu \right )  }  {2} \right ) \left  (   {\tilde r} +\mu \right )^2
	+ \text{sech}^2 \left (  \frac{ \beta \left  (   {\tilde r} - \mu \right )  }  {2} \right ) \left  (   {\tilde r} - \mu \right )^2 
	\Big  ]\nn
%%%%%%%%%%%%%
& =  -\frac{2.16 \, \tau \,\sqrt{a}} 
{2 \,\hbar^2 \,b \,(\pi \, \beta)^{3/2}}
\Big [ 
	\mu^2 \left \lbrace \text{Li}_{3/2}(-e^{-\beta\,\mu)} )  +  \text{Li}_{3/2}(-e^{\beta\,\mu)} )
	\right \rbrace
	+ \frac{5 \,\mu}{\beta} \left \lbrace \text{Li}_{5/2}(-e^{-\beta\,\mu)} )  -  \text{Li}_{5/2}(-e^{\beta\,\mu)} )
	\right \rbrace
	\nn & \hspace{ 3.7 cm }+ \frac{ 35 }{4\,\beta^2 } \left \lbrace \text{Li}_{7/2}(-e^{-\beta\,\mu)} ) 
 +  \text{Li}_{7/2}(-e^{\beta\,\mu)} )
	\right \rbrace \Big ]
\end{align}
%%%%%%%%%%%%%%%
and
\begin{align}
T	\, L_{yy}^{22} &= \frac{\mathcal{L}_{ y y}^{(2)}} {e^2} \nonumber\\
	&=
	\frac{\tau \, b \, \beta}    
{ 32 \, \pi^2 \, \hbar^2 \, \sqrt{a}}
	\int_0^\infty  d {\tilde r} 
\int_0^{2\,\pi} d\theta  \,\sqrt{  {\tilde r} \, |\sec \theta| } \sin^2 \theta  
	\Big [
	\text{sech}^2 \left (  \frac{ \beta \left  (   {\tilde r} + \mu \right )  }  {2} \right ) \left  (   {\tilde r} +\mu \right )^2
	+ \text{sech}^2 \left (  \frac{ \beta \left  (   {\tilde r} - \mu \right )  }  {2} \right ) \left  (   {\tilde r} - \mu \right )^2 
	\Big  ]\nn
	%%%%%%%%%%%%%
& = -\frac{ 3.5 \,e^2 \,\tau \,  b}     
	{ 8  \, \pi^{3/2}\hbar^2 \, \sqrt{ a   \,  \beta}}
	\Big [ 
	\mu^2 \left \lbrace \text{Li}_{1/2}(-e^{-\beta\,\mu)} )  +  \text{Li}_{1/2}(-e^{\beta\,\mu)} )
	\right \rbrace
	+ \frac{ 3 \,\mu}{\beta} \left \lbrace \text{Li}_{ 3/2}(-e^{-\beta\,\mu)} )  -  \text{Li}_{ 3/2}(-e^{\beta\,\mu)} )
	\right \rbrace
	\nn & \hspace{ 3.7 cm }+ \frac{ 15 }{4\,\beta^2 } \left \lbrace \text{Li}_{ 5/2}(-e^{-\beta\,\mu)} )  +  \text{Li}_{ 5/2}(-e^{\beta\,\mu)} )
	\right \rbrace \Big ] .
\end{align}
%%%%%%%%%%%%%%%%%%%%%%
At low temperatures [i.e., $\mu/(k_B\,T)\gg 1$], we obtain
\begin{align}
T\, L _{x x}^{22}
&= \frac{2.88\,\tau \,\sqrt{a}} 
{2 \,\pi^2\,\hbar^2\, b }
\ \left[ \frac{\pi^2\, \mu^{3/2}}{3}(k_B\,T)^2
+ \frac{7\,\pi^4}{40\,\mu^{1/2}}(k_B\,T)^4 \right ],\nn
T\, L_{y y}^{22}
&= \frac{7\,\tau \,b}     { 8\,\pi^2\,\hbar^2\,\sqrt{a}}
\left [\frac{\pi^2\, \mu^{1/2}}{3}(k_B\,T)^2-\frac{7\,\pi^4}
{120\,\mu^{3/2}}(k_B\,T)^4 \right ].
\label{eq:LLlowtemp}
\end{align}
\end{widetext}
%%%%%%%%%%%%%%%%%%%%%%%%%%%%
Together with Eqs.~(\ref{eq:LLlowtemp}) and (\ref{eq:sigmalowtemp}), we recover the Wiedemann-Franz law, $L_{\alpha\alpha}^{22}=\frac{\pi^2\, k_B^2\, T}{3\,e^2}\sigma_{\alpha\alpha} $, up to leading order in $(k_B\,T) $.   Finally, using Eq.~\eqref{eq:kappa}, we get
%%%%%%%%%%%%%%%%5
\begin{align}
& \kappa_{xx}  
=\frac{2.88\, \tau\sqrt{a}} {2\,\pi^2\,\hbar^2 \,b \, T}
\left[\frac{\pi^2 \,\mu^{3/2}\, (k_B\,T)^2} {3}
-\frac{3\, \pi^4\,(k_B\,T)^4} {40\,\mu^{1/2}}\right], \nn
& \kappa_{yy}  
 =\frac{7\,\tau \,b}     { 8\,\pi^2\,\hbar^2\,\sqrt{a}\,T}
\left[\frac{\pi^2 \mu^{1/2}\, (k_B\,T)^2}{3}
-\frac{31\,\pi^4\, (k_B\,T)^4} {360\,\mu^{3/2}}\right].
\end{align}
As expected, to the leading order in $T$, the longitudinal components of the thermal conductivity show a linear-in-temperature dependence for both the $x$- and the $y$-axes. However, their chemical potential dependence differs by a factor of $\mu$ (i.e., $ \kappa_{xx}  /  \kappa_{yy} \sim \mu$) as a result of the inherent anisotropic nature of the Hamiltonian. We would like to point out that we have neglected the phonon-contribution to the thermal conductivity, for the sake of simplicity. A strong contribution from the phonons may lead to a violation of the Wiedemann-Franz law.

Let us also investigate the form of the response in the opposite limit of $\mu/(k_B\,T)\ll 1$. In this high-temperature limit, we get
\begin{align}
&  \sigma _{x x} \simeq
\frac{2.16 \,e^2 \,\tau\sqrt{a}} {2\,\hbar^2\, b\,(\pi\,\beta)^{3/2}}
%\frac{0.19 \,e^2\,\sqrt a\,\tau} 
%{\hbar^2\, b \left(\,k_B\,T \right )^{3/2}}
\left( 1.5303 +\frac{0.3801\,\mu^2}{k_B^2 \, T^2}
\right ),\nn
%%%%%%%%%%%%%%%%%%%%%%%%%%%%%
& \sigma _{yy} \simeq
\frac{ 3.5\, e^2\,\tau\, b}     {8\,\pi^{3/2}\hbar^2\,\sqrt{a\,\beta}}
\left (1.2098 +  \frac{0.1187\,\mu^2} {k_B^2 \, T^2}   \right),\nn 
%%%%%%%%%%%%%%%%%%%%%%%%%%%%%%
& L _{xx}^{21} =-\frac{2.16 \,e^2 \,\tau\sqrt{a}} {2\,\hbar^2\, b\,(\pi \, \beta)^{3/2}}
\times 2.3\,\mu  
\,,\nn
%%%%%%%%%%%%%%%%%%%%%%%%%%%%%%
& L _{yy}^{21} =-
\frac{ 3.5\, e^2\,\tau\, b}     {8\,\pi^{3/2}\hbar^2\,\sqrt{a\,\beta}}
\times 0.60\,\mu \,,\nn
%%%%%%%%%%%%%%%%%
&  L _{x x}^{22} =
\frac{2.16 \,e^2 \,\tau\sqrt{a}\,\beta^2} {2\,T\,\hbar^2\, b\,(\pi\,\beta)^{3/2}}
\left(16.88 + \frac{ 0.6\, \mu^2} {k_B^2 \, T^2}\right ),\nn
%%%%%%%%%%%%%%%%%%
&   L_{yy}^{22}  =
\frac{ 3.5\, e^2\,\tau\, b\,\beta^2}     
{8\,\pi^{3/2} \, \hbar^2\,\sqrt{a\,\beta}}
\left(6.54- \frac{  0.15 \, \mu^2}    {k_B^2 \,T^2}\,
\right ).
\end{align}
It turns out that the prefactors of both Eqs.~\eqref{sigma} and \eqref{Lxx21} give rise to dominant leading order contribution at high temperatures. Thus both $\sigma_{xx}$ and $L^{21}_{xx}$ go as $T^{3/2}$. Consequently, we obtain a thermopower $S_{xx}$ decaying with temperature. This is contrast with an isotropic dispersion, where the leading order behaviour turns out to be $\sigma_{xx} \sim T$ and $L^{21}_{xx}\sim T^{2}$. This leads to a temperature-independent thermopower for graphene in the high-temperature regime \cite{Sarma2009}.
By the term ``high-temperature limit'', we simply refer to the regime $T>>T_F$, where $T_F$ is the Fermi temperature. This is a very standard limit for studying transport properties or thermal coefficients, applicable for generic scattering mechanisms (see, for example, Refs.~\cite{Sarma2009, Gegory2014}). On the other hand, at sufficiently high temperatures, the electron-phonon scatterings will also contribute to the thermoelectric properties. Incorporating such processes will lead to a characteristic energy-dependence of the relaxation time, and, accordingly, it may change the final forms of the thermoelectric coefficients, as has been pointed out by Hwang \textit{et al.} in Ref.~\cite{Sarma2009}. Thus, our formalism as such is not limited to one particular scattering mechanism. But rather, by taking into account the appropriate form of $\tau$, we can apply the same formalism (as exemplified by the cases studied in the two subsequent sections), and derive the appropriate final expressions of the response coefficients.

%%%%%%%%%%%%%%%%%%%%%%%%%%%%%%%%%%%%%%%%%%%%%%%%%%%%%%%%%%%%%%%%%%%%

\section{Thermoelectric response for 2D cases with short-ranged impurity scatterings}

\label{secdisorder}

In this section, we consider the thermoelectric response, caused by scatterings due to short-ranged disorder, for both the graphene and semi-Dirac semimetal cases \cite{ips-kush}. However, this kind of impurities is not very realistic for nodal semimetals because the relatively poor screening of charged impurities usually leads to longer-ranged interactions. Nevertheless, it is useful to investigate the predictions for the thermal properties of gapless isotropic and anisotropic Dirac materials . The short-ranged random disorder potential can be modelled by 
\begin{align}
V(\mathbf r) = V_0\sum_{i}\delta(\mathbf r-\mathbf r_i) \,,
\end{align} 
where $\mathbf r_i$ and $V_0$ denote the random locations and the magnitude of the impurity-potentials, respectively. The scattering time, denoted by $\tau_{\rm{dis}}$, for such a disorder potential can be obtained from the Fermi's golden rule as follows:
\begin{align}
\frac{\hbar}{\tau_{\rm{dis}}} = 2\,\pi
\int \frac{d{\bf k'}^2}{(2\pi)^2}
\, \delta \big (\epsilon({\bf k})-\epsilon({\bf k'}) \big ) 
\,| V_{\bf k, k'}|^2\, ,
\end{align}
where $ V_{\bf k-\bf k'} = \langle{\bf k}| V|{\bf k'}\rangle$ is the scattering matrix element. $ V_{\bf k-\bf k'} \equiv V(\bf k-\bf k')$ is in fact the Fourier transform of $V(\mathbf r)$.

For graphene, using the expression above, we find that
\begin{align}
\tau_{\rm{dis}}=\frac {\hbar} {\pi\,  V_0^2\, n_{\rm imp}\,\rho_g(\epsilon)} \,,
\end{align} 
where $n_{\rm imp}$ is the impurity concentration. Thus, Eq.~\eqref{eqdosg} tells us that $\tau_{\rm{dis}}\sim 1/\epsilon$, resulting in the thermopower to be exponentially suppressed at low temperatures~\cite{Sarma2009}, which is clearly seen from
the expression
\begin{align}
S_{xx}=S_{yy}=- 
\,\frac{ \mu \, e^{-\beta\,\mu} }  {e\,T} \,.
\end{align}
In contrast, the scattering time for the short-ranged impurities in the semi-Dirac semimetal is found to have an angular dependence due to the anisotropic dispersion~\cite{orignac}: 
\begin{align}
\tau_{\rm dis}=\frac{\tau_0(\epsilon)} {1+0.435\cos\theta } \,,
\end{align}
where $\tau_0(\epsilon)=\frac {\hbar} {\pi\,  V_0^2\, n_{\rm imp}\,\rho_{sd}(\epsilon)}
$.
Considering the above energy dependence of the scattering rate (i.e., $\tau_{\rm dis} \sim 1/ {\sqrt{\epsilon}}\,$), the transport coefficients at low temperatures [i.e., $\mu/({k_B T})\gg 1 $] are found to be
\begin{align}
& \sigma _{x x} \simeq \frac{2.88\,e^2 \tau\sqrt{a}} {2\,\pi^2\,\hbar^2\, b}\,\mu\,,~~~~\sigma _{y y} \simeq\frac{7\, e^2\,\tau\, b}     { 8\,\pi^2\,\hbar^2\,\sqrt{a}}\,,\nonumber\\
&  L _{x x}^{21}\simeq-\frac{2.88\, e\, \tau\,\sqrt{a}} {2\,\pi^2\,\hbar^2\, b}\,\frac{(\pi\, k_B\, T)^2}{3},
\quad L _{y y}^{21}\simeq -\frac{7 \,e\,\tau\, b}    
{ 8\,\pi^2\,\hbar^2 \sqrt{a}}\,\mu\,.
\label{eq:diffusive_sc2}
\end{align} 
%%%%%%%%%%
Evidently, the thermopower $S_{xx}$ follows the Mott relation, whereas $S_{yy}$ turns out to be independent of temperature (up to leading order in small $T$).

%%%%%%%%%%%%%%%%%%%%%%%%%%%%%%%%%%%%%%%%%%%%%%%%%%%%%%%%
\section{Thermoelectric response for 2D cases with scatterings caused by charged impurities}
\label{seccoulomb}

The presence of charged impurities in a material acts as dopants, thus shifting the Fermi level away from the nodal points. 
The effective Coulomb potential, generated by such impurities, is given by \cite{hwang2, hwang3, Sarma2009}
%%%%%%%%%%%%%%%%%%%
\begin{align}
\label{eqvq}
V_{cs}(q)= \frac{v_c(q)} {\varepsilon(q)}\,,
\quad v_c(q) = \frac{ 2\,\pi\, e^2} {\kappa \, q} \,,
\quad \varepsilon(q) \simeq 1 + \frac{q_{\text\tiny{\rm TF}}} {q},
\end{align}
where $\kappa $ is the dielectric
constant and $q_{\rm TF}$ is the Thomas-Fermi wavevector. 
We note that the dependence $\sim |q|^{-1}$ in $v_c(q)$
arises from a 2D Fourier transform of the 3d Coulomb interaction (which behaves as $ \propto r^{-1} $ ).
This is because, although the electrons are confined to a 2D plane for our 2D system,
the electromagnetic field lives in three spatial dimensions. The 2D static RPA dielectric (screening) function is given by $\varepsilon (q)$, as derived in Refs.~\cite{hwang2, hwang3}. In the following, we use the symbol $\tilde e \equiv e /\sqrt \kappa $. The relaxation time, within the Born approximation, is given by
\begin{align}
 \frac{1}{\tau(\epsilon(\mathbf k) )}
& = \frac{2\,\pi\,n_{imp} }{\hbar}
\nn 
& \quad \times
\sum \limits_{s = \pm}
\int \frac{d^2 \mathbf{k}'}{(2\,\pi )^2}\,| V_{\bf k-\bf k'}|^2
\, F_{\bf k\,k'}\, \delta \big( \epsilon(\mathbf k) -\epsilon_s (\vec k ')  \big) \, ,
\label{eq:tauscreen}
\end{align}
where $ V_{\bf k-\bf k'} = V_{cs} (|\bf k-\bf k'|)$ [defined in Eq.~\eqref{eqvq}], $F_{\bf k\,k'}=\frac{1-\cos^2\phi_{\vec k  \vec k'}}{2}$, $\phi_{\vec k  \vec k'}$ is the angle between $\vec k$ and $\vec k'$, and $n_{imp}$ is the impurity density.  

For graphene, the screening has an important temperature dependence, which in turns affects the thermopower. In fact, the temperature-dependent screening produces a thermopower quadratic in temperature, which is different from what we obtain by naively applying the Mott formula. In contrast, for unscreened charged impurities, we can immediate find that $\tau\simeq \epsilon$. This in turn leads to thermopower linear in $T$~\cite{Sarma2009}.

For the semi-Dirac semimetal, we use the parametrization introduced before, which leads to
\begin{align}
& \cos \phi_{\vec k  \vec k'}
\nn & = \frac{s_0 
\, \sqrt{\alpha \, |\cos\theta|}\,
\sqrt{\alpha \, |\cos\theta'|} + \sqrt{  {\tilde r}\,  {\tilde r}'}
\sin\theta  \sin\theta'
} 
{\sqrt{\alpha \, |\cos\theta| +   {\tilde r}\, \sin^2 \theta } 
\, \sqrt{ \alpha \, |\cos\theta'| +   {\tilde r}' \sin^2 \theta' }
}\,,
\end{align}
%%%%%%%%%%%%%%%%%%%%%%%%%%%%
where $\alpha=b^2/a$, $s_0=\text{sign}[\cos\theta]\,\text{sign}[\cos\theta']$, and $ {\tilde r}' \geq 0$.
For definiteness, let us consider the case when $\varepsilon (\mathbf k) >0 $, which clicks for $s =+$ in the summation in Eq.~\eqref{eq:tauscreen}. Since $\epsilon_{\vec k}^+ =  {\tilde r}$ is independent of $\theta$, we set $\theta=\frac{\pi}{2}$ without any loss of generality. This leads to 
\begin{align}
F_{\bf k,k'}=\frac{\alpha  \, |\cos(\theta')|}{2\,(\,|\cos\theta'|+  {\tilde r}'\sin^2\theta'\,)}\,.
\label{eq:Ftheta}
\end{align}
Together with Eqs.~(\ref{eq:Ftheta}), (\ref{eq:tauscreen}), and  (\ref{eq:jacobian}), it gives us 
\begin{align}
& \frac{1}{\tau(  {\tilde r})} 
\nn  &
= 
%\nn & \quad \times
\int d\theta'
\frac{
\pi\, n_{imp}\,\tilde e^4\,\alpha\,
\sqrt{\alpha \,|\cos\theta' |}}
{
\hbar\,  {\tilde r}^{3/2}
\left  [
	(1-\sin\theta')^2+\frac{\alpha\, |\cos\theta'|}{   {\tilde r}}\right ]
\left(
\alpha \, |\cos \theta '|+  {\tilde r} \sin^2\theta' \right )	
}\, ,
\label{eq:unscreenedimp}
\end{align}
where we have considered $q_{\rm{TF}} = 0 $ for unscreened charge impurities. 
In this case, Eq.~(\ref{eq:unscreenedimp}) can be further simplified in the various limits as follows (assuming $\alpha\sim 1$):
\begin{align}
&  \frac{1}{\tau(  {\tilde r} )}  
\nn & 
\simeq \frac{ \pi \,\tilde e^4 \,n_{imp}}{ \hbar}
\times
\begin{cases}
\frac{8.0}{  {\tilde r}}  & \text{ for }    {\tilde r}\ll 1\\
& \\
\frac{ 6.0476 }{  {\tilde r}^{5/3}}
+\frac{16.509}{  {\tilde r}^{7/3}}- \frac{10.6889}{  {\tilde r}^{3}}
& \text{ for }    {\tilde r} \gg1
\end{cases} .
\label{eqscat22}
\end{align}
The first limit is found from the leading order contribution of
$2\int_{-\pi/2+  {\tilde r}}^{\pi/2-  {\tilde r}} \frac{ d\theta'\, \sqrt{|\cos\theta' |} } 
{\left (\frac{|\cos\theta'|}{  {\tilde r}}\right )
	\left (|\cos \theta '| \right ) }$, whereas the second limit is found from
the leading order contribution of
$4 \int_{0}^{\pi/2-\left ( {4} / {  {\tilde r}} \right )^{1/3} } 
\frac{ d\theta'\, \sqrt{ |\cos\theta' |} } 
{\left (1-\sin \theta'\right )^2
\,	 {\tilde r}\, \sin^2\theta'}$.

We emphasize that the scattering from the unscreened Coulomb interaction in graphene turns out to be $\tau\sim \epsilon$, irrespective of the values of $\epsilon$. In contrast, the anisotropy in Eq.~(\ref{ham-continuum}) leads to a different expression for energy-dependent scattering for $\epsilon\gg1$.  Considering the leading energy dependent term for $\tau\sim \epsilon^{5/3}$, we find that 
%%%%%%%%%%%%%%%%%
\begin{align}
& \sigma _{x x} = \frac{2.88\, \tilde e^2 \,\tau\,\sqrt{a}} 
{ 4 \,\pi^2\,\hbar^2 \, b}
\left [ \mu^{19/6}+\frac{247 \, \mu^{7/6} \, \pi^2}  {216} \, (k_B\,T)^2\right ],\nn
%%%%
& \sigma _{y y} = \frac{7 \, \tilde e^2\,\tau\, b}     
{  16\,\pi^2\,\hbar^2\,\sqrt{a}}
\left [  \mu^{13/6}+
\frac{91 \, \mu^{1/6} \, \pi^2}{216}
(k_B\,T)^2   \right  ],\nn
%%%%%%%%%%%%%%%%%
& L _{x x}^{21} =-\frac{2.88\,  \tilde e\, \tau\,\sqrt{a}} 
{2\,\sqrt 2 \,\pi^2\,\hbar^2\, b}\times
\frac{19\,\pi^2\mu^{13/6} (k_B\,T)^2}
{18}
,\nn
%%%%%%%%%%%%%%%%%%
& L_{yy}^{21} = -\frac{7 \,  \tilde e\,\tau\, b}    
{ 8\,\sqrt 2 \,\pi^2\,\hbar^2\,\sqrt{a}}\times
\frac{13\,\pi^2\,\mu^{7/6}(k_B\,T)^2}
{18} \, .
\label{eq:diffusive_sc}
\end{align}
%%%%%%%%%%%%%%%%%%%%%%%%%
Thus we recover the Mott relation of $S_{\alpha\alpha}\sim T$. However, the dc conductivities exhibit a nontrivial dependence on the chemical potential due to the energy-dependent relaxation time.

%%%%%%%%%%%%%%%%%%%%%%
\section{Thermopower for 2D cases in the presence of a weak magnetic field}
\label{secmag1}

Having discussed the thermoelectric properties without any external magnetic field, we now turn to the thermopower for finite magnetic field $\mathbf B= B\,\hat z$, in the presence of a finite $\partial_x T$ ( but $\mathbf E$ set to zero), focussing on the case of a semi-Dirac semimetal. Assuming the solution $\delta f_n$ to be independent of time, due to the static nature of $\mathbf B$, the linearized form of the Boltzmann equation [cf. Eq.~(\ref{eqkin4})] is found to be \footnote{
Although at high magnetic field intensities, the Landau quantized levels appear, for small intensities, we can use the semiclassical approximation \cite{mermin,arovas,suh}.}
% using Eqs 26 to 29 of PHYSICAL REVIEW B 93, 035116 (2016)
%%%%%%%%%%%%%%%%%%
\begin{align}
& \frac{e\, B} {\hbar \, c} \left (v_x\,\partial_{k_y} - v_y\, \partial_{k_x} \right )
 \delta f_n\nonumber + v_x \left (\epsilon_n -\mu \right )
\partial_x T\, \frac{\partial f_n^{(0)}} {\partial \epsilon_n }
\nn & = -\delta f_n /\tau.
\end{align}
Following Ref.~\cite{girish1,Gegory2014}, the appropriate ansatz for $\delta f_n$ leads to the thermoelectric coefficients as
% Ref.~\cite{behnia}, we outline here how to find the thermoelectric coefficients for this case.

%. In the absence of a magnetic field, we have solved Eq.~\eqref{eqkin4} to find the response of the system, which is set by the relaxation time and the magnitude of the deviation-from-equilibrium function $\delta f_n $.

% use https://academic.oup.com/book/12037/chapter/161343799

%Considering all these ingredients, the thermoelectric coefficients can be expressed as [see Eq.~(\ref{eq:transcoeff})]: 
\begin{align}
\label{eq:lowfield}
&L_{\alpha\beta}^{11}=-\int d\epsilon\, 
\frac{\partial  f^{(0)} (\epsilon) } {\partial \epsilon } 
\, \tilde \sigma_{\alpha\beta}(\epsilon)\,, \nonumber\\
&L_{\alpha\beta}^{21} = -  e
\int d\epsilon\,  
\frac{\partial  f^{(0)} (\epsilon) } {\partial \epsilon } 
\, (\epsilon-\mu)\,\tilde \sigma_{\alpha\beta}(\epsilon)\,, 
\end{align} 
where the energy-dependent $\tilde \sigma_{\alpha\beta} (\epsilon)$'s are the components of the matrix
%%%%%%%%%%%%%%%%%%%
\begin{align}
\tilde \sigma (\epsilon) & =
\sigma_0 (\epsilon)
\begin{pmatrix} 
v_{x}^2 (\epsilon)
&-B\, \tau \,\tilde v_x (\epsilon)
\cr B\,\tau  \, \tilde v_y (\epsilon)
&v_y^2(\epsilon)
\end{pmatrix} , \nn & \nn
%%%%%%%%%
\sigma_0 (\epsilon) & = e^2\,\rho_{sd}(\epsilon) \,\tau \,,
\quad v_x^2(\epsilon)= {2.88\,a \, \epsilon} / {\pi^2} \,,
\quad
v_y^2(\epsilon)=7\,b^2 \,,
\nn
\tilde v_x (\epsilon) 
& = v_x^2(\epsilon)\,\,
\partial^2_{k_y} \epsilon 
- v_x(\epsilon)\, v_y(\epsilon)\,
\partial_{k_x} \partial_{k_y}\epsilon \,,\nn
%%%%%%%%
\tilde v_y  (\epsilon)  & =
v_y^2(\epsilon)\, \,{\partial_{k_x}^2 \epsilon}
- v_x(\epsilon)\, v_y(\epsilon)\,\,
\partial_{k_x} \partial_{k_y} \epsilon \,.
\end{align} 
%%%%%%%%%%%%%%%%%%%%%
We note that the diagonal elements of $ \tilde \sigma(\epsilon)$ are taken up to zeroth-order in $B$ and, for the off-diagonal components, we retain the terms upto linear order in $B$. For simplicity, we assume $\tau$ to be independent of the energy. For $k_B \,T\ll \mu$ and $ \hbar \, \omega_c\ll \mu$, Eq.~(\ref{eq:lowfield}) can be further simplified as
\begin{align}
\label{eq:lowfieldS}
L_{\alpha\beta}^{11}\simeq  \tilde \sigma_{\alpha\beta}(\mu)\,,
\quad 
L_{\alpha\beta}^{21}
\simeq \frac{\pi^2 \left(k_B \, T\right)^2 } {3 \, e} 
\frac{d  \tilde \sigma_{\alpha\beta}(\epsilon)}   {d\epsilon}
\Bigg |_{\epsilon=\mu}\,. 
\end{align} 
These allow us to compute the longitudinal components of the thermopower, which take the following forms:
\begin{align}
S_{xx}
& \simeq \frac{\pi^2\, k_B^2\, T}{3\,e\,\mu} 
\nn &  \, \times
\frac{1.5\, v^2_x(\epsilon)\,v^2_y(\epsilon)
	+ B^2\,\tau^2\, \tilde v_x(\epsilon)
	\left[0.5\,\tilde v_{y}(\epsilon)+\mu\,\partial_\epsilon \tilde  v_y(\epsilon)\right]}
{v^2_x(\epsilon)\,v^2_y(\epsilon)
	+ B^2\,\tau^2\, \tilde v^2_x(\epsilon)\,
	\tilde v^2_y(\epsilon)} 
\Bigg |_{\epsilon=\mu} ,
\nonumber\\
%%%%%%%%%%%%%%%%%%%%
S_{yy} 
& \simeq 
\frac{\pi^2\, k_B^2 \,T} {3\,e} 
\nn & \,  \times
\frac{ \frac{0.5\, v^2_x(\epsilon)\,v^2_y(\epsilon)} {\mu}
	+ B^2\,\tau^2\,\tilde v_y(\epsilon)
	\left[0.5\,\tilde v_{x}(\epsilon)
+ \mu\,  \partial_\epsilon \tilde v_x (\epsilon)\right]}
{v^2_x(\epsilon)\,v^2_y(\epsilon) +
B^2\,\tau^2\,\tilde v^2_x(\epsilon)\,\tilde v^2_y(\epsilon)}
\Bigg |_{\epsilon=\mu} .
\end{align}
Evidently, in the limit of $B\rightarrow 0$, these expressions coincide with those shown in Eq.~\eqref{eqthermozerob}.

%%%%%%%%%%%%%%%%%%%%%%%%%%%%%%%%%%%%%%%
\section{Thermopower for 2D cases in the presence of a strong magnetic field}
\label{secmag2}

In this section, we take the same set-up as considered in Sec.~\ref{secmag1}, but elucidate the characteristics of the thermopower in presence of a quantizing (i.e., strong) magnetic field along the $z$-direction, such that Landau levels are formed. Our strategy is to use the entropy ($ \mathcal S$) in determining the thermopower. We observe that, since the heat current is the electronic contribution to the heat transport in an electronic system, the Seebeck coefficient can be expressed in terms of the entropy of the system. We motivate this strategy by considering the case of a one-dimensional wire oriented along the $x$-direction, which then deals with the case of the longitudinal Seebeck coefficient $S_{xx}$ only. An electrostatic potential gradient ($\delta V$) appears on applying a temperature gradient, such that $\delta V = S_{xx} \, \partial_x T$. Using thermodynamics, we have the relation $n_0 \, e \, \delta V \sim {\mathcal S } \, \delta T$, where $n_0$ is the number density of the carriers, which in turn gives $S_{xx}= {\mathcal S }/ \left(  e\,n_0 \right ) $.

In order to derive the explicit relation between the entropy and the Seebeck coefficient, in the presence of a sufficiently {\it strong magnetic field}, we begin with the general expression of thermoelectric coefficients given by~\cite{Girvin_1982}
\begin{align}
\label{eq:Hall}
L^{11}_{xy}=-\frac{e^2}{h}\sum_{l}\int_{\epsilon_l-\mu}^{\infty} d\epsilon\, 
{\partial f^{(0)}(\epsilon)\over\partial \epsilon} \,, \nonumber\\
%%%%%%
L^{12}_{xy}=\frac{k_B\,e\,\beta}{h} \sum_{l}
\int_{\epsilon_l-\mu}^{\infty}  d\epsilon\, \epsilon\,
{\partial f^{(0)}(\epsilon)\over\partial \epsilon} \,,
\end{align} 
where $\epsilon_{l}$ here denotes the Landau level energies and $f^{(0)}(\epsilon)= {1} / \left( 1+e^{\beta \,\epsilon}\right )$ [cf. Eq.~\eqref{eqfd}]. We would like to point out that the macroscopic transport properties are independent of the specific details of the confining potential in the sample, although the microscopic currents do depend on it. Eq.~\eqref{eq:Hall} can be further simplified by changing variables from $\epsilon\rightarrow f^{(0)} $, leading to
\cite{Skinner2018,Bergman2010}
%%%%%%%%%%%%%%%%%%%%%%%%%%
\begin{align}
L^{11}_{xy}&=-\frac{e^2}{h}\sum_{l}f_l \,,
\nonumber\\
L^{12}_{xy}&=\frac{k_B\,e}{h} \sum_{l}  \int_{\epsilon_l-\mu}^{\infty}\,
f_l\, \left [\ln (1- f_l)-\ln f_l  \right ]
\nn & = \frac{e\, \mathcal S} {h}  \,.
\end{align}  	
Here, $f_l=f^{(0)}(\epsilon_l-\mu)$, and 
\begin{align}
\mathcal S=-k_B\sum_l \left [ \, f_l \ln f_l+(1-f_l)\ln (1-f_l) \,\right ]
\end{align}
is the total entropy of the carriers. This leads to the form $S_{xx}= {\mathcal S}  /\left( e\,n_0 \right ) $ for the longitudinal thermopower.

%%%%%%%%%%%%%%%%%%%%%%%%%%%%%%%%%%%%%%%%%%%%%%%%%%%%%%%%%
\begin{figure}[t]
\includegraphics[width=0.99 \linewidth]{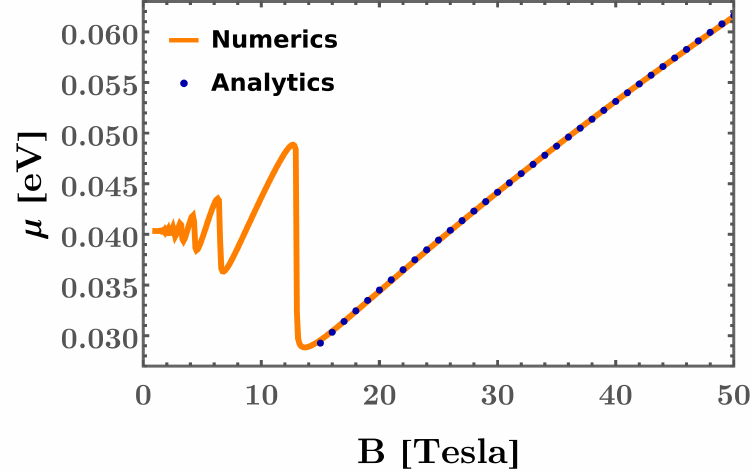}
\caption{\label{fig:chempot}
Plot (solid orange line) of the chemical potential as a function of the strength of the magnetic field for electron density $n_0=5\times 10^{11}$ cm$^{-2}$, at temperature $T=5 $ K. The behaviour in the strong field regime, represented by the blue dots, is obtained from the approximate analytical form shown in Eq.~\eqref{eq: mufit}, using the values $ b_0=0.0017$ eV, $b_1=0.0006$ eV Tesla$^{-2/3}$, $b_2 = 0.0014 $ eV, and $ b_3 = 1 $ Tesla$^{-1}$ as the fitting parameters.
Considering parameter values characteristic of Dirac semimetals, we have used $v=5\times 10^5 $ m/s and $m=3.1\times m_e$, where $m_e$ is the electron mass.}
\end{figure}
%%%%%%%%%%%%%%%%%%%%

For a semi-Dirac semimetal, in the presence of a quantizing magnetic magnetic field $\mathbf B = B \,\hat{\mathbf z}$ [using the Landau gauge $\boldsymbol{A} = (- B\,y,0, 0)$], the energies of the Landau levels are found to be \cite{landau-level}
%%%%%%%%%%%%%%%%%%%%%%%%%%%%%%%%%%%%%%%%%%%%%%%%%%%%%%%%%%%
\begin{align}
\epsilon_l= & \pm 1.17325
\left (m \, v^2\right )^{1/3}
\left  [\left (l+1/2\right )\hbar \,\omega_c\right ]^{2/3} \nn
& \text{with }
l \in \lbrace 0, {\mathbb{Z}}^+ \rbrace \,,
\label{WKB1} 
\end{align}
%%%%%%%%%%
where $\omega_c =  { e\, B} /{m}$ is the effective cyclotron frequency.
%%%%%%%%%%%%%%%%%%%%% 
With this, we obtain
\begin{align}
S_{xx}=\frac{k_B}{2\, \pi\,n_0\,e\, l_b^2}\sum_l \left[\ln(1+e^{\tilde x_l})
-\frac{{\tilde x_l}\, e^{\tilde x_l}}{e^{\tilde x_l}+1}\right],
\label{eq:alphaxx}
\end{align} 
%%%%%%%%%%%%%%%%%%%%%%%%%%%%
where $\tilde x_l = \beta\,(\epsilon_l-\mu)$, 
$l_b=\sqrt{\frac{\hbar}{e\,B}}$ is the magnetic length, and $n_0$ fixes the Fermi energy through
\begin{align}
n_0=2 \times \frac{1}{2\,\pi \,l_b^2}\sum_{l=0}^{\infty} f_{l}
\Big \vert_{\epsilon_l>0} \,.
\label{eq:edos}
\end{align} 
The factor of $2$ in the above equation accounts for the contributions coming from $ \epsilon_l <0$. 

For a reasonably strong magnetic field [i.e., for $\hbar \,\omega_c \gg \mu $], the system enters into a strong quantum limit, and the quasiparticles occupy only the $l=0$ Landau levels. In this regime, we can approximate Eq.~\eqref{eq:edos} as
\begin{align}
n_0\simeq   \, \frac{1}{\pi \,l_b^2}
\times \frac{1}{1+e^{\beta \, \left  (|\epsilon_0|-\mu \right )} }\,.
\label{eq:density}
\end{align}
This leads to $\mu= |\epsilon_0|-\beta^{-1}
\ln(\frac{1}
{n_0 \, \pi \, \l_b^2}-1)$, with the leading-order field dependence
captured by
\begin{align}
\mu=b_0 + b_1 \, B^{2/3}+ b_2\ln(b_3 \,B-1) \,,
\label{eq: mufit}
\end{align}
%%%%%%%%%%%%%%%%%%%%%%%%
where $ \lbrace b_0, \, b_1, \, b_2, \, b_3 \rbrace $ can be readily obtained from the approximate analytical solution of $\mu$. This approximate analytical result fits reasonably well with the numerical solution obtained from Eq.~(\ref{eq:density}), as corroborated by the curves in Fig.~\ref{fig:chempot}, which confirms it validity. Notably, Eq.~\eqref{eq: mufit} differs from the 3D version (which is also known as the double-Weyl semimetal) exhibiting $\mu\sim {B}^{-1}$. This difference, of course, originates from the specific magnetic-field-dependence of the Landau level spectrum. We note that, for weak-enough values of the magnetic field strength, which refers to the limit $ \hbar \, \omega_c\ll\mu$, the chemical potential is mostly unaffected by the magnetic field. As we increase the strength of the magnetic field, we start to observe quantum oscillations in the chemical potential, which in turn leads to oscillations in the thermopower as well.

%%%%%%%%%%%%%%%
\begin{figure}[t]	
\includegraphics[width=0.98\linewidth]{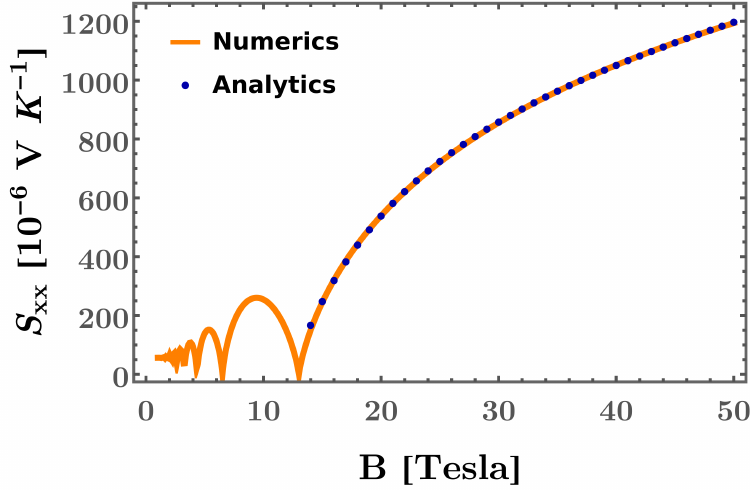}
\caption{\label{fig:Seebeck}
Plot (solid orange line) of $S_{xx}$, as shown in Eq.~\eqref{eq:alphaxx}, as a function of the strength of the magnetic field for an electron density of $n_0 = 5\times10^{11} \,\rm{cm}^{-2}$, at temperature $T=5$ K. The blue dots represent the approximate analytical results obtained using Eq.~(\ref{eq:sxxanaly}). The values of the parameters $v$ and $m$ are the same as in Fig.~\ref{fig:chempot}.}
\end{figure}
%%%%%%%%%%%%%%%%%%%%%%%%%%%%%%%%%%%%%%%%%%%%

In order to find the approximate high-field dependence of the thermopower, we plug in the expression for $\mu $ [given by Eq.~(\ref{eq: mufit})] into Eq.~\eqref{eq:alphaxx} after setting $l=0$. This results in
\begin{align}
S_{xx}&=\frac{k_B}{e}
\left[\left(\pi \,n_0 \,l_b^2\right)
\ln\left(1-\pi \,n_0 \, l_b^2 \right)
-\ln\left(\frac{1}{\pi \, n_0 \, l_b^2}-1\right)\right]\nonumber\\
&=\frac{k_B}{e}\left[\left(1-\frac{B}{\alpha_0}\right)
\ln\left(1-\frac{\alpha_0}{B}\right)-\ln\left(\frac{\alpha_0}{B}\right)\right], \nn
\alpha_0 & =\frac{n_0  \,h}{2 \, e} \,.
\label{eq:sxxanaly}
\end{align}
%%%%%%%%%%%
In order to verify this complex $ B $-dependence in our analytical approximation, we numerically evaluate Eq.~\eqref{eq:alphaxx} using the numerical solution of $\mu (B)$. Fig.~\ref{fig:Seebeck} shows the behaviour of $S_{xx}$ as a function of $B$. Clearly, the approximate large-field dependence of $S_{xx}$ (represented by the blue dots), obtained from the formula shown above, fits well with the numerical solution (shown by the solid orange line) of Eq.~\eqref{eq:alphaxx}. Interestingly, the part of the thermopower curve obtained in the low-field region agrees very well with the experimental results available for $\alpha$-(BEDT-TTF)$_2$I$_3$ \cite{osada}.

In contrast to the semi-Dirac case, the Landau levels in graphene take the form
\begin{align}
\epsilon_l\propto 
\begin{cases}
\pm  \,\sqrt{l} &\text{for } l>0 \\
0 &\text{for } l =0
\end{cases}  \,,
\end{align}
showing a square root dependence on $l \in {\mathbb{Z}}$. Subsequently, the Seebeck coefficients are found to oscillate at low temperature, as functions of the Landau level index $l$, for a fixed magnetic field~\cite{wei_PRL2009}. Moreover, $S_{xx}$ has been found to increase slowly as a function of the magnetic field~\cite{Checkelsky2009}. For the 3D Dirac and Weyl nodes, the Landau levels disperse in one of the momentum directions. This, in turn, leads to a strong field dependence of the form $S_{xx}\sim B^2$~\cite{Liang2017}.

Finally, we comment on the behaviour of the transverse thermoelectric coefficient $S_{xy}$, also dubbed as the magneto-thermoelectric Nernst-Ettinghausen effect. Using the relation $S_{xy}=L^{12}_{xy}/L_{xy}^{11} $, this can be found using Eq.~\eqref{eq:Hall}.
It turns out that $S_{xy}$ oscillates as a function of the chemical potential. The maximum value of $S_{xy}$ is given by $ k_B \, e^{-1} \ln 2 $, corroborating the universal behavior pointed out in Ref.~\cite{Girvin_1982}. However, the peak positions differ from graphene or typical semiconductors. We again attribute this difference to the difference in the Landau spectrum, as discussed before.

%%%%%%%%%%%%%%%%%%%%%%%
\section{Thermoelectric response for 3D Weyl semimetals in planar Hall and planar thermal Hall set-ups}
\label{secweyl}

Using the formalism reviewed in Sec.~\ref{secboltz2}, in this section, we show the expressions for the response related to PHE and PTHE for a Weyl semimetal (WSM). A single Weyl node with chirality $\chi$ (taking the values of $\pm$ 1) is described by the Hamiltonian
\begin{equation}
H=\chi \, \hbar\, v_F \,
\boldsymbol {\sigma}\cdot \mathbf {k},
\end{equation}
where $v_F$ is the Fermi velocity and $\vec{k}$ is the 3D wavevector. In the following, we will use natural units, thus setting $\hbar  $, $c$, and $k_B $ to unity. Furthermore, we will consider the low-temperature regime, satisfying $\beta \mu \gg 1$, such that Sommerfeld expansion is applicable for evaluating the response involving integrands containing functions of the Fermi-Dirac distribution.

The eigenvalues of the Hamiltonian are given by
\begin{align} 
\label{eqeval}
\epsilon_{n} ({ \mathbf k})= 
 - (-1)^{n} \,  v_F \, k\ \,, \quad
n \in \lbrace 1,2 \rbrace\,,
\end{align}
where the value $1$($2$) for the band index $n$ refers to the conduction(valence) band.

The Berry curvature expression for the node is captured by
\begin{align}
\mathbf \Omega_{\chi, n }({ \mathbf k})= 
 \frac{ \chi \,(-1)^n }
{2 \,k^3
} 
\left
\lbrace k_x, \, k_y, \,  k_z \right \rbrace ,
\end{align}
%%%%%%%%%%%%%%%%%%%%%%%%%
while the band velocity vector for the quasiparticles is given by
\begin{align}
{\boldsymbol{v}}_{n } ( \mathbf k) =
 \nabla_{ \mathbf k} \,  \epsilon_{n} 
 ({ \mathbf k}) = - \frac{ (-1)^n \,v_F
 }
 { k } 
 \left \lbrace  k_x, \,  k_y, \,  k_z
\right \rbrace .
\end{align}
%%%%%%%%%%%%%%%%%%%%%%%%%%%%%%%%%%
Henceforth, we will consider the situation when the chemical potential cuts the conduction band, such that the quasiparticles in that band participate in the transport.

%%%%%%%%%%%%%%%%%%%%%%%%%%%%%%%%%%%%%%%%%
\subsection{Expressions neglecting the OMM-contributions}

 The longitudinal magnetoconductivity tensor [cf. Eq.~\eqref{eqsigmatot}] evaluates to \cite{ips-rahul-ph}
\begin{align} 
\label{eqn:sigmaxxfinal}
 \bar  \sigma_{xx}^{\chi} & =   \frac{  \tau \, e^2  \, \mu^2  }
    {6\, \pi^2 \, v_F} 
    \left(  1 + \frac{\pi ^2} {3 \, \beta^2 \, \mu^2} \right )
\nn & \hspace{ 0.5 cm }
+  
  \frac{ \tau \, e^4 \,  v_F^{3} 
  }
   { 64 \, \pi^{\frac{3}{2}} \, \mu^2
 \, \Gamma (\frac{7}{2})  
   } 
\left( 1 +  \frac{ \pi^2 } { \beta^2 \, \mu^2 } \right )
\left (  8\,  B_x^2   +  B_y^2 \right ).
\end{align}
%%%%%%%%%%%%
The first term in $\bar \sigma_{xx}^{\chi} $ is independent of the magnetic field, and has a nonzero value even at zero temperature. This $ \mathbf B$-independent part is usually referred to as the Drude contribution. 
The planar Hall conductivity tensor [cf. Eq.~\eqref{eqsigmatot}] takes the form \cite{ips-rahul-ph}
\begin{align} 
\label{eqn:sigmaxyfinal}
\bar  \sigma_{y x }^{\chi} =
  \bar  \sigma_{x y }^{\chi} =  
  \frac{ 7\,\tau \, e^4 \,  v_F^{3} 
  }
 {64 \, \pi^{\frac{3}{2}} \, \mu^{2}
\, \Gamma (\frac{7}{2}) 
 }    
\left ( 1 +  \frac{ \pi^2}
 { \beta^2 \, \mu^2} \right ) B_{x} \,  B_{y} . 
\end{align}
%%%%%%%%%%%%%%%%%

The expression for the longitudinal thermoelectric coefficient [cf. Eq.~\eqref{eqalphatot}] turns out to be \cite{ips-rahul-ph}
\begin{align}
\bar \Upsilon_{xx}^\chi  =
 -\frac{  \tau \, e \,  \mu}
 {9 \, v_F \, \beta} 
 + \frac{ \tau \, e^3 \,v_F^3 \,  {\sqrt \pi}\, \mu^{-3}
 }
 { 96 \,\beta \, \Gamma (\frac{7}{2})
 }
\left (8 \,   B_x^2 +  B_y  ^2 \right ).
\end{align}
Comparing with Eq.~\eqref{eqn:sigmaxxfinal}, we observe that $
  \partial_\mu \bar \sigma_{xx}^{\chi} = - \frac  {3\, e\, \beta }  {\pi^2} \,\bar \Upsilon_{xx}^\chi
  + \mathcal{O} (\beta^{-2}) $. Hence, the Mott relation $ L_{\alpha \gamma}^{12}  =  - \frac{\pi^2} {3\, e\, \beta } \,\partial_\mu L_{\alpha \gamma}^{11} $, which holds in the $\beta \rightarrow \infty $ limit, is satisfied \cite{di}. 
%%%%%%%%%%%%%%%%%%
Finally, the expression for the longitudinal thermoelectric coefficient [cf. Eq.~\eqref{eqalphatot}] is found to be \cite{ips-rahul-ph}
\begin{align}
  \bar   \Upsilon^\chi_{yx}
  = \bar  \Upsilon^\chi_{xy}
  = \frac{ 7\, \tau \, e^3 \,v_F^3 \, 
 {\sqrt \pi} \, \mu^{-3}  \,  B_{x} \,  B_{y} 
 }   
  {96 \, \beta \, \Gamma (\frac{7}{2})} \,  .
\end{align}
Comparing with Eq.~\eqref{eqn:sigmaxyfinal}, we observe that $
  \partial_\mu \bar \sigma_{yx}^{\chi} = - \frac  {3\, e\, \beta }  {\pi^2} \,\bar \Upsilon^\chi_{yx}
  + \mathcal{O} (\beta^{-2}) $. Therefore, in this case too, we find that the Mott relation $ L_{\alpha \gamma}^{12}  =  - \frac{\pi^2} {3\, e\, \beta } \,\partial_\mu L_{\alpha \gamma}^{11} $ (valid in the $\beta \rightarrow \infty $ limit) is satisfied \cite{di}. 
  
In a similar spirit, the final expressions for the coefficient $\bar \ell^\chi $ (which have been explicitly derived in Ref.~\cite{ips-ruiz}) can be inferred from the expressions above, as they satisfy the relations arising from the Wiedemann-Franz law, viz.
\begin{align}
\bar \sigma^{\chi} _{\alpha \gamma} =
 \frac{ 3 \,e^2 } {\pi^2\, T}   \,\bar \ell^{ \chi} _{\alpha \beta}  
  + \order{\beta ^{-2}}.
\label{eqwf}
\end{align}

  \subsection{Expressions considering the OMM-contributions}
  
Although in the earlier parts of the literature, many papers ignored the OMM-contributed parts, recent works \cite{ips-rahul-ph, ips-rahul-tilt, ips-ruiz} have shown that they necessarily do not have negligible contributions.
When the OMM is considered, the longitudinal magnetoconductivity tensor gets corrected to
\begin{align}
\bar \sigma^{\chi, \rm{full}}_{xx} &= 
\bar \sigma^{\chi}_{xx}
+  \sigma^{m, \chi}_{xx}  \,,
\end{align}
where
\begin{align}
\label{eqsigxxfin}
& \sigma^{m, \chi}_{xx} 
  = -\,  \frac{ \tau \, e^4 \,  v_F^{3} 
  }
   { 64 \, \pi^{\frac{3}{2}} \, \mu^2
 \, \Gamma (\frac{7}{2})  
   } 
\left( 1 +  \frac{ \pi^2 }
{ \beta^2 \, \mu^2 } \right )
\left (  4\,  B_x^2   + 3\, B_y^2 \right ).
\end{align}
Similarly, the in-plane transverse components need to be corrected to
\begin{align}
\bar \sigma^{\chi, \rm{full}}_{yx} &= \bar \sigma^{\chi, \rm{full}}_{xy} 
=
\bar \sigma^{\chi}_{xy}
+  \sigma^{m, \chi}_{xy}  \,,
\end{align}
where
\begin{align}
\label{eqsigxxfin}
& \sigma^{m, \chi}_{xy} 
  =   -\, \frac{ \tau \, e^4 \,  v_F^{3} 
  }
 {64 \, \pi^{\frac{3}{2}} \, \mu^{2}
\, \Gamma (\frac{7}{2}) 
 }    
\left ( 1 +  \frac{ \pi^2}
 { \beta^2 \, \mu^2} \right ) B_{x} \,  B_{y}\,.
\end{align}
Computing the tensor components of $\bar \Upsilon^\chi$ and $\bar \ell^\chi$ explicitly \cite{ips-ruiz}, one finds that they are related to the components of $\bar \sigma^{\chi}$ via the Mott relation and the Wiedemann-Franz law quoted in the earlier subsection, since these relations should continue to hold even in the presence of the OMM \cite{niu_prl}.

\subsection{Further generalizations}

Although we have quoted here the results for an isotropic Weyl node, there have been corresponding studies for the double-Weyl and triple-Weyl semimetals, collectively called multi-Weyl semimetals (mWSMs). These are generalizations of the WSM  \cite{bradlyn, bernevig2}, with a dispersion which is linear along one direction and quadratic/cubic in the plane perpendicular to it. Anisotropy arising from tilting of the dispersion \cite{emil_tilted, trescher17_tilted} is expected to be present in WSMs/mWSMs in generic situations, because this is allowed by the generic nature of nodal points in 3D \cite{herring}.\footnote{For example, in a WSM with broken time-reversal symmetry $ \mathcal T $ \cite{emil_tilted}), is made possible by their topological stability.} In the PHE and PTHE set-ups, tilting leads to the presence of terms linearly dependent on $ B $, when we look at the theoretical expressions of the longitudinal and transverse components of the magnetoelectric conductivity tensor \cite{ma19_planar, kundu20_magnetotransport, konye21_microscopic, shao22_plane}. This can explain the resistivity behaviour in experimental observations \cite{li18_giant}. In Ref.~\cite{ips-rahul-tilt}, one of us has demonstrated how tilt parameters, in conjunction with the intrinsic mixed linear-nonlinear dispersion of mWSMs, lead to strongly direction-dependent novel signatures in PHE. It has been shown that a time-reversal-symmetry-breaking is induced by the tilt of nodes, thereby producing linear-in-$ B $ terms, which depend on the direction of the tilt.

A supplementary way of generating linear-in-$B$ terms in the response spectra is to introduce artificial pseudoelectromagnetic fields in the systems, with the help of mechanical strain \cite{onofre, ips-rahul-ph, ips-ruiz}. The form of these emergent gauge fields show that they couple to the quasiparticles of opposite chiralities with opposite signs \cite{guinea10_energy, guinea10_generating, low10_strain, landsteiner_gaguge,  liu_gauge, pikulin_gauge, arjona18_rotational, ghosh20_chirality, girish2023, onofre}. Due to the chiral nature of the coupling between the pseudoelectromagnetic fields and the itinerant fermionic carriers, the effective picture provides an example of axial gauge fields in three spatial dimensions.

\subsection{Experimental realizations}

In the planar Hall response, the $B^2$-dependence and the $\pi$-periodic behaviour (with respect to $\theta$) of the magnetoelectric conductivity in the absence of tilt have been observed in numerous experiments. These involve materials like ZrTe$_{5}$ \cite{li_2016}, TaAs \cite{cheng-long}, NbP and NbAs \cite{li_nmr17}, and Co$_3$Sn$_2$S$_2$ \cite{shama}, which are known to host Weyl nodes in their bandstructures. Furthermore, the magnetothermal coefficient has also been measured in materials such as NdAlSi \cite{marcin}, which again shows the expected $B^2$-dependence.
The natural consequence of the emergence of the terms that are linearly-dependent on $B$ is the change of periodicity of the response curves, as functions of $\theta $, from $\pi$ to $2\pi$. Such a change in the angular dependence has been reported in various experimental data \cite{thete_dep}.

%%%%%%%%%%%%%
\section{Summary and outlook}
\label{conclusion}

In this review, we have presented the methods to compute the thermoelectric coefficients in nodal-point semimetals, covering both the 2D and 3D cases.

In 2D, we have provided the explicit expressions of the response tensors, taking the specific examples of graphene and semi-Dirac semimetal, which represent the cases with isotropic and anisotropic dispersions, respectively. We have investigated these properties both in the absence and in the presence of an external magnetic field. In the latter case, the field is applied along the direction perpendicular to the plane of the semimetals. For the zero and weak (nonquantizing) magnetic field regimes, we have employed the semiclassical Boltzmann formalism (along with a relation time approximation) to derive the expressions for the elements of the response matrix (built out of the thermoelectric transport coefficients). In order to demonstrate how the response depends on the primary mechanism responsible for the scatterings of the emergent quasiparticles under a given situation, in addition to the simplest case of a constant relaxation time, we have considered the appropriate forms of the relaxation time arising due to the influence of short-ranged disorder potential and screened charged impurities. In all the cases considered, we have articulated that an anisotropy in the band-spectrum invariably affects the response quite significantly. We have illustrated this point by comparing the results for a semi-Dirac semimetal with those for graphene. Some of these observations include the following:
\begin{enumerate}

\item  In the low-temperature regime, the components of the dc conductivity tensor for a semi-Dirac semimetal have a different dependence on the chemical potential (or Fermi level) compared to the case of graphene. 

\item In the high-temperature limits, the thermoelectric response decays with temperature in the semi-Dirac semimetal. On the contrary, the response is independent of temperature in graphene.

\item The relaxation rates due to the distinct scenarios of the short-ranged impurity potential and the screened Coulomb interactions (caused by charged impurities) lead to distinct expressions for the dc and thermal conductivities. 

\item For the strong magnetic field case, when we need to consider the Landau levels, we have outlined a procedure to compute the response using the entropy of the carriers. Using the formula derived through this pathway, we have illustrated the results for the response via numerical data, in addition to providing analytical approximations. While the thermopower shows a strong unsaturating behaviour for the semi-Dirac semimetal [cf. Fig.~\ref{fig:Seebeck}], $S_{xx}$ in graphene increases with $B$ at a much slower rate.

\end{enumerate}
%%%%%%%%%%%%%%%%%
The results for various distinct cases, as summarized above, prove beyond doubt that the interplay of anisotropy and strengths of externally applied electromagnetic fields might lead to useful technological applications, such as achievement of a high value of the thermoelectric figure-of-merit \cite{Skinner2018}.
%%%%%%%%
From the ubiquitous use of graphene in uncountable experiments, it is needless to say that graphene-based set-ups have emerged as the primary 2D platform for observing the thermopower and related coefficients, discussed in this review. The experimental data of magnetohermopower and Nernst coefficients in graphene have turned out to agree well with the Mott relations \cite{Philip2009, Wei2009, Checkelsky2009}. Thermoelectric properties of  bilayer graphene are also found to satisfy Mott relations \cite{nam_PRB2010}.

For the 3D nodal phases, we have shown the explicit expressions for the various conductivity tensors, applicable in planar Hall and planar thermal Hall set-ups, considering an untilted Weyl semimetal node. Such expressions for anisotropic and/or tilted nodal-point semimetals are available in the literature \cite{pal22b_berry, 2021nag_nandy, ips-serena, ips-rahul-ph, ips-rahul-tilt, ips-ruiz}.

The results presented here can also be easily generalized to the cases of multiple nodes, and band-crossings involving more than two bands (e.g., 2D pseudospin-1 Dirac–Weyl semimetal \cite{bradlyn, claudia-multifold} and 3D quadratic band-touching semimetal \cite{abrikosov1996}). 
In presence of a long-ranged Coulomb potential, the nodal-point semimetals may transition into non-Fermi liquid phases when the Fermi level cuts a band-crossing point \cite{Abrikosov,Moon2013PRL,broy,ips-biref}. In such a scenario, the techniques discussed in this review will fail. The way out is to use various strong-coupling techniques, for example (a) the Kubo formalism, with the theory regulated using methods like large-N expansion and/or dimensional regularization \cite{Abrikosov,Moon2013PRL,ips-freire1,ips-freire-raman,ips-biref, ips-hermann-review}; and (b) the memory-matrix approach \cite{ips-freire1,ips-freire-thermo,ips-freire-raman, ips-hermann-review}.
Another possibility even in the presence of screened Coulomb interactions is the emergence of the plasmons depending on the parameter regime, which will then drastically affect the relaxation time \cite{kozii_plasmon,ips-plasmon,ips-jing}, necessitating the use of quantum field theoretic methods to compute any transport characteristics.
The full-fledged analysis of the effects of strong disorder on the transport properties is another aspect worthy of careful investigations, which also requires strong-coupling approaches \cite{ips-klaus,rahul-sid,ips-rahul,ips-qbt-sc,ips-biref}.
The same applies for short-ranged strong interactions \cite{kozii-cpge, ips-cpge, ips-hermann-review}.

 %=======================================================================
\section*{Acknowledgments}

We thank Rahul Ghosh for participating in various projects related to the parts involving three-dimensional semimetals. IM's research has received funding from the European Union's Horizon 2020 research and innovation
programme under the Marie Skłodowska-Curie grant agreement number 754340.

%%%%%%%%%%%%%%%%%%%%%%%%%%%%
\bibliography{biblio}

\begin{thebibliography}{137}%
\makeatletter
\providecommand \@ifxundefined [1]{%
 \@ifx{#1\undefined}
}%
\providecommand \@ifnum [1]{%
 \ifnum #1\expandafter \@firstoftwo
 \else \expandafter \@secondoftwo
 \fi
}%
\providecommand \@ifx [1]{%
 \ifx #1\expandafter \@firstoftwo
 \else \expandafter \@secondoftwo
 \fi
}%
\providecommand \natexlab [1]{#1}%
\providecommand \enquote  [1]{``#1''}%
\providecommand \bibnamefont  [1]{#1}%
\providecommand \bibfnamefont [1]{#1}%
\providecommand \citenamefont [1]{#1}%
\providecommand \href@noop [0]{\@secondoftwo}%
\providecommand \href [0]{\begingroup \@sanitize@url \@href}%
\providecommand \@href[1]{\@@startlink{#1}\@@href}%
\providecommand \@@href[1]{\endgroup#1\@@endlink}%
\providecommand \@sanitize@url [0]{\catcode `\\12\catcode `\$12\catcode
  `\&12\catcode `\#12\catcode `\^12\catcode `\_12\catcode `\%12\relax}%
\providecommand \@@startlink[1]{}%
\providecommand \@@endlink[0]{}%
\providecommand \url  [0]{\begingroup\@sanitize@url \@url }%
\providecommand \@url [1]{\endgroup\@href {#1}{\urlprefix }}%
\providecommand \urlprefix  [0]{URL }%
\providecommand \Eprint [0]{\href }%
\providecommand \doibase [0]{https://doi.org/}%
\providecommand \selectlanguage [0]{\@gobble}%
\providecommand \bibinfo  [0]{\@secondoftwo}%
\providecommand \bibfield  [0]{\@secondoftwo}%
\providecommand \translation [1]{[#1]}%
\providecommand \BibitemOpen [0]{}%
\providecommand \bibitemStop [0]{}%
\providecommand \bibitemNoStop [0]{.\EOS\space}%
\providecommand \EOS [0]{\spacefactor3000\relax}%
\providecommand \BibitemShut  [1]{\csname bibitem#1\endcsname}%
\let\auto@bib@innerbib\@empty
%</preamble>
\bibitem [{\citenamefont {Skinner}\ and\ \citenamefont
  {Fu}(2018)}]{Skinner2018}%
  \BibitemOpen
  \bibfield  {author} {\bibinfo {author} {\bibfnamefont {B.}~\bibnamefont
  {Skinner}}\ and\ \bibinfo {author} {\bibfnamefont {L.}~\bibnamefont {Fu}},\
  }\bibfield  {title} {\bibinfo {title} {Large, nonsaturating thermopower in a
  quantizing magnetic field},\ }\href {https://doi.org/DOI:
  10.1126/sciadv.aat2621} {\bibfield  {journal} {\bibinfo  {journal} {Science
  Advances}\ }\textbf {\bibinfo {volume} {4}},\ \bibinfo {pages} {2621}
  (\bibinfo {year} {2018})}\BibitemShut {NoStop}%
\bibitem [{\citenamefont {Bergman}\ and\ \citenamefont
  {Oganesyan}(2010)}]{Bergman2010}%
  \BibitemOpen
  \bibfield  {author} {\bibinfo {author} {\bibfnamefont {D.~L.}\ \bibnamefont
  {Bergman}}\ and\ \bibinfo {author} {\bibfnamefont {V.}~\bibnamefont
  {Oganesyan}},\ }\bibfield  {title} {\bibinfo {title} {Theory of
  dissipationless {N}ernst effects},\ }\href
  {https://doi.org/10.1103/PhysRevLett.104.066601} {\bibfield  {journal}
  {\bibinfo  {journal} {Phys. Rev. Lett.}\ }\textbf {\bibinfo {volume} {104}},\
  \bibinfo {pages} {066601} (\bibinfo {year} {2010})}\BibitemShut {NoStop}%
\bibitem [{\citenamefont {Wei}\ \emph {et~al.}(2009{\natexlab{a}})\citenamefont
  {Wei}, \citenamefont {Bao}, \citenamefont {Pu}, \citenamefont {Lau},\ and\
  \citenamefont {Shi}}]{Wei2009}%
  \BibitemOpen
  \bibfield  {author} {\bibinfo {author} {\bibfnamefont {P.}~\bibnamefont
  {Wei}}, \bibinfo {author} {\bibfnamefont {W.}~\bibnamefont {Bao}}, \bibinfo
  {author} {\bibfnamefont {Y.}~\bibnamefont {Pu}}, \bibinfo {author}
  {\bibfnamefont {C.~N.}\ \bibnamefont {Lau}},\ and\ \bibinfo {author}
  {\bibfnamefont {J.}~\bibnamefont {Shi}},\ }\bibfield  {title} {\bibinfo
  {title} {Anomalous thermoelectric transport of {D}irac particles in
  graphene},\ }\href {https://doi.org/10.1103/PhysRevLett.102.166808}
  {\bibfield  {journal} {\bibinfo  {journal} {Phys. Rev. Lett.}\ }\textbf
  {\bibinfo {volume} {102}},\ \bibinfo {pages} {166808} (\bibinfo {year}
  {2009}{\natexlab{a}})}\BibitemShut {NoStop}%
\bibitem [{\citenamefont {Zhu}\ \emph {et~al.}(2010)\citenamefont {Zhu},
  \citenamefont {Ma}, \citenamefont {Sheng}, \citenamefont {Liu},\ and\
  \citenamefont {Sheng}}]{zhu2010}%
  \BibitemOpen
  \bibfield  {author} {\bibinfo {author} {\bibfnamefont {L.}~\bibnamefont
  {Zhu}}, \bibinfo {author} {\bibfnamefont {R.}~\bibnamefont {Ma}}, \bibinfo
  {author} {\bibfnamefont {L.}~\bibnamefont {Sheng}}, \bibinfo {author}
  {\bibfnamefont {M.}~\bibnamefont {Liu}},\ and\ \bibinfo {author}
  {\bibfnamefont {D.-N.}\ \bibnamefont {Sheng}},\ }\bibfield  {title} {\bibinfo
  {title} {Universal thermoelectric effect of {D}irac fermions in graphene},\
  }\href {https://doi.org/10.1103/PhysRevLett.104.076804} {\bibfield  {journal}
  {\bibinfo  {journal} {Phys. Rev. Lett.}\ }\textbf {\bibinfo {volume} {104}},\
  \bibinfo {pages} {076804} (\bibinfo {year} {2010})}\BibitemShut {NoStop}%
\bibitem [{\citenamefont {Sharma}\ \emph {et~al.}(2016)\citenamefont {Sharma},
  \citenamefont {Goswami},\ and\ \citenamefont {Tewari}}]{girish1}%
  \BibitemOpen
  \bibfield  {author} {\bibinfo {author} {\bibfnamefont {G.}~\bibnamefont
  {Sharma}}, \bibinfo {author} {\bibfnamefont {P.}~\bibnamefont {Goswami}},\
  and\ \bibinfo {author} {\bibfnamefont {S.}~\bibnamefont {Tewari}},\
  }\bibfield  {title} {\bibinfo {title} {{N}ernst and magnetothermal
  conductivity in a lattice model of {W}eyl fermions},\ }\href
  {https://doi.org/10.1103/PhysRevB.93.035116} {\bibfield  {journal} {\bibinfo
  {journal} {Phys. Rev. B}\ }\textbf {\bibinfo {volume} {93}},\ \bibinfo
  {pages} {035116} (\bibinfo {year} {2016})}\BibitemShut {NoStop}%
\bibitem [{\citenamefont {Sharma}\ \emph
  {et~al.}(2017{\natexlab{a}})\citenamefont {Sharma}, \citenamefont {Moore},
  \citenamefont {Saha},\ and\ \citenamefont {Tewari}}]{Girish2017}%
  \BibitemOpen
  \bibfield  {author} {\bibinfo {author} {\bibfnamefont {G.}~\bibnamefont
  {Sharma}}, \bibinfo {author} {\bibfnamefont {C.}~\bibnamefont {Moore}},
  \bibinfo {author} {\bibfnamefont {S.}~\bibnamefont {Saha}},\ and\ \bibinfo
  {author} {\bibfnamefont {S.}~\bibnamefont {Tewari}},\ }\bibfield  {title}
  {\bibinfo {title} {{N}ernst effect in {D}irac and inversion-asymmetric {W}eyl
  semimetals},\ }\href {https://doi.org/10.1103/PhysRevB.96.195119} {\bibfield
  {journal} {\bibinfo  {journal} {Phys. Rev. B}\ }\textbf {\bibinfo {volume}
  {96}},\ \bibinfo {pages} {195119} (\bibinfo {year}
  {2017}{\natexlab{a}})}\BibitemShut {NoStop}%
\bibitem [{\citenamefont {Lundgren}\ \emph
  {et~al.}(2014{\natexlab{a}})\citenamefont {Lundgren}, \citenamefont
  {Laurell},\ and\ \citenamefont {Fiete}}]{Gegory2014}%
  \BibitemOpen
  \bibfield  {author} {\bibinfo {author} {\bibfnamefont {R.}~\bibnamefont
  {Lundgren}}, \bibinfo {author} {\bibfnamefont {P.}~\bibnamefont {Laurell}},\
  and\ \bibinfo {author} {\bibfnamefont {G.~A.}\ \bibnamefont {Fiete}},\
  }\bibfield  {title} {\bibinfo {title} {Thermoelectric properties of {W}eyl
  and {D}irac semimetals},\ }\href {https://doi.org/10.1103/PhysRevB.90.165115}
  {\bibfield  {journal} {\bibinfo  {journal} {Phys. Rev. B}\ }\textbf {\bibinfo
  {volume} {90}},\ \bibinfo {pages} {165115} (\bibinfo {year}
  {2014}{\natexlab{a}})}\BibitemShut {NoStop}%
\bibitem [{\citenamefont {Sharma}\ \emph
  {et~al.}(2017{\natexlab{b}})\citenamefont {Sharma}, \citenamefont {Goswami},\
  and\ \citenamefont {Tewari}}]{GirishTiwari2017}%
  \BibitemOpen
  \bibfield  {author} {\bibinfo {author} {\bibfnamefont {G.}~\bibnamefont
  {Sharma}}, \bibinfo {author} {\bibfnamefont {P.}~\bibnamefont {Goswami}},\
  and\ \bibinfo {author} {\bibfnamefont {S.}~\bibnamefont {Tewari}},\
  }\bibfield  {title} {\bibinfo {title} {Chiral anomaly and longitudinal
  magnetotransport in type-{II} {W}eyl semimetals},\ }\href
  {https://doi.org/10.1103/PhysRevB.96.045112} {\bibfield  {journal} {\bibinfo
  {journal} {Phys. Rev. B}\ }\textbf {\bibinfo {volume} {96}},\ \bibinfo
  {pages} {045112} (\bibinfo {year} {2017}{\natexlab{b}})}\BibitemShut
  {NoStop}%
\bibitem [{\citenamefont {Liang}\ \emph {et~al.}(2017)\citenamefont {Liang},
  \citenamefont {Lin}, \citenamefont {Gibson}, \citenamefont {Gao},
  \citenamefont {Hirschberger}, \citenamefont {Liu}, \citenamefont {Cava},\
  and\ \citenamefont {Ong}}]{Liang2017}%
  \BibitemOpen
  \bibfield  {author} {\bibinfo {author} {\bibfnamefont {T.}~\bibnamefont
  {Liang}}, \bibinfo {author} {\bibfnamefont {J.}~\bibnamefont {Lin}}, \bibinfo
  {author} {\bibfnamefont {Q.}~\bibnamefont {Gibson}}, \bibinfo {author}
  {\bibfnamefont {T.}~\bibnamefont {Gao}}, \bibinfo {author} {\bibfnamefont
  {M.}~\bibnamefont {Hirschberger}}, \bibinfo {author} {\bibfnamefont
  {M.}~\bibnamefont {Liu}}, \bibinfo {author} {\bibfnamefont {R.~J.}\
  \bibnamefont {Cava}},\ and\ \bibinfo {author} {\bibfnamefont {N.~P.}\
  \bibnamefont {Ong}},\ }\bibfield  {title} {\bibinfo {title} {Anomalous
  {N}ernst effect in the {D}irac semimetal {C}d$_3${A}s$_{2}$},\ }\href
  {https://doi.org/10.1103/PhysRevLett.118.136601} {\bibfield  {journal}
  {\bibinfo  {journal} {Phys. Rev. Lett.}\ }\textbf {\bibinfo {volume} {118}},\
  \bibinfo {pages} {136601} (\bibinfo {year} {2017})}\BibitemShut {NoStop}%
\bibitem [{\citenamefont {Chernodub}\ \emph {et~al.}(2018)\citenamefont
  {Chernodub}, \citenamefont {Cortijo},\ and\ \citenamefont
  {Vozmediano}}]{Vozmediano2018}%
  \BibitemOpen
  \bibfield  {author} {\bibinfo {author} {\bibfnamefont {M.~N.}\ \bibnamefont
  {Chernodub}}, \bibinfo {author} {\bibfnamefont {A.}~\bibnamefont {Cortijo}},\
  and\ \bibinfo {author} {\bibfnamefont {M.~A.~H.}\ \bibnamefont
  {Vozmediano}},\ }\bibfield  {title} {\bibinfo {title} {Generation of a
  {N}ernst current from the conformal anomaly in {D}irac and {W}eyl
  semimetals},\ }\href {https://doi.org/10.1103/PhysRevLett.120.206601}
  {\bibfield  {journal} {\bibinfo  {journal} {Phys. Rev. Lett.}\ }\textbf
  {\bibinfo {volume} {120}},\ \bibinfo {pages} {206601} (\bibinfo {year}
  {2018})}\BibitemShut {NoStop}%
\bibitem [{\citenamefont {{Nag}}\ and\ \citenamefont
  {{Nandy}}(2021)}]{2021nag_nandy}%
  \BibitemOpen
  \bibfield  {author} {\bibinfo {author} {\bibfnamefont {T.}~\bibnamefont
  {{Nag}}}\ and\ \bibinfo {author} {\bibfnamefont {S.}~\bibnamefont
  {{Nandy}}},\ }\bibfield  {title} {\bibinfo {title} {{Magneto-transport
  phenomena of type-I multi-Weyl semimetals in co-planar setups}},\ }\href
  {https://doi.org/10.1088/1361-648X/abc310} {\bibfield  {journal} {\bibinfo
  {journal} {Journal of Physics Condensed Matter}\ }\textbf {\bibinfo {volume}
  {33}},\ \bibinfo {eid} {075504} (\bibinfo {year} {2021})}\BibitemShut
  {NoStop}%
\bibitem [{\citenamefont {{Yadav}}\ \emph {et~al.}(2022)\citenamefont
  {{Yadav}}, \citenamefont {{Fazzini}},\ and\ \citenamefont
  {{Mandal}}}]{ips-serena}%
  \BibitemOpen
  \bibfield  {author} {\bibinfo {author} {\bibfnamefont {S.}~\bibnamefont
  {{Yadav}}}, \bibinfo {author} {\bibfnamefont {S.}~\bibnamefont {{Fazzini}}},\
  and\ \bibinfo {author} {\bibfnamefont {I.}~\bibnamefont {{Mandal}}},\
  }\bibfield  {title} {\bibinfo {title} {{Magneto-transport signatures in
  periodically-driven Weyl and multi-Weyl semimetals}},\ }\href
  {https://doi.org/10.1016/j.physe.2022.115444} {\bibfield  {journal} {\bibinfo
   {journal} {Physica E Low-Dimensional Systems and Nanostructures}\ }\textbf
  {\bibinfo {volume} {144}},\ \bibinfo {eid} {115444} (\bibinfo {year}
  {2022})}\BibitemShut {NoStop}%
\bibitem [{\citenamefont {Mandal}\ \emph {et~al.}(2020)\citenamefont {Mandal},
  \citenamefont {Das},\ and\ \citenamefont {Agarwal}}]{amit-magnus}%
  \BibitemOpen
  \bibfield  {author} {\bibinfo {author} {\bibfnamefont {D.}~\bibnamefont
  {Mandal}}, \bibinfo {author} {\bibfnamefont {K.}~\bibnamefont {Das}},\ and\
  \bibinfo {author} {\bibfnamefont {A.}~\bibnamefont {Agarwal}},\ }\bibfield
  {title} {\bibinfo {title} {Magnus {N}ernst and thermal {H}all effect},\
  }\href {https://doi.org/10.1103/PhysRevB.102.205414} {\bibfield  {journal}
  {\bibinfo  {journal} {Phys. Rev. B}\ }\textbf {\bibinfo {volume} {102}},\
  \bibinfo {pages} {205414} (\bibinfo {year} {2020})}\BibitemShut {NoStop}%
\bibitem [{\citenamefont {Papaj}\ and\ \citenamefont
  {Fu}(2019)}]{papaj_magnus}%
  \BibitemOpen
  \bibfield  {author} {\bibinfo {author} {\bibfnamefont {M.}~\bibnamefont
  {Papaj}}\ and\ \bibinfo {author} {\bibfnamefont {L.}~\bibnamefont {Fu}},\
  }\bibfield  {title} {\bibinfo {title} {Magnus {H}all effect},\ }\href
  {https://doi.org/10.1103/PhysRevLett.123.216802} {\bibfield  {journal}
  {\bibinfo  {journal} {Phys. Rev. Lett.}\ }\textbf {\bibinfo {volume} {123}},\
  \bibinfo {pages} {216802} (\bibinfo {year} {2019})}\BibitemShut {NoStop}%
\bibitem [{\citenamefont {{Sekh, Sajid}}\ and\ \citenamefont {{Mandal,
  Ipsita}}(2022)}]{sajid_magnus}%
  \BibitemOpen
  \bibfield  {author} {\bibinfo {author} {\bibnamefont {{Sekh, Sajid}}}\ and\
  \bibinfo {author} {\bibnamefont {{Mandal, Ipsita}}},\ }\bibfield  {title}
  {\bibinfo {title} {Magnus {H}all effect in three-dimensional topological
  semimetals},\ }\href {https://doi.org/10.1140/epjp/s13360-022-02840-2}
  {\bibfield  {journal} {\bibinfo  {journal} {Eur. Phys. J. Plus}\ }\textbf
  {\bibinfo {volume} {137}},\ \bibinfo {pages} {736} (\bibinfo {year}
  {2022})}\BibitemShut {NoStop}%
\bibitem [{\citenamefont {Sbierski}\ \emph {et~al.}(2014)\citenamefont
  {Sbierski}, \citenamefont {Pohl}, \citenamefont {Bergholtz},\ and\
  \citenamefont {Brouwer}}]{piet2014}%
  \BibitemOpen
  \bibfield  {author} {\bibinfo {author} {\bibfnamefont {B.}~\bibnamefont
  {Sbierski}}, \bibinfo {author} {\bibfnamefont {G.}~\bibnamefont {Pohl}},
  \bibinfo {author} {\bibfnamefont {E.~J.}\ \bibnamefont {Bergholtz}},\ and\
  \bibinfo {author} {\bibfnamefont {P.~W.}\ \bibnamefont {Brouwer}},\
  }\bibfield  {title} {\bibinfo {title} {Quantum transport of disordered {W}eyl
  semimetals at the nodal point},\ }\href
  {https://doi.org/10.1103/PhysRevLett.113.026602} {\bibfield  {journal}
  {\bibinfo  {journal} {Phys. Rev. Lett.}\ }\textbf {\bibinfo {volume} {113}},\
  \bibinfo {pages} {026602} (\bibinfo {year} {2014})}\BibitemShut {NoStop}%
\bibitem [{\citenamefont {Huang}\ \emph {et~al.}(2013)\citenamefont {Huang},
  \citenamefont {Arovas},\ and\ \citenamefont {Balatsky}}]{Huang2013}%
  \BibitemOpen
  \bibfield  {author} {\bibinfo {author} {\bibfnamefont {Z.}~\bibnamefont
  {Huang}}, \bibinfo {author} {\bibfnamefont {D.~P.}\ \bibnamefont {Arovas}},\
  and\ \bibinfo {author} {\bibfnamefont {A.~V.}\ \bibnamefont {Balatsky}},\
  }\bibfield  {title} {\bibinfo {title} {Impurity scattering in {W}eyl
  semimetals and their stability classification},\ }\href
  {https://doi.org/10.1088/1367-2630/15/12/123019} {\bibfield  {journal}
  {\bibinfo  {journal} {New Journal of Physics}\ }\textbf {\bibinfo {volume}
  {15}},\ \bibinfo {pages} {123019} (\bibinfo {year} {2013})}\BibitemShut
  {NoStop}%
\bibitem [{\citenamefont {Ominato}\ and\ \citenamefont
  {Koshino}(2014)}]{mikito2014}%
  \BibitemOpen
  \bibfield  {author} {\bibinfo {author} {\bibfnamefont {Y.}~\bibnamefont
  {Ominato}}\ and\ \bibinfo {author} {\bibfnamefont {M.}~\bibnamefont
  {Koshino}},\ }\bibfield  {title} {\bibinfo {title} {Quantum transport in a
  three-dimensional {W}eyl electron system},\ }\href
  {https://doi.org/10.1103/PhysRevB.89.054202} {\bibfield  {journal} {\bibinfo
  {journal} {Phys. Rev. B}\ }\textbf {\bibinfo {volume} {89}},\ \bibinfo
  {pages} {054202} (\bibinfo {year} {2014})}\BibitemShut {NoStop}%
\bibitem [{\citenamefont {Hosur}\ \emph {et~al.}(2012)\citenamefont {Hosur},
  \citenamefont {Parameswaran},\ and\ \citenamefont {Vishwanath}}]{hosur2012}%
  \BibitemOpen
  \bibfield  {author} {\bibinfo {author} {\bibfnamefont {P.}~\bibnamefont
  {Hosur}}, \bibinfo {author} {\bibfnamefont {S.~A.}\ \bibnamefont
  {Parameswaran}},\ and\ \bibinfo {author} {\bibfnamefont {A.}~\bibnamefont
  {Vishwanath}},\ }\bibfield  {title} {\bibinfo {title} {Charge transport in
  {W}eyl semimetals},\ }\href {https://doi.org/10.1103/PhysRevLett.108.046602}
  {\bibfield  {journal} {\bibinfo  {journal} {Phys. Rev. Lett.}\ }\textbf
  {\bibinfo {volume} {108}},\ \bibinfo {pages} {046602} (\bibinfo {year}
  {2012})}\BibitemShut {NoStop}%
\bibitem [{\citenamefont {Landsteiner}(2014)}]{karl2014}%
  \BibitemOpen
  \bibfield  {author} {\bibinfo {author} {\bibfnamefont {K.}~\bibnamefont
  {Landsteiner}},\ }\bibfield  {title} {\bibinfo {title} {Anomalous transport
  of {W}eyl fermions in {W}eyl semimetals},\ }\href
  {https://doi.org/10.1103/PhysRevB.89.075124} {\bibfield  {journal} {\bibinfo
  {journal} {Phys. Rev. B}\ }\textbf {\bibinfo {volume} {89}},\ \bibinfo
  {pages} {075124} (\bibinfo {year} {2014})}\BibitemShut {NoStop}%
\bibitem [{\citenamefont {Fauqu\'e}\ \emph {et~al.}(2013)\citenamefont
  {Fauqu\'e}, \citenamefont {Butch}, \citenamefont {Syers}, \citenamefont
  {Paglione}, \citenamefont {Wiedmann}, \citenamefont {Collaudin},
  \citenamefont {Grena}, \citenamefont {Zeitler},\ and\ \citenamefont
  {Behnia}}]{kamran2013}%
  \BibitemOpen
  \bibfield  {author} {\bibinfo {author} {\bibfnamefont {B.}~\bibnamefont
  {Fauqu\'e}}, \bibinfo {author} {\bibfnamefont {N.~P.}\ \bibnamefont {Butch}},
  \bibinfo {author} {\bibfnamefont {P.}~\bibnamefont {Syers}}, \bibinfo
  {author} {\bibfnamefont {J.}~\bibnamefont {Paglione}}, \bibinfo {author}
  {\bibfnamefont {S.}~\bibnamefont {Wiedmann}}, \bibinfo {author}
  {\bibfnamefont {A.}~\bibnamefont {Collaudin}}, \bibinfo {author}
  {\bibfnamefont {B.}~\bibnamefont {Grena}}, \bibinfo {author} {\bibfnamefont
  {U.}~\bibnamefont {Zeitler}},\ and\ \bibinfo {author} {\bibfnamefont
  {K.}~\bibnamefont {Behnia}},\ }\bibfield  {title} {\bibinfo {title}
  {Magnetothermoelectric properties of {B}i$_{2}${S}e$_{3}$},\ }\href
  {https://doi.org/10.1103/PhysRevB.87.035133} {\bibfield  {journal} {\bibinfo
  {journal} {Phys. Rev. B}\ }\textbf {\bibinfo {volume} {87}},\ \bibinfo
  {pages} {035133} (\bibinfo {year} {2013})}\BibitemShut {NoStop}%
\bibitem [{\citenamefont {Xiao}\ \emph {et~al.}(2005)\citenamefont {Xiao},
  \citenamefont {Shi},\ and\ \citenamefont {Niu}}]{prl_niu}%
  \BibitemOpen
  \bibfield  {author} {\bibinfo {author} {\bibfnamefont {D.}~\bibnamefont
  {Xiao}}, \bibinfo {author} {\bibfnamefont {J.}~\bibnamefont {Shi}},\ and\
  \bibinfo {author} {\bibfnamefont {Q.}~\bibnamefont {Niu}},\ }\bibfield
  {title} {\bibinfo {title} {Berry phase correction to electron density of
  states in solids},\ }\href {https://doi.org/10.1103/PhysRevLett.95.137204}
  {\bibfield  {journal} {\bibinfo  {journal} {Phys. Rev. Lett.}\ }\textbf
  {\bibinfo {volume} {95}},\ \bibinfo {pages} {137204} (\bibinfo {year}
  {2005})}\BibitemShut {NoStop}%
\bibitem [{\citenamefont {Zhu}\ \emph {et~al.}(2015)\citenamefont {Zhu},
  \citenamefont {Lin}, \citenamefont {Liu}, \citenamefont {Fauqu\'e},
  \citenamefont {Tao}, \citenamefont {Yang}, \citenamefont {Shi},\ and\
  \citenamefont {Behnia}}]{kamran2015}%
  \BibitemOpen
  \bibfield  {author} {\bibinfo {author} {\bibfnamefont {Z.}~\bibnamefont
  {Zhu}}, \bibinfo {author} {\bibfnamefont {X.}~\bibnamefont {Lin}}, \bibinfo
  {author} {\bibfnamefont {J.}~\bibnamefont {Liu}}, \bibinfo {author}
  {\bibfnamefont {B.}~\bibnamefont {Fauqu\'e}}, \bibinfo {author}
  {\bibfnamefont {Q.}~\bibnamefont {Tao}}, \bibinfo {author} {\bibfnamefont
  {C.}~\bibnamefont {Yang}}, \bibinfo {author} {\bibfnamefont {Y.}~\bibnamefont
  {Shi}},\ and\ \bibinfo {author} {\bibfnamefont {K.}~\bibnamefont {Behnia}},\
  }\bibfield  {title} {\bibinfo {title} {Quantum oscillations, thermoelectric
  coefficients, and the {F}ermi surface of semimetallic {WT}e$_{2}$},\ }\href
  {https://doi.org/10.1103/PhysRevLett.114.176601} {\bibfield  {journal}
  {\bibinfo  {journal} {Phys. Rev. Lett.}\ }\textbf {\bibinfo {volume} {114}},\
  \bibinfo {pages} {176601} (\bibinfo {year} {2015})}\BibitemShut {NoStop}%
\bibitem [{\citenamefont {Ferreiros}\ \emph {et~al.}(2017)\citenamefont
  {Ferreiros}, \citenamefont {Zyuzin},\ and\ \citenamefont
  {Bardarson}}]{Bardarson2017}%
  \BibitemOpen
  \bibfield  {author} {\bibinfo {author} {\bibfnamefont {Y.}~\bibnamefont
  {Ferreiros}}, \bibinfo {author} {\bibfnamefont {A.~A.}\ \bibnamefont
  {Zyuzin}},\ and\ \bibinfo {author} {\bibfnamefont {J.~H.}\ \bibnamefont
  {Bardarson}},\ }\bibfield  {title} {\bibinfo {title} {Anomalous {N}ernst and
  thermal {H}all effects in tilted {W}eyl semimetals},\ }\href
  {https://doi.org/10.1103/PhysRevB.96.115202} {\bibfield  {journal} {\bibinfo
  {journal} {Phys. Rev. B}\ }\textbf {\bibinfo {volume} {96}},\ \bibinfo
  {pages} {115202} (\bibinfo {year} {2017})}\BibitemShut {NoStop}%
\bibitem [{\citenamefont {Gorbar}\ \emph {et~al.}(2017)\citenamefont {Gorbar},
  \citenamefont {Miransky}, \citenamefont {Shovkovy},\ and\ \citenamefont
  {Sukhachov}}]{gorbar2017}%
  \BibitemOpen
  \bibfield  {author} {\bibinfo {author} {\bibfnamefont {E.~V.}\ \bibnamefont
  {Gorbar}}, \bibinfo {author} {\bibfnamefont {V.~A.}\ \bibnamefont
  {Miransky}}, \bibinfo {author} {\bibfnamefont {I.~A.}\ \bibnamefont
  {Shovkovy}},\ and\ \bibinfo {author} {\bibfnamefont {P.~O.}\ \bibnamefont
  {Sukhachov}},\ }\bibfield  {title} {\bibinfo {title} {Anomalous
  thermoelectric phenomena in lattice models of multi-{W}eyl semimetals},\
  }\href {https://doi.org/10.1103/PhysRevB.96.155138} {\bibfield  {journal}
  {\bibinfo  {journal} {Phys. Rev. B}\ }\textbf {\bibinfo {volume} {96}},\
  \bibinfo {pages} {155138} (\bibinfo {year} {2017})}\BibitemShut {NoStop}%
\bibitem [{\citenamefont {McCormick}\ \emph {et~al.}(2017)\citenamefont
  {McCormick}, \citenamefont {McKay},\ and\ \citenamefont
  {Trivedi}}]{trivedi2017}%
  \BibitemOpen
  \bibfield  {author} {\bibinfo {author} {\bibfnamefont {T.~M.}\ \bibnamefont
  {McCormick}}, \bibinfo {author} {\bibfnamefont {R.~C.}\ \bibnamefont
  {McKay}},\ and\ \bibinfo {author} {\bibfnamefont {N.}~\bibnamefont
  {Trivedi}},\ }\bibfield  {title} {\bibinfo {title} {Semiclassical theory of
  anomalous transport in type-{II} topological {W}eyl semimetals},\ }\href
  {https://doi.org/10.1103/PhysRevB.96.235116} {\bibfield  {journal} {\bibinfo
  {journal} {Phys. Rev. B}\ }\textbf {\bibinfo {volume} {96}},\ \bibinfo
  {pages} {235116} (\bibinfo {year} {2017})}\BibitemShut {NoStop}%
\bibitem [{\citenamefont {St\aa{}lhammar}\ \emph {et~al.}(2020)\citenamefont
  {St\aa{}lhammar}, \citenamefont {Larana-Aragon}, \citenamefont {Knolle},\
  and\ \citenamefont {Bergholtz}}]{emil-magneto}%
  \BibitemOpen
  \bibfield  {author} {\bibinfo {author} {\bibfnamefont {M.}~\bibnamefont
  {St\aa{}lhammar}}, \bibinfo {author} {\bibfnamefont {J.}~\bibnamefont
  {Larana-Aragon}}, \bibinfo {author} {\bibfnamefont {J.}~\bibnamefont
  {Knolle}},\ and\ \bibinfo {author} {\bibfnamefont {E.~J.}\ \bibnamefont
  {Bergholtz}},\ }\bibfield  {title} {\bibinfo {title} {{Magneto-optical
  conductivity in generic Weyl semimetals}},\ }\href
  {https://doi.org/10.1103/PhysRevB.102.235134} {\bibfield  {journal} {\bibinfo
   {journal} {Phys. Rev. B}\ }\textbf {\bibinfo {volume} {102}},\ \bibinfo
  {pages} {235134} (\bibinfo {year} {2020})}\BibitemShut {NoStop}%
\bibitem [{\citenamefont {{Yadav}}\ \emph {et~al.}(2023)\citenamefont
  {{Yadav}}, \citenamefont {{Sekh}},\ and\ \citenamefont
  {{Mandal}}}]{ips-magneto}%
  \BibitemOpen
  \bibfield  {author} {\bibinfo {author} {\bibfnamefont {S.}~\bibnamefont
  {{Yadav}}}, \bibinfo {author} {\bibfnamefont {S.}~\bibnamefont {{Sekh}}},\
  and\ \bibinfo {author} {\bibfnamefont {I.}~\bibnamefont {{Mandal}}},\
  }\bibfield  {title} {\bibinfo {title} {{Magneto-optical conductivity in the
  type-I and type-II phases of Weyl/multi-Weyl semimetals}},\ }\href
  {https://doi.org/10.1016/j.physb.2023.414765} {\bibfield  {journal} {\bibinfo
   {journal} {Physica B Condensed Matter}\ }\textbf {\bibinfo {volume} {656}},\
  \bibinfo {eid} {414765} (\bibinfo {year} {2023})}\BibitemShut {NoStop}%
\bibitem [{\citenamefont {Pardo}\ and\ \citenamefont {Pickett}(2009)}]{pardo}%
  \BibitemOpen
  \bibfield  {author} {\bibinfo {author} {\bibfnamefont {V.}~\bibnamefont
  {Pardo}}\ and\ \bibinfo {author} {\bibfnamefont {W.~E.}\ \bibnamefont
  {Pickett}},\ }\bibfield  {title} {\bibinfo {title} {Half-metallic
  semi-{D}irac-point generated by quantum confinement in {T}i{O}$_2/${VO}$_2$
  nanostructures},\ }\href {https://doi.org/10.1103/PhysRevLett.102.166803}
  {\bibfield  {journal} {\bibinfo  {journal} {Phys. Rev. Lett.}\ }\textbf
  {\bibinfo {volume} {102}},\ \bibinfo {pages} {166803} (\bibinfo {year}
  {2009})}\BibitemShut {NoStop}%
\bibitem [{\citenamefont {Pardo}\ and\ \citenamefont {Pickett}(2010)}]{pardo2}%
  \BibitemOpen
  \bibfield  {author} {\bibinfo {author} {\bibfnamefont {V.}~\bibnamefont
  {Pardo}}\ and\ \bibinfo {author} {\bibfnamefont {W.~E.}\ \bibnamefont
  {Pickett}},\ }\bibfield  {title} {\bibinfo {title} {Metal-insulator
  transition through a semi-{D}irac point in oxide nanostructures: {VO}$_{2}$
  (001) layers confined within {T}i{O}$_2$},\ }\href
  {https://doi.org/10.1103/PhysRevB.81.035111} {\bibfield  {journal} {\bibinfo
  {journal} {Phys. Rev. B}\ }\textbf {\bibinfo {volume} {81}},\ \bibinfo
  {pages} {035111} (\bibinfo {year} {2010})}\BibitemShut {NoStop}%
\bibitem [{\citenamefont {Banerjee}\ \emph {et~al.}(2009)\citenamefont
  {Banerjee}, \citenamefont {Singh}, \citenamefont {Pardo},\ and\ \citenamefont
  {Pickett}}]{banerjee}%
  \BibitemOpen
  \bibfield  {author} {\bibinfo {author} {\bibfnamefont {S.}~\bibnamefont
  {Banerjee}}, \bibinfo {author} {\bibfnamefont {R.~R.~P.}\ \bibnamefont
  {Singh}}, \bibinfo {author} {\bibfnamefont {V.}~\bibnamefont {Pardo}},\ and\
  \bibinfo {author} {\bibfnamefont {W.~E.}\ \bibnamefont {Pickett}},\
  }\bibfield  {title} {\bibinfo {title} {Tight-binding modeling and low-energy
  behavior of the semi-{D}irac point},\ }\href
  {https://doi.org/10.1103/PhysRevLett.103.016402} {\bibfield  {journal}
  {\bibinfo  {journal} {Phys. Rev. Lett.}\ }\textbf {\bibinfo {volume} {103}},\
  \bibinfo {pages} {016402} (\bibinfo {year} {2009})}\BibitemShut {NoStop}%
\bibitem [{\citenamefont {Kobayashi}\ \emph {et~al.}(2011)\citenamefont
  {Kobayashi}, \citenamefont {Suzumura}, \citenamefont {Pi\'echon},\ and\
  \citenamefont {Montambaux}}]{kobayashi}%
  \BibitemOpen
  \bibfield  {author} {\bibinfo {author} {\bibfnamefont {A.}~\bibnamefont
  {Kobayashi}}, \bibinfo {author} {\bibfnamefont {Y.}~\bibnamefont {Suzumura}},
  \bibinfo {author} {\bibfnamefont {F.}~\bibnamefont {Pi\'echon}},\ and\
  \bibinfo {author} {\bibfnamefont {G.}~\bibnamefont {Montambaux}},\ }\bibfield
   {title} {\bibinfo {title} {Emergence of {D}irac electron pair in the
  charge-ordered state of the organic conductor
  ${\alpha}$-({BEDT}-{TTF})$_2$i$_3$},\ }\href
  {https://doi.org/10.1103/PhysRevB.84.075450} {\bibfield  {journal} {\bibinfo
  {journal} {Phys. Rev. B}\ }\textbf {\bibinfo {volume} {84}},\ \bibinfo
  {pages} {075450} (\bibinfo {year} {2011})}\BibitemShut {NoStop}%
\bibitem [{\citenamefont {Suzumura}\ \emph {et~al.}(2013)\citenamefont
  {Suzumura}, \citenamefont {Morinari},\ and\ \citenamefont
  {Piéchon}}]{suzumura}%
  \BibitemOpen
  \bibfield  {author} {\bibinfo {author} {\bibfnamefont {Y.}~\bibnamefont
  {Suzumura}}, \bibinfo {author} {\bibfnamefont {T.}~\bibnamefont {Morinari}},\
  and\ \bibinfo {author} {\bibfnamefont {F.}~\bibnamefont {Piéchon}},\
  }\bibfield  {title} {\bibinfo {title} {Mechanism of {D}irac point in $\alpha$
  type organic conductor under pressure},\ }\href
  {https://doi.org/10.7566/JPSJ.82.023708} {\bibfield  {journal} {\bibinfo
  {journal} {Journal of the Physical Society of Japan}\ }\textbf {\bibinfo
  {volume} {82}},\ \bibinfo {pages} {023708} (\bibinfo {year}
  {2013})}\BibitemShut {NoStop}%
\bibitem [{\citenamefont {Hasegawa}\ \emph {et~al.}(2006)\citenamefont
  {Hasegawa}, \citenamefont {Konno}, \citenamefont {Nakano},\ and\
  \citenamefont {Kohmoto}}]{hasegawa}%
  \BibitemOpen
  \bibfield  {author} {\bibinfo {author} {\bibfnamefont {Y.}~\bibnamefont
  {Hasegawa}}, \bibinfo {author} {\bibfnamefont {R.}~\bibnamefont {Konno}},
  \bibinfo {author} {\bibfnamefont {H.}~\bibnamefont {Nakano}},\ and\ \bibinfo
  {author} {\bibfnamefont {M.}~\bibnamefont {Kohmoto}},\ }\bibfield  {title}
  {\bibinfo {title} {Zero modes of tight-binding electrons on the honeycomb
  lattice},\ }\href {https://doi.org/10.1103/PhysRevB.74.033413} {\bibfield
  {journal} {\bibinfo  {journal} {Phys. Rev. B}\ }\textbf {\bibinfo {volume}
  {74}},\ \bibinfo {pages} {033413} (\bibinfo {year} {2006})}\BibitemShut
  {NoStop}%
\bibitem [{\citenamefont {Adroguer}\ \emph {et~al.}(2016)\citenamefont
  {Adroguer}, \citenamefont {Carpentier}, \citenamefont {Montambaux},\ and\
  \citenamefont {Orignac}}]{orignac}%
  \BibitemOpen
  \bibfield  {author} {\bibinfo {author} {\bibfnamefont {P.}~\bibnamefont
  {Adroguer}}, \bibinfo {author} {\bibfnamefont {D.}~\bibnamefont
  {Carpentier}}, \bibinfo {author} {\bibfnamefont {G.}~\bibnamefont
  {Montambaux}},\ and\ \bibinfo {author} {\bibfnamefont {E.}~\bibnamefont
  {Orignac}},\ }\bibfield  {title} {\bibinfo {title} {Diffusion of {D}irac
  fermions across a topological merging transition in two dimensions},\ }\href
  {https://doi.org/10.1103/PhysRevB.93.125113} {\bibfield  {journal} {\bibinfo
  {journal} {Phys. Rev. B}\ }\textbf {\bibinfo {volume} {93}},\ \bibinfo
  {pages} {125113} (\bibinfo {year} {2016})}\BibitemShut {NoStop}%
\bibitem [{\citenamefont {Montambaux}\ \emph
  {et~al.}(2009{\natexlab{a}})\citenamefont {Montambaux}, \citenamefont
  {Pi\'echon}, \citenamefont {Fuchs},\ and\ \citenamefont
  {Goerbig}}]{montambaux1}%
  \BibitemOpen
  \bibfield  {author} {\bibinfo {author} {\bibfnamefont {G.}~\bibnamefont
  {Montambaux}}, \bibinfo {author} {\bibfnamefont {F.}~\bibnamefont
  {Pi\'echon}}, \bibinfo {author} {\bibfnamefont {J.-N.}\ \bibnamefont
  {Fuchs}},\ and\ \bibinfo {author} {\bibfnamefont {M.~O.}\ \bibnamefont
  {Goerbig}},\ }\bibfield  {title} {\bibinfo {title} {Merging of {D}irac points
  in a two-dimensional crystal},\ }\href
  {https://doi.org/10.1103/PhysRevB.80.153412} {\bibfield  {journal} {\bibinfo
  {journal} {Phys. Rev. B}\ }\textbf {\bibinfo {volume} {80}},\ \bibinfo
  {pages} {153412} (\bibinfo {year} {2009}{\natexlab{a}})}\BibitemShut
  {NoStop}%
\bibitem [{\citenamefont {Montambaux}\ \emph
  {et~al.}(2009{\natexlab{b}})\citenamefont {Montambaux}, \citenamefont
  {Pi{\'e}chon}, \citenamefont {Fuchs},\ and\ \citenamefont
  {Goerbig}}]{montambaux2}%
  \BibitemOpen
  \bibfield  {author} {\bibinfo {author} {\bibfnamefont {G.}~\bibnamefont
  {Montambaux}}, \bibinfo {author} {\bibfnamefont {F.}~\bibnamefont
  {Pi{\'e}chon}}, \bibinfo {author} {\bibfnamefont {J.-N.}\ \bibnamefont
  {Fuchs}},\ and\ \bibinfo {author} {\bibfnamefont {M.~O.}\ \bibnamefont
  {Goerbig}},\ }\bibfield  {title} {\bibinfo {title} {A universal {H}amiltonian
  for motion and merging of {D}irac points in a two-dimensional crystal},\
  }\href {https://doi.org/10.1140/epjb/e2009-00383-0} {\bibfield  {journal}
  {\bibinfo  {journal} {The European Physical Journal B}\ }\textbf {\bibinfo
  {volume} {72}},\ \bibinfo {pages} {509} (\bibinfo {year}
  {2009}{\natexlab{b}})}\BibitemShut {NoStop}%
\bibitem [{\citenamefont {Mandal}\ and\ \citenamefont {Saha}(2020)}]{ips-kush}%
  \BibitemOpen
  \bibfield  {author} {\bibinfo {author} {\bibfnamefont {I.}~\bibnamefont
  {Mandal}}\ and\ \bibinfo {author} {\bibfnamefont {K.}~\bibnamefont {Saha}},\
  }\bibfield  {title} {\bibinfo {title} {{Thermopower in an anisotropic
  two-dimensional Weyl semimetal}},\ }\href
  {https://doi.org/10.1103/PhysRevB.101.045101} {\bibfield  {journal} {\bibinfo
   {journal} {Phys. Rev. B}\ }\textbf {\bibinfo {volume} {101}},\ \bibinfo
  {pages} {045101} (\bibinfo {year} {2020})}\BibitemShut {NoStop}%
\bibitem [{\citenamefont {Dietl}\ \emph {et~al.}(2008)\citenamefont {Dietl},
  \citenamefont {Pi\'echon},\ and\ \citenamefont {Montambaux}}]{landau-level}%
  \BibitemOpen
  \bibfield  {author} {\bibinfo {author} {\bibfnamefont {P.}~\bibnamefont
  {Dietl}}, \bibinfo {author} {\bibfnamefont {F.}~\bibnamefont {Pi\'echon}},\
  and\ \bibinfo {author} {\bibfnamefont {G.}~\bibnamefont {Montambaux}},\
  }\bibfield  {title} {\bibinfo {title} {New magnetic field dependence of
  {L}andau levels in a graphenelike structure},\ }\href
  {https://doi.org/10.1103/PhysRevLett.100.236405} {\bibfield  {journal}
  {\bibinfo  {journal} {Phys. Rev. Lett.}\ }\textbf {\bibinfo {volume} {100}},\
  \bibinfo {pages} {236405} (\bibinfo {year} {2008})}\BibitemShut {NoStop}%
\bibitem [{\citenamefont {{Cho}}\ and\ \citenamefont {{Moon}}(2016)}]{moon}%
  \BibitemOpen
  \bibfield  {author} {\bibinfo {author} {\bibfnamefont {G.~Y.}\ \bibnamefont
  {{Cho}}}\ and\ \bibinfo {author} {\bibfnamefont {E.-G.}\ \bibnamefont
  {{Moon}}},\ }\bibfield  {title} {\bibinfo {title} {Novel quantum criticality
  in two dimensional topological phase transitions},\ }\href
  {https://doi.org/10.1038/srep19198} {\bibfield  {journal} {\bibinfo
  {journal} {Scientific Reports}\ }\textbf {\bibinfo {volume} {6}},\ \bibinfo
  {eid} {19198} (\bibinfo {year} {2016})}\BibitemShut {NoStop}%
\bibitem [{\citenamefont {Chen}\ and\ \citenamefont
  {Fiete}(2016{\natexlab{a}})}]{Gegory2016}%
  \BibitemOpen
  \bibfield  {author} {\bibinfo {author} {\bibfnamefont {Q.}~\bibnamefont
  {Chen}}\ and\ \bibinfo {author} {\bibfnamefont {G.~A.}\ \bibnamefont
  {Fiete}},\ }\bibfield  {title} {\bibinfo {title} {Thermoelectric transport in
  double-{W}eyl semimetals},\ }\href
  {https://doi.org/10.1103/PhysRevB.93.155125} {\bibfield  {journal} {\bibinfo
  {journal} {Phys. Rev. B}\ }\textbf {\bibinfo {volume} {93}},\ \bibinfo
  {pages} {155125} (\bibinfo {year} {2016}{\natexlab{a}})}\BibitemShut
  {NoStop}%
\bibitem [{\citenamefont {Ashcroft}\ and\ \citenamefont
  {Mermin}(2011)}]{mermin}%
  \BibitemOpen
  \bibfield  {author} {\bibinfo {author} {\bibfnamefont {N.}~\bibnamefont
  {Ashcroft}}\ and\ \bibinfo {author} {\bibfnamefont {N.}~\bibnamefont
  {Mermin}},\ }\href {https://books.google.de/books?id=x\_s\_YAAACAAJ} {\emph
  {\bibinfo {title} {Solid State Physics}}}\ (\bibinfo  {publisher} {Cengage
  Learning},\ \bibinfo {year} {2011})\BibitemShut {NoStop}%
\bibitem [{\citenamefont {Tong}(2012)}]{tong}%
  \BibitemOpen
  \bibfield  {author} {\bibinfo {author} {\bibfnamefont {D.}~\bibnamefont
  {Tong}},\ }\href {http://www.damtp.cam.ac.uk/user/tong/kinetic.html} {\emph
  {\bibinfo {title} {Lectures on Kinetic Theory}}}\ (\bibinfo {year}
  {2012})\BibitemShut {NoStop}%
\bibitem [{\citenamefont {Arovas}(2014)}]{arovas}%
  \BibitemOpen
  \bibfield  {author} {\bibinfo {author} {\bibfnamefont {D.}~\bibnamefont
  {Arovas}},\ }\href {https://books.google.co.in/books?id=dj7brQEACAAJ} {\emph
  {\bibinfo {title} {Lecture Notes on Condensed Matter Physics}}}\ (\bibinfo
  {publisher} {CreateSpace Independent Publishing Platform},\ \bibinfo {year}
  {2014})\BibitemShut {NoStop}%
\bibitem [{\citenamefont {Soto}(2016)}]{soto}%
  \BibitemOpen
  \bibfield  {author} {\bibinfo {author} {\bibfnamefont {R.}~\bibnamefont
  {Soto}},\ }\href {https://books.google.co.in/books?id=FclLDQAAQBAJ} {\emph
  {\bibinfo {title} {Kinetic Theory and Transport Phenomena}}},\ Oxford master
  series in condensed matter physics\ (\bibinfo  {publisher} {Oxford University
  Press},\ \bibinfo {year} {2016})\BibitemShut {NoStop}%
\bibitem [{\citenamefont {Onsager}(1931)}]{onsager}%
  \BibitemOpen
  \bibfield  {author} {\bibinfo {author} {\bibfnamefont {L.}~\bibnamefont
  {Onsager}},\ }\bibfield  {title} {\bibinfo {title} {Reciprocal relations in
  irreversible processes. {I}.},\ }\href
  {https://doi.org/10.1103/PhysRev.37.405} {\bibfield  {journal} {\bibinfo
  {journal} {Phys. Rev.}\ }\textbf {\bibinfo {volume} {37}},\ \bibinfo {pages}
  {405} (\bibinfo {year} {1931})}\BibitemShut {NoStop}%
\bibitem [{\citenamefont {Nielsen}\ and\ \citenamefont
  {Ninomiya}(1981)}]{nielsen}%
  \BibitemOpen
  \bibfield  {author} {\bibinfo {author} {\bibfnamefont {H.}~\bibnamefont
  {Nielsen}}\ and\ \bibinfo {author} {\bibfnamefont {M.}~\bibnamefont
  {Ninomiya}},\ }\bibfield  {title} {\bibinfo {title} {A no-go theorem for
  regularizing chiral fermions},\ }\href
  {https://doi.org/https://doi.org/10.1016/0370-2693(81)91026-1} {\bibfield
  {journal} {\bibinfo  {journal} {Physics Letters B}\ }\textbf {\bibinfo
  {volume} {105}},\ \bibinfo {pages} {219} (\bibinfo {year}
  {1981})}\BibitemShut {NoStop}%
\bibitem [{\citenamefont {Sundaram}\ and\ \citenamefont
  {Niu}(1999)}]{Sundurum:1999}%
  \BibitemOpen
  \bibfield  {author} {\bibinfo {author} {\bibfnamefont {G.}~\bibnamefont
  {Sundaram}}\ and\ \bibinfo {author} {\bibfnamefont {Q.}~\bibnamefont {Niu}},\
  }\bibfield  {title} {\bibinfo {title} {{Wave-packet dynamics in slowly
  perturbed crystals: Gradient corrections and Berry-phase effects}},\ }\href
  {https://doi.org/10.1103/PhysRevB.59.14915} {\bibfield  {journal} {\bibinfo
  {journal} {Phys. Rev. B}\ }\textbf {\bibinfo {volume} {59}},\ \bibinfo
  {pages} {14915} (\bibinfo {year} {1999})}\BibitemShut {NoStop}%
\bibitem [{\citenamefont {Son}\ and\ \citenamefont
  {Spivak}(2013)}]{son13_chiral}%
  \BibitemOpen
  \bibfield  {author} {\bibinfo {author} {\bibfnamefont {D.~T.}\ \bibnamefont
  {Son}}\ and\ \bibinfo {author} {\bibfnamefont {B.~Z.}\ \bibnamefont
  {Spivak}},\ }\bibfield  {title} {\bibinfo {title} {{Chiral anomaly and
  classical negative magnetoresistance of Weyl metals}},\ }\href
  {https://doi.org/10.1103/PhysRevB.88.104412} {\bibfield  {journal} {\bibinfo
  {journal} {Phys. Rev. B}\ }\textbf {\bibinfo {volume} {88}},\ \bibinfo
  {pages} {104412} (\bibinfo {year} {2013})}\BibitemShut {NoStop}%
\bibitem [{\citenamefont {Xiao}\ \emph {et~al.}(2010)\citenamefont {Xiao},
  \citenamefont {Chang},\ and\ \citenamefont {Niu}}]{xiao10_Berry}%
  \BibitemOpen
  \bibfield  {author} {\bibinfo {author} {\bibfnamefont {D.}~\bibnamefont
  {Xiao}}, \bibinfo {author} {\bibfnamefont {M.-C.}\ \bibnamefont {Chang}},\
  and\ \bibinfo {author} {\bibfnamefont {Q.}~\bibnamefont {Niu}},\ }\bibfield
  {title} {\bibinfo {title} {Berry phase effects on electronic properties},\
  }\href {https://doi.org/10.1103/RevModPhys.82.1959} {\bibfield  {journal}
  {\bibinfo  {journal} {Rev. Mod. Phys.}\ }\textbf {\bibinfo {volume} {82}},\
  \bibinfo {pages} {1959} (\bibinfo {year} {2010})}\BibitemShut {NoStop}%
\bibitem [{\citenamefont {Duval}\ \emph {et~al.}(2006)\citenamefont {Duval},
  \citenamefont {Horv{\'a}th}, \citenamefont {Horvathy}, \citenamefont
  {Martina},\ and\ \citenamefont {Stichel}}]{duval06_Berry}%
  \BibitemOpen
  \bibfield  {author} {\bibinfo {author} {\bibfnamefont {C.}~\bibnamefont
  {Duval}}, \bibinfo {author} {\bibfnamefont {Z.}~\bibnamefont {Horv{\'a}th}},
  \bibinfo {author} {\bibfnamefont {P.~A.}\ \bibnamefont {Horvathy}}, \bibinfo
  {author} {\bibfnamefont {L.}~\bibnamefont {Martina}},\ and\ \bibinfo {author}
  {\bibfnamefont {P.}~\bibnamefont {Stichel}},\ }\bibfield  {title} {\bibinfo
  {title} {Berry phase correction to electron density in solids and ``exotic''
  dynamics},\ }\href {https://doi.org/10.1142/S0217984906010573} {\bibfield
  {journal} {\bibinfo  {journal} {Mod. Phys. Lett. B}\ }\textbf {\bibinfo
  {volume} {20}},\ \bibinfo {pages} {373} (\bibinfo {year} {2006})}\BibitemShut
  {NoStop}%
\bibitem [{\citenamefont {Son}\ and\ \citenamefont
  {Yamamoto}(2012)}]{Son:2012}%
  \BibitemOpen
  \bibfield  {author} {\bibinfo {author} {\bibfnamefont {D.~T.}\ \bibnamefont
  {Son}}\ and\ \bibinfo {author} {\bibfnamefont {N.}~\bibnamefont {Yamamoto}},\
  }\bibfield  {title} {\bibinfo {title} {Berry curvature, triangle anomalies,
  and the chiral magnetic effect in fermi liquids},\ }\href
  {https://doi.org/10.1103/PhysRevLett.109.181602} {\bibfield  {journal}
  {\bibinfo  {journal} {Phys. Rev. Lett.}\ }\textbf {\bibinfo {volume} {109}},\
  \bibinfo {pages} {181602} (\bibinfo {year} {2012})}\BibitemShut {NoStop}%
\bibitem [{\citenamefont {Lundgren}\ \emph
  {et~al.}(2014{\natexlab{b}})\citenamefont {Lundgren}, \citenamefont
  {Laurell},\ and\ \citenamefont {Fiete}}]{lundgren14_thermoelectric}%
  \BibitemOpen
  \bibfield  {author} {\bibinfo {author} {\bibfnamefont {R.}~\bibnamefont
  {Lundgren}}, \bibinfo {author} {\bibfnamefont {P.}~\bibnamefont {Laurell}},\
  and\ \bibinfo {author} {\bibfnamefont {G.~A.}\ \bibnamefont {Fiete}},\
  }\bibfield  {title} {\bibinfo {title} {Thermoelectric properties of {W}eyl
  and {D}irac semimetals},\ }\href {https://doi.org/10.1103/PhysRevB.90.165115}
  {\bibfield  {journal} {\bibinfo  {journal} {Phys. Rev. B}\ }\textbf {\bibinfo
  {volume} {90}},\ \bibinfo {pages} {165115} (\bibinfo {year}
  {2014}{\natexlab{b}})}\BibitemShut {NoStop}%
\bibitem [{\citenamefont {Das}\ and\ \citenamefont
  {Agarwal}(2019)}]{das19_linear}%
  \BibitemOpen
  \bibfield  {author} {\bibinfo {author} {\bibfnamefont {K.}~\bibnamefont
  {Das}}\ and\ \bibinfo {author} {\bibfnamefont {A.}~\bibnamefont {Agarwal}},\
  }\bibfield  {title} {\bibinfo {title} {Linear magnetochiral transport in
  tilted type-{I} and type-{II} {W}eyl semimetals},\ }\href
  {https://doi.org/10.1103/PhysRevB.99.085405} {\bibfield  {journal} {\bibinfo
  {journal} {Phys. Rev. B}\ }\textbf {\bibinfo {volume} {99}},\ \bibinfo
  {pages} {085405} (\bibinfo {year} {2019})}\BibitemShut {NoStop}%
\bibitem [{\citenamefont {Nandy}\ \emph {et~al.}(2017)\citenamefont {Nandy},
  \citenamefont {Sharma}, \citenamefont {Taraphder},\ and\ \citenamefont
  {Tewari}}]{nandy_2017_chiral}%
  \BibitemOpen
  \bibfield  {author} {\bibinfo {author} {\bibfnamefont {S.}~\bibnamefont
  {Nandy}}, \bibinfo {author} {\bibfnamefont {G.}~\bibnamefont {Sharma}},
  \bibinfo {author} {\bibfnamefont {A.}~\bibnamefont {Taraphder}},\ and\
  \bibinfo {author} {\bibfnamefont {S.}~\bibnamefont {Tewari}},\ }\bibfield
  {title} {\bibinfo {title} {Chiral anomaly as the origin of the planar {H}all
  effect in {W}eyl semimetals},\ }\href
  {https://doi.org/10.1103/PhysRevLett.119.176804} {\bibfield  {journal}
  {\bibinfo  {journal} {Phys. Rev. Lett.}\ }\textbf {\bibinfo {volume} {119}},\
  \bibinfo {pages} {176804} (\bibinfo {year} {2017})}\BibitemShut {NoStop}%
\bibitem [{\citenamefont {Nandy}\ \emph {et~al.}(2019)\citenamefont {Nandy},
  \citenamefont {Taraphder},\ and\ \citenamefont
  {Tewari}}]{nandy_thermal_hall}%
  \BibitemOpen
  \bibfield  {author} {\bibinfo {author} {\bibfnamefont {S.}~\bibnamefont
  {Nandy}}, \bibinfo {author} {\bibfnamefont {A.}~\bibnamefont {Taraphder}},\
  and\ \bibinfo {author} {\bibfnamefont {S.}~\bibnamefont {Tewari}},\
  }\bibfield  {title} {\bibinfo {title} {{Planar thermal Hall effect in Weyl
  semimetals}},\ }\href {https://doi.org/10.1103/PhysRevB.100.115139}
  {\bibfield  {journal} {\bibinfo  {journal} {Phys. Rev. B}\ }\textbf {\bibinfo
  {volume} {100}},\ \bibinfo {pages} {115139} (\bibinfo {year}
  {2019})}\BibitemShut {NoStop}%
\bibitem [{\citenamefont {Xiao}\ and\ \citenamefont
  {Niu}(2020)}]{prb101235430}%
  \BibitemOpen
  \bibfield  {author} {\bibinfo {author} {\bibfnamefont {C.}~\bibnamefont
  {Xiao}}\ and\ \bibinfo {author} {\bibfnamefont {Q.}~\bibnamefont {Niu}},\
  }\bibfield  {title} {\bibinfo {title} {Unified bulk semiclassical theory for
  intrinsic thermal transport and magnetization currents},\ }\href
  {https://doi.org/10.1103/PhysRevB.101.235430} {\bibfield  {journal} {\bibinfo
   {journal} {Phys. Rev. B}\ }\textbf {\bibinfo {volume} {101}},\ \bibinfo
  {pages} {235430} (\bibinfo {year} {2020})}\BibitemShut {NoStop}%
\bibitem [{\citenamefont {Qin}\ \emph {et~al.}(2011)\citenamefont {Qin},
  \citenamefont {Niu},\ and\ \citenamefont {Shi}}]{tao}%
  \BibitemOpen
  \bibfield  {author} {\bibinfo {author} {\bibfnamefont {T.}~\bibnamefont
  {Qin}}, \bibinfo {author} {\bibfnamefont {Q.}~\bibnamefont {Niu}},\ and\
  \bibinfo {author} {\bibfnamefont {J.}~\bibnamefont {Shi}},\ }\bibfield
  {title} {\bibinfo {title} {Energy magnetization and the thermal {H}all
  effect},\ }\href {https://doi.org/10.1103/PhysRevLett.107.236601} {\bibfield
  {journal} {\bibinfo  {journal} {Phys. Rev. Lett.}\ }\textbf {\bibinfo
  {volume} {107}},\ \bibinfo {pages} {236601} (\bibinfo {year}
  {2011})}\BibitemShut {NoStop}%
\bibitem [{\citenamefont {Zhang}(2016)}]{Zhang_2016}%
  \BibitemOpen
  \bibfield  {author} {\bibinfo {author} {\bibfnamefont {L.}~\bibnamefont
  {Zhang}},\ }\bibfield  {title} {\bibinfo {title} {{Berry curvature and
  various thermal Hall effects}},\ }\href
  {https://doi.org/10.1088/1367-2630/18/10/103039} {\bibfield  {journal}
  {\bibinfo  {journal} {New Journal of Physics}\ }\textbf {\bibinfo {volume}
  {18}},\ \bibinfo {pages} {103039} (\bibinfo {year} {2016})}\BibitemShut
  {NoStop}%
\bibitem [{\citenamefont {Cooper}\ \emph {et~al.}(1997)\citenamefont {Cooper},
  \citenamefont {Halperin},\ and\ \citenamefont {Ruzin}}]{cooper_omm}%
  \BibitemOpen
  \bibfield  {author} {\bibinfo {author} {\bibfnamefont {N.~R.}\ \bibnamefont
  {Cooper}}, \bibinfo {author} {\bibfnamefont {B.~I.}\ \bibnamefont
  {Halperin}},\ and\ \bibinfo {author} {\bibfnamefont {I.~M.}\ \bibnamefont
  {Ruzin}},\ }\bibfield  {title} {\bibinfo {title} {Thermoelectric response of
  an interacting two-dimensional electron gas in a quantizing magnetic field},\
  }\href {https://doi.org/10.1103/PhysRevB.55.2344} {\bibfield  {journal}
  {\bibinfo  {journal} {Phys. Rev. B}\ }\textbf {\bibinfo {volume} {55}},\
  \bibinfo {pages} {2344} (\bibinfo {year} {1997})}\BibitemShut {NoStop}%
\bibitem [{\citenamefont {Burkov}(2017)}]{burkov17_giant}%
  \BibitemOpen
  \bibfield  {author} {\bibinfo {author} {\bibfnamefont {A.~A.}\ \bibnamefont
  {Burkov}},\ }\bibfield  {title} {\bibinfo {title} {Giant planar {H}all effect
  in topological metals},\ }\href {https://doi.org/10.1103/PhysRevB.96.041110}
  {\bibfield  {journal} {\bibinfo  {journal} {Phys. Rev. B}\ }\textbf {\bibinfo
  {volume} {96}},\ \bibinfo {pages} {041110} (\bibinfo {year}
  {2017})}\BibitemShut {NoStop}%
\bibitem [{\citenamefont {Li}\ \emph {et~al.}(2017)\citenamefont {Li},
  \citenamefont {Wang}, \citenamefont {Li}, \citenamefont {Yang}, \citenamefont
  {Shen}, \citenamefont {Sheng}, \citenamefont {Li}, \citenamefont {Lu},
  \citenamefont {Zheng},\ and\ \citenamefont {Xu}}]{li_nmr17}%
  \BibitemOpen
  \bibfield  {author} {\bibinfo {author} {\bibfnamefont {Y.}~\bibnamefont
  {Li}}, \bibinfo {author} {\bibfnamefont {Z.}~\bibnamefont {Wang}}, \bibinfo
  {author} {\bibfnamefont {P.}~\bibnamefont {Li}}, \bibinfo {author}
  {\bibfnamefont {X.}~\bibnamefont {Yang}}, \bibinfo {author} {\bibfnamefont
  {Z.}~\bibnamefont {Shen}}, \bibinfo {author} {\bibfnamefont {F.}~\bibnamefont
  {Sheng}}, \bibinfo {author} {\bibfnamefont {X.}~\bibnamefont {Li}}, \bibinfo
  {author} {\bibfnamefont {Y.}~\bibnamefont {Lu}}, \bibinfo {author}
  {\bibfnamefont {Y.}~\bibnamefont {Zheng}},\ and\ \bibinfo {author}
  {\bibfnamefont {Z.-A.}\ \bibnamefont {Xu}},\ }\bibfield  {title} {\bibinfo
  {title} {{Negative magnetoresistance in {W}eyl semimetals NbAs and NbP:
  Intrinsic chiral anomaly and extrinsic effects}},\ }\href
  {https://doi.org/10.1007/s11467-016-0636-8} {\bibfield  {journal} {\bibinfo
  {journal} {Frontiers of Physics}\ }\textbf {\bibinfo {volume} {12}},\
  \bibinfo {pages} {127205} (\bibinfo {year} {2017})}\BibitemShut {NoStop}%
\bibitem [{\citenamefont {Nandy}\ \emph {et~al.}(2018)\citenamefont {Nandy},
  \citenamefont {Taraphder},\ and\ \citenamefont {Tewari}}]{nandy18_Berry}%
  \BibitemOpen
  \bibfield  {author} {\bibinfo {author} {\bibfnamefont {S.}~\bibnamefont
  {Nandy}}, \bibinfo {author} {\bibfnamefont {A.}~\bibnamefont {Taraphder}},\
  and\ \bibinfo {author} {\bibfnamefont {S.}~\bibnamefont {Tewari}},\
  }\bibfield  {title} {\bibinfo {title} {Berry phase theory of planar {H}all
  effect in topological insulators},\ }\href
  {https://doi.org/10.1038/s41598-018-33258-5} {\bibfield  {journal} {\bibinfo
  {journal} {Scientific Reports}\ }\textbf {\bibinfo {volume} {8}},\ \bibinfo
  {pages} {14983} (\bibinfo {year} {2018})}\BibitemShut {NoStop}%
\bibitem [{\citenamefont {Zhang}\ \emph {et~al.}(2016)\citenamefont {Zhang},
  \citenamefont {Lu},\ and\ \citenamefont {Shen}}]{zhang16_linear}%
  \BibitemOpen
  \bibfield  {author} {\bibinfo {author} {\bibfnamefont {S.-B.}\ \bibnamefont
  {Zhang}}, \bibinfo {author} {\bibfnamefont {H.-Z.}\ \bibnamefont {Lu}},\ and\
  \bibinfo {author} {\bibfnamefont {S.-Q.}\ \bibnamefont {Shen}},\ }\bibfield
  {title} {\bibinfo {title} {{Linear magnetoconductivity in an intrinsic
  topological Weyl semimetal}},\ }\href
  {https://doi.org/10.1088/1367-2630/18/5/053039} {\bibfield  {journal}
  {\bibinfo  {journal} {New Journal of Physics}\ }\textbf {\bibinfo {volume}
  {18}},\ \bibinfo {pages} {053039} (\bibinfo {year} {2016})}\BibitemShut
  {NoStop}%
\bibitem [{\citenamefont {Chen}\ and\ \citenamefont
  {Fiete}(2016{\natexlab{b}})}]{chen16_thermoelectric}%
  \BibitemOpen
  \bibfield  {author} {\bibinfo {author} {\bibfnamefont {Q.}~\bibnamefont
  {Chen}}\ and\ \bibinfo {author} {\bibfnamefont {G.~A.}\ \bibnamefont
  {Fiete}},\ }\bibfield  {title} {\bibinfo {title} {{Thermoelectric transport
  in double-Weyl semimetals}},\ }\href
  {https://doi.org/10.1103/PhysRevB.93.155125} {\bibfield  {journal} {\bibinfo
  {journal} {Phys. Rev. B}\ }\textbf {\bibinfo {volume} {93}},\ \bibinfo
  {pages} {155125} (\bibinfo {year} {2016}{\natexlab{b}})}\BibitemShut
  {NoStop}%
\bibitem [{\citenamefont {Das}\ and\ \citenamefont
  {Agarwal}(2020)}]{das20_thermal}%
  \BibitemOpen
  \bibfield  {author} {\bibinfo {author} {\bibfnamefont {K.}~\bibnamefont
  {Das}}\ and\ \bibinfo {author} {\bibfnamefont {A.}~\bibnamefont {Agarwal}},\
  }\bibfield  {title} {\bibinfo {title} {Thermal and gravitational chiral
  anomaly induced magneto-transport in {W}eyl semimetals},\ }\href
  {https://doi.org/10.1103/PhysRevResearch.2.013088} {\bibfield  {journal}
  {\bibinfo  {journal} {Phys. Rev. Res.}\ }\textbf {\bibinfo {volume} {2}},\
  \bibinfo {pages} {013088} (\bibinfo {year} {2020})}\BibitemShut {NoStop}%
\bibitem [{\citenamefont {Das}\ \emph {et~al.}(2022)\citenamefont {Das},
  \citenamefont {Das},\ and\ \citenamefont {Agarwal}}]{das22_nonlinear}%
  \BibitemOpen
  \bibfield  {author} {\bibinfo {author} {\bibfnamefont {S.}~\bibnamefont
  {Das}}, \bibinfo {author} {\bibfnamefont {K.}~\bibnamefont {Das}},\ and\
  \bibinfo {author} {\bibfnamefont {A.}~\bibnamefont {Agarwal}},\ }\bibfield
  {title} {\bibinfo {title} {{Nonlinear magnetoconductivity in Weyl and
  multi-Weyl semimetals in quantizing magnetic field}},\ }\href
  {https://doi.org/10.1103/PhysRevB.105.235408} {\bibfield  {journal} {\bibinfo
   {journal} {Phys. Rev. B}\ }\textbf {\bibinfo {volume} {105}},\ \bibinfo
  {pages} {235408} (\bibinfo {year} {2022})}\BibitemShut {NoStop}%
\bibitem [{\citenamefont {Pal}\ \emph {et~al.}(2022{\natexlab{a}})\citenamefont
  {Pal}, \citenamefont {Dey},\ and\ \citenamefont {Ghosh}}]{pal22a_berry}%
  \BibitemOpen
  \bibfield  {author} {\bibinfo {author} {\bibfnamefont {O.}~\bibnamefont
  {Pal}}, \bibinfo {author} {\bibfnamefont {B.}~\bibnamefont {Dey}},\ and\
  \bibinfo {author} {\bibfnamefont {T.~K.}\ \bibnamefont {Ghosh}},\ }\bibfield
  {title} {\bibinfo {title} {Berry curvature induced magnetotransport in 3d
  noncentrosymmetric metals},\ }\href
  {https://doi.org/10.1088/1361-648X/ac2fd4} {\bibfield  {journal} {\bibinfo
  {journal} {Journal of Physics: Condensed Matter}\ }\textbf {\bibinfo {volume}
  {34}},\ \bibinfo {pages} {025702} (\bibinfo {year}
  {2022}{\natexlab{a}})}\BibitemShut {NoStop}%
\bibitem [{\citenamefont {Pal}\ \emph {et~al.}(2022{\natexlab{b}})\citenamefont
  {Pal}, \citenamefont {Dey},\ and\ \citenamefont {Ghosh}}]{pal22b_berry}%
  \BibitemOpen
  \bibfield  {author} {\bibinfo {author} {\bibfnamefont {O.}~\bibnamefont
  {Pal}}, \bibinfo {author} {\bibfnamefont {B.}~\bibnamefont {Dey}},\ and\
  \bibinfo {author} {\bibfnamefont {T.~K.}\ \bibnamefont {Ghosh}},\ }\bibfield
  {title} {\bibinfo {title} {Berry curvature induced anisotropic
  magnetotransport in a quadratic triple-component fermionic system},\ }\href
  {https://doi.org/10.1088/1361-648X/ac4cee} {\bibfield  {journal} {\bibinfo
  {journal} {Journal of Physics: Condensed Matter}\ }\textbf {\bibinfo {volume}
  {34}},\ \bibinfo {pages} {155702} (\bibinfo {year}
  {2022}{\natexlab{b}})}\BibitemShut {NoStop}%
\bibitem [{\citenamefont {Fu}\ and\ \citenamefont
  {Wang}(2022)}]{fu22_thermoelectric}%
  \BibitemOpen
  \bibfield  {author} {\bibinfo {author} {\bibfnamefont {L.~X.}\ \bibnamefont
  {Fu}}\ and\ \bibinfo {author} {\bibfnamefont {C.~M.}\ \bibnamefont {Wang}},\
  }\bibfield  {title} {\bibinfo {title} {{Thermoelectric transport of
  multi-Weyl semimetals in the quantum limit}},\ }\href
  {https://doi.org/10.1103/PhysRevB.105.035201} {\bibfield  {journal} {\bibinfo
   {journal} {Phys. Rev. B}\ }\textbf {\bibinfo {volume} {105}},\ \bibinfo
  {pages} {035201} (\bibinfo {year} {2022})}\BibitemShut {NoStop}%
\bibitem [{\citenamefont {{Araki}}(2020)}]{araki20_magnetic}%
  \BibitemOpen
  \bibfield  {author} {\bibinfo {author} {\bibfnamefont {Y.}~\bibnamefont
  {{Araki}}},\ }\bibfield  {title} {\bibinfo {title} {Magnetic textures and
  dynamics in magnetic {W}eyl semimetals},\ }\href
  {https://doi.org/10.1002/andp.201900287} {\bibfield  {journal} {\bibinfo
  {journal} {Annalen der Physik}\ }\textbf {\bibinfo {volume} {532}},\ \bibinfo
  {pages} {1900287} (\bibinfo {year} {2020})}\BibitemShut {NoStop}%
\bibitem [{\citenamefont {Mizuta}\ and\ \citenamefont
  {Ishii}(2014)}]{mizuta14_contribution}%
  \BibitemOpen
  \bibfield  {author} {\bibinfo {author} {\bibfnamefont {Y.~P.}\ \bibnamefont
  {Mizuta}}\ and\ \bibinfo {author} {\bibfnamefont {F.}~\bibnamefont {Ishii}},\
  }\bibfield  {title} {\bibinfo {title} {Contribution of {B}erry curvature to
  thermoelectric effects},\ }\href {https://doi.org/10.7566/JPSCP.3.017035}
  {\bibfield  {journal} {\bibinfo  {journal} {Proceedings of the International
  Conference on Strongly Correlated Electron Systems (SCES2013)}\ }\textbf
  {\bibinfo {volume} {3}},\ \bibinfo {pages} {017035} (\bibinfo {year}
  {2014})}\BibitemShut {NoStop}%
\bibitem [{\citenamefont {Medel~Onofre}\ and\ \citenamefont
  {Mart\'{\i}n-Ruiz}(2023)}]{onofre}%
  \BibitemOpen
  \bibfield  {author} {\bibinfo {author} {\bibfnamefont {L.}~\bibnamefont
  {Medel~Onofre}}\ and\ \bibinfo {author} {\bibfnamefont {A.}~\bibnamefont
  {Mart\'{\i}n-Ruiz}},\ }\bibfield  {title} {\bibinfo {title} {{Planar Hall
  effect in Weyl semimetals induced by pseudoelectromagnetic fields}},\ }\href
  {https://doi.org/10.1103/PhysRevB.108.155132} {\bibfield  {journal} {\bibinfo
   {journal} {Phys. Rev. B}\ }\textbf {\bibinfo {volume} {108}},\ \bibinfo
  {pages} {155132} (\bibinfo {year} {2023})}\BibitemShut {NoStop}%
\bibitem [{\citenamefont {Ghosh}\ and\ \citenamefont
  {Mandal}(2024)}]{ips-rahul-ph}%
  \BibitemOpen
  \bibfield  {author} {\bibinfo {author} {\bibfnamefont {R.}~\bibnamefont
  {Ghosh}}\ and\ \bibinfo {author} {\bibfnamefont {I.}~\bibnamefont {Mandal}},\
  }\bibfield  {title} {\bibinfo {title} {{Electric and thermoelectric response
  for Weyl and multi-Weyl semimetals in planar Hall configurations including
  the effects of strain}},\ }\href
  {https://doi.org/https://doi.org/10.1016/j.physe.2024.115914} {\bibfield
  {journal} {\bibinfo  {journal} {Physica E: Low-dimensional Systems and
  Nanostructures}\ }\textbf {\bibinfo {volume} {159}},\ \bibinfo {pages}
  {115914} (\bibinfo {year} {2024})}\BibitemShut {NoStop}%
\bibitem [{\citenamefont {Haldane}(2004)}]{haldane}%
  \BibitemOpen
  \bibfield  {author} {\bibinfo {author} {\bibfnamefont {F.~D.~M.}\
  \bibnamefont {Haldane}},\ }\bibfield  {title} {\bibinfo {title} {Berry
  curvature on the {F}ermi surface: {A}nomalous {H}all effect as a topological
  {F}ermi-liquid property},\ }\href
  {https://doi.org/10.1103/PhysRevLett.93.206602} {\bibfield  {journal}
  {\bibinfo  {journal} {Phys. Rev. Lett.}\ }\textbf {\bibinfo {volume} {93}},\
  \bibinfo {pages} {206602} (\bibinfo {year} {2004})}\BibitemShut {NoStop}%
\bibitem [{\citenamefont {Goswami}\ and\ \citenamefont
  {Tewari}(2013)}]{pallab_axionic}%
  \BibitemOpen
  \bibfield  {author} {\bibinfo {author} {\bibfnamefont {P.}~\bibnamefont
  {Goswami}}\ and\ \bibinfo {author} {\bibfnamefont {S.}~\bibnamefont
  {Tewari}},\ }\bibfield  {title} {\bibinfo {title} {{Axionic field theory of
  $(3+1)$-dimensional Weyl semimetals}},\ }\href
  {https://doi.org/10.1103/PhysRevB.88.245107} {\bibfield  {journal} {\bibinfo
  {journal} {Phys. Rev. B}\ }\textbf {\bibinfo {volume} {88}},\ \bibinfo
  {pages} {245107} (\bibinfo {year} {2013})}\BibitemShut {NoStop}%
\bibitem [{\citenamefont {Burkov}(2014)}]{burkov_intrinsic_hall}%
  \BibitemOpen
  \bibfield  {author} {\bibinfo {author} {\bibfnamefont {A.~A.}\ \bibnamefont
  {Burkov}},\ }\bibfield  {title} {\bibinfo {title} {Anomalous {H}all effect in
  {W}eyl metals},\ }\href {https://doi.org/10.1103/PhysRevLett.113.187202}
  {\bibfield  {journal} {\bibinfo  {journal} {Phys. Rev. Lett.}\ }\textbf
  {\bibinfo {volume} {113}},\ \bibinfo {pages} {187202} (\bibinfo {year}
  {2014})}\BibitemShut {NoStop}%
\bibitem [{\citenamefont {{Medel}}\ \emph {et~al.}(2024)\citenamefont
  {{Medel}}, \citenamefont {{Ghosh}}, \citenamefont {{Mart{\'\i}n-Ruiz}},\ and\
  \citenamefont {{Mandal}}}]{ips-ruiz}%
  \BibitemOpen
  \bibfield  {author} {\bibinfo {author} {\bibfnamefont {L.}~\bibnamefont
  {{Medel}}}, \bibinfo {author} {\bibfnamefont {R.}~\bibnamefont {{Ghosh}}},
  \bibinfo {author} {\bibfnamefont {A.}~\bibnamefont {{Mart{\'\i}n-Ruiz}}},\
  and\ \bibinfo {author} {\bibfnamefont {I.}~\bibnamefont {{Mandal}}},\
  }\bibfield  {title} {\bibinfo {title} {{Electric, thermal, and thermoelectric
  magnetoconductivity for Weyl/multi-Weyl semimetals in planar Hall set-ups
  induced by the combined effects of topology and strain}},\ }\href
  {https://doi.org/10.1038/s41598-024-68615-0} {\bibfield  {journal} {\bibinfo
  {journal} {Scientific Reports}\ }\textbf {\bibinfo {volume} {14}},\ \bibinfo
  {eid} {21390} (\bibinfo {year} {2024})}\BibitemShut {NoStop}%
\bibitem [{\citenamefont {Slonczewski}\ and\ \citenamefont
  {Weiss}(1958)}]{weiss}%
  \BibitemOpen
  \bibfield  {author} {\bibinfo {author} {\bibfnamefont {J.~C.}\ \bibnamefont
  {Slonczewski}}\ and\ \bibinfo {author} {\bibfnamefont {P.~R.}\ \bibnamefont
  {Weiss}},\ }\bibfield  {title} {\bibinfo {title} {Band structure of
  graphite},\ }\href {https://doi.org/10.1103/PhysRev.109.272} {\bibfield
  {journal} {\bibinfo  {journal} {Phys. Rev.}\ }\textbf {\bibinfo {volume}
  {109}},\ \bibinfo {pages} {272} (\bibinfo {year} {1958})}\BibitemShut
  {NoStop}%
\bibitem [{\citenamefont {{Mandal}}(2024)}]{ips_cd}%
  \BibitemOpen
  \bibfield  {author} {\bibinfo {author} {\bibfnamefont {I.}~\bibnamefont
  {{Mandal}}},\ }\bibfield  {title} {\bibinfo {title} {{Signatures of two- and
  three-dimensional semimetals from circular dichroism}},\ }\href
  {https://doi.org/10.1142/S0217979224502163} {\bibfield  {journal} {\bibinfo
  {journal} {International Journal of Modern Physics B}\ }\textbf {\bibinfo
  {volume} {38}},\ \bibinfo {eid} {2450216} (\bibinfo {year}
  {2024})}\BibitemShut {NoStop}%
\bibitem [{\citenamefont {Park}\ \emph {et~al.}(2017)\citenamefont {Park},
  \citenamefont {Woo}, \citenamefont {Mele},\ and\ \citenamefont {Min}}]{park}%
  \BibitemOpen
  \bibfield  {author} {\bibinfo {author} {\bibfnamefont {S.}~\bibnamefont
  {Park}}, \bibinfo {author} {\bibfnamefont {S.}~\bibnamefont {Woo}}, \bibinfo
  {author} {\bibfnamefont {E.~J.}\ \bibnamefont {Mele}},\ and\ \bibinfo
  {author} {\bibfnamefont {H.}~\bibnamefont {Min}},\ }\bibfield  {title}
  {\bibinfo {title} {Semiclassical {B}oltzmann transport theory for
  multi-{W}eyl semimetals},\ }\href
  {https://doi.org/10.1103/PhysRevB.95.161113} {\bibfield  {journal} {\bibinfo
  {journal} {Phys. Rev. B}\ }\textbf {\bibinfo {volume} {95}},\ \bibinfo
  {pages} {161113} (\bibinfo {year} {2017})}\BibitemShut {NoStop}%
\bibitem [{\citenamefont {Hwang}\ \emph {et~al.}(2009)\citenamefont {Hwang},
  \citenamefont {Rossi},\ and\ \citenamefont {Das~Sarma}}]{Sarma2009}%
  \BibitemOpen
  \bibfield  {author} {\bibinfo {author} {\bibfnamefont {E.~H.}\ \bibnamefont
  {Hwang}}, \bibinfo {author} {\bibfnamefont {E.}~\bibnamefont {Rossi}},\ and\
  \bibinfo {author} {\bibfnamefont {S.}~\bibnamefont {Das~Sarma}},\ }\bibfield
  {title} {\bibinfo {title} {Theory of thermopower in two-dimensional
  graphene},\ }\href {https://doi.org/10.1103/PhysRevB.80.235415} {\bibfield
  {journal} {\bibinfo  {journal} {Phys. Rev. B}\ }\textbf {\bibinfo {volume}
  {80}},\ \bibinfo {pages} {235415} (\bibinfo {year} {2009})}\BibitemShut
  {NoStop}%
\bibitem [{\citenamefont {Hwang}\ \emph {et~al.}(2007)\citenamefont {Hwang},
  \citenamefont {Adam},\ and\ \citenamefont {Sarma}}]{hwang2}%
  \BibitemOpen
  \bibfield  {author} {\bibinfo {author} {\bibfnamefont {E.~H.}\ \bibnamefont
  {Hwang}}, \bibinfo {author} {\bibfnamefont {S.}~\bibnamefont {Adam}},\ and\
  \bibinfo {author} {\bibfnamefont {S.~D.}\ \bibnamefont {Sarma}},\ }\bibfield
  {title} {\bibinfo {title} {Carrier transport in two-dimensional graphene
  layers},\ }\href {https://doi.org/10.1103/PhysRevLett.98.186806} {\bibfield
  {journal} {\bibinfo  {journal} {Phys. Rev. Lett.}\ }\textbf {\bibinfo
  {volume} {98}},\ \bibinfo {pages} {186806} (\bibinfo {year}
  {2007})}\BibitemShut {NoStop}%
\bibitem [{\citenamefont {Hwang}\ and\ \citenamefont
  {Das~Sarma}(2007)}]{hwang3}%
  \BibitemOpen
  \bibfield  {author} {\bibinfo {author} {\bibfnamefont {E.~H.}\ \bibnamefont
  {Hwang}}\ and\ \bibinfo {author} {\bibfnamefont {S.}~\bibnamefont
  {Das~Sarma}},\ }\bibfield  {title} {\bibinfo {title} {Dielectric function,
  screening, and plasmons in two-dimensional graphene},\ }\href
  {https://doi.org/10.1103/PhysRevB.75.205418} {\bibfield  {journal} {\bibinfo
  {journal} {Phys. Rev. B}\ }\textbf {\bibinfo {volume} {75}},\ \bibinfo
  {pages} {205418} (\bibinfo {year} {2007})}\BibitemShut {NoStop}%
\bibitem [{\citenamefont {Suh}\ \emph {et~al.}(2023)\citenamefont {Suh},
  \citenamefont {Park},\ and\ \citenamefont {Min}}]{suh}%
  \BibitemOpen
  \bibfield  {author} {\bibinfo {author} {\bibfnamefont {J.}~\bibnamefont
  {Suh}}, \bibinfo {author} {\bibfnamefont {S.}~\bibnamefont {Park}},\ and\
  \bibinfo {author} {\bibfnamefont {H.}~\bibnamefont {Min}},\ }\bibfield
  {title} {\bibinfo {title} {Semiclassical {B}oltzmann magnetotransport theory
  in anisotropic systems with a nonvanishing {B}erry curvature},\ }\href
  {https://doi.org/10.1088/1367-2630/acc122} {\bibfield  {journal} {\bibinfo
  {journal} {New Journal of Physics}\ }\textbf {\bibinfo {volume} {25}},\
  \bibinfo {pages} {033021} (\bibinfo {year} {2023})}\BibitemShut {NoStop}%
\bibitem [{\citenamefont {Girvin}\ and\ \citenamefont
  {Jonson}(1982)}]{Girvin_1982}%
  \BibitemOpen
  \bibfield  {author} {\bibinfo {author} {\bibfnamefont {S.~M.}\ \bibnamefont
  {Girvin}}\ and\ \bibinfo {author} {\bibfnamefont {M.}~\bibnamefont
  {Jonson}},\ }\bibfield  {title} {\bibinfo {title} {Inversion layer
  thermopower in high magnetic field},\ }\href
  {https://doi.org/10.1088/0022-3719/15/32/006} {\bibfield  {journal} {\bibinfo
   {journal} {Journal of Physics C: Solid State Physics}\ }\textbf {\bibinfo
  {volume} {15}},\ \bibinfo {pages} {L1147} (\bibinfo {year}
  {1982})}\BibitemShut {NoStop}%
\bibitem [{\citenamefont {Konoike}\ \emph {et~al.}(2013)\citenamefont
  {Konoike}, \citenamefont {Sato}, \citenamefont {Uchida},\ and\ \citenamefont
  {Osada}}]{osada}%
  \BibitemOpen
  \bibfield  {author} {\bibinfo {author} {\bibfnamefont {T.}~\bibnamefont
  {Konoike}}, \bibinfo {author} {\bibfnamefont {M.}~\bibnamefont {Sato}},
  \bibinfo {author} {\bibfnamefont {K.}~\bibnamefont {Uchida}},\ and\ \bibinfo
  {author} {\bibfnamefont {T.}~\bibnamefont {Osada}},\ }\bibfield  {title}
  {\bibinfo {title} {Anomalous thermoelectric transport and giant {N}ernst
  effect in multilayered massless {D}irac fermion system},\ }\href
  {https://doi.org/10.7566/JPSJ.82.073601} {\bibfield  {journal} {\bibinfo
  {journal} {Journal of the Physical Society of Japan}\ }\textbf {\bibinfo
  {volume} {82}},\ \bibinfo {pages} {073601} (\bibinfo {year}
  {2013})}\BibitemShut {NoStop}%
\bibitem [{\citenamefont {Wei}\ \emph {et~al.}(2009{\natexlab{b}})\citenamefont
  {Wei}, \citenamefont {Bao}, \citenamefont {Pu}, \citenamefont {Lau},\ and\
  \citenamefont {Shi}}]{wei_PRL2009}%
  \BibitemOpen
  \bibfield  {author} {\bibinfo {author} {\bibfnamefont {P.}~\bibnamefont
  {Wei}}, \bibinfo {author} {\bibfnamefont {W.}~\bibnamefont {Bao}}, \bibinfo
  {author} {\bibfnamefont {Y.}~\bibnamefont {Pu}}, \bibinfo {author}
  {\bibfnamefont {C.~N.}\ \bibnamefont {Lau}},\ and\ \bibinfo {author}
  {\bibfnamefont {J.}~\bibnamefont {Shi}},\ }\bibfield  {title} {\bibinfo
  {title} {{Anomalous thermoelectric transport of Dirac particles in
  graphene}},\ }\href {https://doi.org/10.1103/PhysRevLett.102.166808}
  {\bibfield  {journal} {\bibinfo  {journal} {Phys. Rev. Lett.}\ }\textbf
  {\bibinfo {volume} {102}},\ \bibinfo {pages} {166808} (\bibinfo {year}
  {2009}{\natexlab{b}})}\BibitemShut {NoStop}%
\bibitem [{\citenamefont {Checkelsky}\ and\ \citenamefont
  {Ong}(2009)}]{Checkelsky2009}%
  \BibitemOpen
  \bibfield  {author} {\bibinfo {author} {\bibfnamefont {J.~G.}\ \bibnamefont
  {Checkelsky}}\ and\ \bibinfo {author} {\bibfnamefont {N.~P.}\ \bibnamefont
  {Ong}},\ }\bibfield  {title} {\bibinfo {title} {Thermopower and {N}ernst
  effect in graphene in a magnetic field},\ }\href
  {https://doi.org/10.1103/PhysRevB.80.081413} {\bibfield  {journal} {\bibinfo
  {journal} {Phys. Rev. B}\ }\textbf {\bibinfo {volume} {80}},\ \bibinfo
  {pages} {081413} (\bibinfo {year} {2009})}\BibitemShut {NoStop}%
\bibitem [{\citenamefont {Xiao}\ \emph {et~al.}(2006)\citenamefont {Xiao},
  \citenamefont {Yao}, \citenamefont {Fang},\ and\ \citenamefont {Niu}}]{di}%
  \BibitemOpen
  \bibfield  {author} {\bibinfo {author} {\bibfnamefont {D.}~\bibnamefont
  {Xiao}}, \bibinfo {author} {\bibfnamefont {Y.}~\bibnamefont {Yao}}, \bibinfo
  {author} {\bibfnamefont {Z.}~\bibnamefont {Fang}},\ and\ \bibinfo {author}
  {\bibfnamefont {Q.}~\bibnamefont {Niu}},\ }\bibfield  {title} {\bibinfo
  {title} {Berry-phase effect in anomalous thermoelectric transport},\ }\href
  {https://doi.org/10.1103/PhysRevLett.97.026603} {\bibfield  {journal}
  {\bibinfo  {journal} {Phys. Rev. Lett.}\ }\textbf {\bibinfo {volume} {97}},\
  \bibinfo {pages} {026603} (\bibinfo {year} {2006})}\BibitemShut {NoStop}%
\bibitem [{\citenamefont {{Ghosh}}\ and\ \citenamefont
  {{Mandal}}(2024)}]{ips-rahul-tilt}%
  \BibitemOpen
  \bibfield  {author} {\bibinfo {author} {\bibfnamefont {R.}~\bibnamefont
  {{Ghosh}}}\ and\ \bibinfo {author} {\bibfnamefont {I.}~\bibnamefont
  {{Mandal}}},\ }\bibfield  {title} {\bibinfo {title} {{Direction-dependent
  conductivity in planar Hall set-ups with tilted Weyl/multi-Weyl
  semimetals}},\ }\href {https://doi.org/10.1088/1361-648X/ad38fa} {\bibfield
  {journal} {\bibinfo  {journal} {Journal of Physics Condensed Matter}\
  }\textbf {\bibinfo {volume} {36}},\ \bibinfo {eid} {275501} (\bibinfo {year}
  {2024})}\BibitemShut {NoStop}%
\bibitem [{\citenamefont {Dong}\ \emph {et~al.}(2020)\citenamefont {Dong},
  \citenamefont {Xiao}, \citenamefont {Xiong},\ and\ \citenamefont
  {Niu}}]{niu_prl}%
  \BibitemOpen
  \bibfield  {author} {\bibinfo {author} {\bibfnamefont {L.}~\bibnamefont
  {Dong}}, \bibinfo {author} {\bibfnamefont {C.}~\bibnamefont {Xiao}}, \bibinfo
  {author} {\bibfnamefont {B.}~\bibnamefont {Xiong}},\ and\ \bibinfo {author}
  {\bibfnamefont {Q.}~\bibnamefont {Niu}},\ }\bibfield  {title} {\bibinfo
  {title} {Berry phase effects in dipole density and the mott relation},\
  }\href {https://doi.org/10.1103/PhysRevLett.124.066601} {\bibfield  {journal}
  {\bibinfo  {journal} {Phys. Rev. Lett.}\ }\textbf {\bibinfo {volume} {124}},\
  \bibinfo {pages} {066601} (\bibinfo {year} {2020})}\BibitemShut {NoStop}%
\bibitem [{\citenamefont {{Bradlyn}}\ \emph {et~al.}(2016)\citenamefont
  {{Bradlyn}}, \citenamefont {{Cano}}, \citenamefont {{Wang}}, \citenamefont
  {{Vergniory}}, \citenamefont {{Felser}}, \citenamefont {{Cava}},\ and\
  \citenamefont {{Bernevig}}}]{bradlyn}%
  \BibitemOpen
  \bibfield  {author} {\bibinfo {author} {\bibfnamefont {B.}~\bibnamefont
  {{Bradlyn}}}, \bibinfo {author} {\bibfnamefont {J.}~\bibnamefont {{Cano}}},
  \bibinfo {author} {\bibfnamefont {Z.}~\bibnamefont {{Wang}}}, \bibinfo
  {author} {\bibfnamefont {M.~G.}\ \bibnamefont {{Vergniory}}}, \bibinfo
  {author} {\bibfnamefont {C.}~\bibnamefont {{Felser}}}, \bibinfo {author}
  {\bibfnamefont {R.~J.}\ \bibnamefont {{Cava}}},\ and\ \bibinfo {author}
  {\bibfnamefont {B.~A.}\ \bibnamefont {{Bernevig}}},\ }\bibfield  {title}
  {\bibinfo {title} {{Beyond Dirac and Weyl fermions: Unconventional
  quasiparticles in conventional crystals}},\ }\href
  {https://doi.org/10.1126/science.aaf5037} {\bibfield  {journal} {\bibinfo
  {journal} {Science}\ }\textbf {\bibinfo {volume} {353}},\ \bibinfo {pages}
  {aaf5037} (\bibinfo {year} {2016})}\BibitemShut {NoStop}%
\bibitem [{\citenamefont {Fang}\ \emph {et~al.}(2012)\citenamefont {Fang},
  \citenamefont {Gilbert}, \citenamefont {Dai},\ and\ \citenamefont
  {Bernevig}}]{bernevig2}%
  \BibitemOpen
  \bibfield  {author} {\bibinfo {author} {\bibfnamefont {C.}~\bibnamefont
  {Fang}}, \bibinfo {author} {\bibfnamefont {M.~J.}\ \bibnamefont {Gilbert}},
  \bibinfo {author} {\bibfnamefont {X.}~\bibnamefont {Dai}},\ and\ \bibinfo
  {author} {\bibfnamefont {B.~A.}\ \bibnamefont {Bernevig}},\ }\bibfield
  {title} {\bibinfo {title} {Multi-{W}eyl topological semimetals stabilized by
  point group symmetry},\ }\href
  {https://doi.org/10.1103/PhysRevLett.108.266802} {\bibfield  {journal}
  {\bibinfo  {journal} {Phys. Rev. Lett.}\ }\textbf {\bibinfo {volume} {108}},\
  \bibinfo {pages} {266802} (\bibinfo {year} {2012})}\BibitemShut {NoStop}%
\bibitem [{\citenamefont {Trescher}\ \emph {et~al.}(2015)\citenamefont
  {Trescher}, \citenamefont {Sbierski}, \citenamefont {Brouwer},\ and\
  \citenamefont {Bergholtz}}]{emil_tilted}%
  \BibitemOpen
  \bibfield  {author} {\bibinfo {author} {\bibfnamefont {M.}~\bibnamefont
  {Trescher}}, \bibinfo {author} {\bibfnamefont {B.}~\bibnamefont {Sbierski}},
  \bibinfo {author} {\bibfnamefont {P.~W.}\ \bibnamefont {Brouwer}},\ and\
  \bibinfo {author} {\bibfnamefont {E.~J.}\ \bibnamefont {Bergholtz}},\
  }\bibfield  {title} {\bibinfo {title} {Quantum transport in dirac materials:
  Signatures of tilted and anisotropic dirac and weyl cones},\ }\href
  {https://doi.org/10.1103/PhysRevB.91.115135} {\bibfield  {journal} {\bibinfo
  {journal} {Phys. Rev. B}\ }\textbf {\bibinfo {volume} {91}},\ \bibinfo
  {pages} {115135} (\bibinfo {year} {2015})}\BibitemShut {NoStop}%
\bibitem [{\citenamefont {Trescher}\ \emph {et~al.}(2017)\citenamefont
  {Trescher}, \citenamefont {Sbierski}, \citenamefont {Brouwer},\ and\
  \citenamefont {Bergholtz}}]{trescher17_tilted}%
  \BibitemOpen
  \bibfield  {author} {\bibinfo {author} {\bibfnamefont {M.}~\bibnamefont
  {Trescher}}, \bibinfo {author} {\bibfnamefont {B.}~\bibnamefont {Sbierski}},
  \bibinfo {author} {\bibfnamefont {P.~W.}\ \bibnamefont {Brouwer}},\ and\
  \bibinfo {author} {\bibfnamefont {E.~J.}\ \bibnamefont {Bergholtz}},\
  }\bibfield  {title} {\bibinfo {title} {{Tilted disordered Weyl semimetals}},\
  }\href {https://doi.org/10.1103/PhysRevB.95.045139} {\bibfield  {journal}
  {\bibinfo  {journal} {Phys. Rev. B}\ }\textbf {\bibinfo {volume} {95}},\
  \bibinfo {pages} {045139} (\bibinfo {year} {2017})}\BibitemShut {NoStop}%
\bibitem [{\citenamefont {Herring}(1937)}]{herring}%
  \BibitemOpen
  \bibfield  {author} {\bibinfo {author} {\bibfnamefont {C.}~\bibnamefont
  {Herring}},\ }\bibfield  {title} {\bibinfo {title} {Accidental degeneracy in
  the energy bands of crystals},\ }\href
  {https://doi.org/10.1103/PhysRev.52.365} {\bibfield  {journal} {\bibinfo
  {journal} {Phys. Rev.}\ }\textbf {\bibinfo {volume} {52}},\ \bibinfo {pages}
  {365} (\bibinfo {year} {1937})}\BibitemShut {NoStop}%
\bibitem [{\citenamefont {Ma}\ \emph {et~al.}(2019)\citenamefont {Ma},
  \citenamefont {Jiang}, \citenamefont {Liu},\ and\ \citenamefont
  {Xie}}]{ma19_planar}%
  \BibitemOpen
  \bibfield  {author} {\bibinfo {author} {\bibfnamefont {D.}~\bibnamefont
  {Ma}}, \bibinfo {author} {\bibfnamefont {H.}~\bibnamefont {Jiang}}, \bibinfo
  {author} {\bibfnamefont {H.}~\bibnamefont {Liu}},\ and\ \bibinfo {author}
  {\bibfnamefont {X.~C.}\ \bibnamefont {Xie}},\ }\bibfield  {title} {\bibinfo
  {title} {{Planar Hall effect in tilted Weyl semimetals}},\ }\href
  {https://doi.org/10.1103/PhysRevB.99.115121} {\bibfield  {journal} {\bibinfo
  {journal} {Phys. Rev. B}\ }\textbf {\bibinfo {volume} {99}},\ \bibinfo
  {pages} {115121} (\bibinfo {year} {2019})}\BibitemShut {NoStop}%
\bibitem [{\citenamefont {Kundu}\ \emph {et~al.}(2020)\citenamefont {Kundu},
  \citenamefont {Siu}, \citenamefont {Yang},\ and\ \citenamefont
  {Jalil}}]{kundu20_magnetotransport}%
  \BibitemOpen
  \bibfield  {author} {\bibinfo {author} {\bibfnamefont {A.}~\bibnamefont
  {Kundu}}, \bibinfo {author} {\bibfnamefont {Z.~B.}\ \bibnamefont {Siu}},
  \bibinfo {author} {\bibfnamefont {H.}~\bibnamefont {Yang}},\ and\ \bibinfo
  {author} {\bibfnamefont {M.~B.}\ \bibnamefont {Jalil}},\ }\bibfield  {title}
  {\bibinfo {title} {{Magnetotransport of Weyl semimetals with tilted Dirac
  cones}},\ }\href {https://doi.org/10.1088/1367-2630/aba98d} {\bibfield
  {journal} {\bibinfo  {journal} {New Journal of Phys.}\ }\textbf {\bibinfo
  {volume} {22}},\ \bibinfo {pages} {083081} (\bibinfo {year}
  {2020})}\BibitemShut {NoStop}%
\bibitem [{\citenamefont {K\"onye}\ and\ \citenamefont
  {Ogata}(2021)}]{konye21_microscopic}%
  \BibitemOpen
  \bibfield  {author} {\bibinfo {author} {\bibfnamefont {V.}~\bibnamefont
  {K\"onye}}\ and\ \bibinfo {author} {\bibfnamefont {M.}~\bibnamefont
  {Ogata}},\ }\bibfield  {title} {\bibinfo {title} {Microscopic theory of
  magnetoconductivity at low magnetic fields in terms of {B}erry curvature and
  orbital magnetic moment},\ }\href
  {https://doi.org/10.1103/PhysRevResearch.3.033076} {\bibfield  {journal}
  {\bibinfo  {journal} {Phys. Rev. Res.}\ }\textbf {\bibinfo {volume} {3}},\
  \bibinfo {pages} {033076} (\bibinfo {year} {2021})}\BibitemShut {NoStop}%
\bibitem [{\citenamefont {Shao}\ and\ \citenamefont
  {Yan}(2022)}]{shao22_plane}%
  \BibitemOpen
  \bibfield  {author} {\bibinfo {author} {\bibfnamefont {J.}~\bibnamefont
  {Shao}}\ and\ \bibinfo {author} {\bibfnamefont {L.}~\bibnamefont {Yan}},\
  }\bibfield  {title} {\bibinfo {title} {{In-plane magnetotransport phenomena
  in tilted Weyl semimetals}},\ }\href
  {https://doi.org/10.1088/1361-648X/ac9e35} {\bibfield  {journal} {\bibinfo
  {journal} {Journal of Phys.: Condensed Matter}\ }\textbf {\bibinfo {volume}
  {51}},\ \bibinfo {pages} {025401} (\bibinfo {year} {2022})}\BibitemShut
  {NoStop}%
\bibitem [{\citenamefont {Li}\ \emph {et~al.}(2018)\citenamefont {Li},
  \citenamefont {Zhang}, \citenamefont {Zhang}, \citenamefont {Wen},\ and\
  \citenamefont {Zhang}}]{li18_giant}%
  \BibitemOpen
  \bibfield  {author} {\bibinfo {author} {\bibfnamefont {P.}~\bibnamefont
  {Li}}, \bibinfo {author} {\bibfnamefont {C.~H.}\ \bibnamefont {Zhang}},
  \bibinfo {author} {\bibfnamefont {J.~W.}\ \bibnamefont {Zhang}}, \bibinfo
  {author} {\bibfnamefont {Y.}~\bibnamefont {Wen}},\ and\ \bibinfo {author}
  {\bibfnamefont {X.~X.}\ \bibnamefont {Zhang}},\ }\bibfield  {title} {\bibinfo
  {title} {{Giant planar Hall effect in the Dirac semimetal {Z}r{T}e$_{5}$}},\
  }\href {https://doi.org/10.1103/PhysRevB.98.121108} {\bibfield  {journal}
  {\bibinfo  {journal} {Phys. Rev. B}\ }\textbf {\bibinfo {volume} {98}},\
  \bibinfo {pages} {121108} (\bibinfo {year} {2018})}\BibitemShut {NoStop}%
\bibitem [{\citenamefont {Guinea}\ \emph
  {et~al.}(2010{\natexlab{a}})\citenamefont {Guinea}, \citenamefont
  {Katsnelson},\ and\ \citenamefont {Geim}}]{guinea10_energy}%
  \BibitemOpen
  \bibfield  {author} {\bibinfo {author} {\bibfnamefont {F.}~\bibnamefont
  {Guinea}}, \bibinfo {author} {\bibfnamefont {M.~I.}\ \bibnamefont
  {Katsnelson}},\ and\ \bibinfo {author} {\bibfnamefont {A.}~\bibnamefont
  {Geim}},\ }\bibfield  {title} {\bibinfo {title} {Energy gaps and a zero-field
  quantum {H}all effect in graphene by strain engineering},\ }\href
  {https://doi.org/10.1038/nphys1420} {\bibfield  {journal} {\bibinfo
  {journal} {Nature Physics}\ }\textbf {\bibinfo {volume} {6}},\ \bibinfo
  {pages} {30} (\bibinfo {year} {2010}{\natexlab{a}})}\BibitemShut {NoStop}%
\bibitem [{\citenamefont {Guinea}\ \emph
  {et~al.}(2010{\natexlab{b}})\citenamefont {Guinea}, \citenamefont {Geim},
  \citenamefont {Katsnelson},\ and\ \citenamefont
  {Novoselov}}]{guinea10_generating}%
  \BibitemOpen
  \bibfield  {author} {\bibinfo {author} {\bibfnamefont {F.}~\bibnamefont
  {Guinea}}, \bibinfo {author} {\bibfnamefont {A.~K.}\ \bibnamefont {Geim}},
  \bibinfo {author} {\bibfnamefont {M.~I.}\ \bibnamefont {Katsnelson}},\ and\
  \bibinfo {author} {\bibfnamefont {K.~S.}\ \bibnamefont {Novoselov}},\
  }\bibfield  {title} {\bibinfo {title} {Generating quantizing pseudomagnetic
  fields by bending graphene ribbons},\ }\href
  {https://doi.org/10.1103/PhysRevB.81.035408} {\bibfield  {journal} {\bibinfo
  {journal} {Phys. Rev. B}\ }\textbf {\bibinfo {volume} {81}},\ \bibinfo
  {pages} {035408} (\bibinfo {year} {2010}{\natexlab{b}})}\BibitemShut
  {NoStop}%
\bibitem [{\citenamefont {Low}\ and\ \citenamefont
  {Guinea}(2010)}]{low10_strain}%
  \BibitemOpen
  \bibfield  {author} {\bibinfo {author} {\bibfnamefont {T.}~\bibnamefont
  {Low}}\ and\ \bibinfo {author} {\bibfnamefont {F.}~\bibnamefont {Guinea}},\
  }\bibfield  {title} {\bibinfo {title} {Strain-induced pseudomagnetic field
  for novel graphene electronics},\ }\href {https://doi.org/10.1021/nl1018063}
  {\bibfield  {journal} {\bibinfo  {journal} {Nano letters}\ }\textbf {\bibinfo
  {volume} {10}},\ \bibinfo {pages} {3551} (\bibinfo {year}
  {2010})}\BibitemShut {NoStop}%
\bibitem [{\citenamefont {Cortijo}\ \emph {et~al.}(2015)\citenamefont
  {Cortijo}, \citenamefont {Ferreir\'os}, \citenamefont {Landsteiner},\ and\
  \citenamefont {Vozmediano}}]{landsteiner_gaguge}%
  \BibitemOpen
  \bibfield  {author} {\bibinfo {author} {\bibfnamefont {A.}~\bibnamefont
  {Cortijo}}, \bibinfo {author} {\bibfnamefont {Y.}~\bibnamefont
  {Ferreir\'os}}, \bibinfo {author} {\bibfnamefont {K.}~\bibnamefont
  {Landsteiner}},\ and\ \bibinfo {author} {\bibfnamefont {M.~A.~H.}\
  \bibnamefont {Vozmediano}},\ }\bibfield  {title} {\bibinfo {title} {Elastic
  gauge fields in {W}eyl semimetals},\ }\href
  {https://doi.org/10.1103/PhysRevLett.115.177202} {\bibfield  {journal}
  {\bibinfo  {journal} {Phys. Rev. Lett.}\ }\textbf {\bibinfo {volume} {115}},\
  \bibinfo {pages} {177202} (\bibinfo {year} {2015})}\BibitemShut {NoStop}%
\bibitem [{\citenamefont {Liu}\ \emph {et~al.}(2013)\citenamefont {Liu},
  \citenamefont {Ye},\ and\ \citenamefont {Qi}}]{liu_gauge}%
  \BibitemOpen
  \bibfield  {author} {\bibinfo {author} {\bibfnamefont {C.-X.}\ \bibnamefont
  {Liu}}, \bibinfo {author} {\bibfnamefont {P.}~\bibnamefont {Ye}},\ and\
  \bibinfo {author} {\bibfnamefont {X.-L.}\ \bibnamefont {Qi}},\ }\bibfield
  {title} {\bibinfo {title} {{Chiral gauge field and axial anomaly in a Weyl
  semimetal}},\ }\href {https://doi.org/10.1103/PhysRevB.87.235306} {\bibfield
  {journal} {\bibinfo  {journal} {Phys. Rev. B}\ }\textbf {\bibinfo {volume}
  {87}},\ \bibinfo {pages} {235306} (\bibinfo {year} {2013})}\BibitemShut
  {NoStop}%
\bibitem [{\citenamefont {Pikulin}\ \emph {et~al.}(2016)\citenamefont
  {Pikulin}, \citenamefont {Chen},\ and\ \citenamefont
  {Franz}}]{pikulin_gauge}%
  \BibitemOpen
  \bibfield  {author} {\bibinfo {author} {\bibfnamefont {D.~I.}\ \bibnamefont
  {Pikulin}}, \bibinfo {author} {\bibfnamefont {A.}~\bibnamefont {Chen}},\ and\
  \bibinfo {author} {\bibfnamefont {M.}~\bibnamefont {Franz}},\ }\bibfield
  {title} {\bibinfo {title} {Chiral anomaly from strain-induced gauge fields in
  {D}irac and {W}eyl semimetals},\ }\href
  {https://doi.org/10.1103/PhysRevX.6.041021} {\bibfield  {journal} {\bibinfo
  {journal} {Phys. Rev. X}\ }\textbf {\bibinfo {volume} {6}},\ \bibinfo {pages}
  {041021} (\bibinfo {year} {2016})}\BibitemShut {NoStop}%
\bibitem [{\citenamefont {Arjona}\ and\ \citenamefont
  {Vozmediano}(2018)}]{arjona18_rotational}%
  \BibitemOpen
  \bibfield  {author} {\bibinfo {author} {\bibfnamefont {V.}~\bibnamefont
  {Arjona}}\ and\ \bibinfo {author} {\bibfnamefont {M.~A.}\ \bibnamefont
  {Vozmediano}},\ }\bibfield  {title} {\bibinfo {title} {{Rotational strain in
  Weyl semimetals: A continuum approach}},\ }\href
  {https://doi.org/10.1103/PhysRevB.97.201404} {\bibfield  {journal} {\bibinfo
  {journal} {Physical Review B}\ }\textbf {\bibinfo {volume} {97}},\ \bibinfo
  {pages} {201404} (\bibinfo {year} {2018})}\BibitemShut {NoStop}%
\bibitem [{\citenamefont {Ghosh}\ \emph {et~al.}(2020)\citenamefont {Ghosh},
  \citenamefont {Sinha}, \citenamefont {Nandy},\ and\ \citenamefont
  {Taraphder}}]{ghosh20_chirality}%
  \BibitemOpen
  \bibfield  {author} {\bibinfo {author} {\bibfnamefont {S.}~\bibnamefont
  {Ghosh}}, \bibinfo {author} {\bibfnamefont {D.}~\bibnamefont {Sinha}},
  \bibinfo {author} {\bibfnamefont {S.}~\bibnamefont {Nandy}},\ and\ \bibinfo
  {author} {\bibfnamefont {A.}~\bibnamefont {Taraphder}},\ }\bibfield  {title}
  {\bibinfo {title} {Chirality-dependent planar {H}all effect in inhomogeneous
  {W}eyl semimetals},\ }\href {https://doi.org/10.1103/PhysRevB.102.121105}
  {\bibfield  {journal} {\bibinfo  {journal} {Phys. Rev. B}\ }\textbf {\bibinfo
  {volume} {102}},\ \bibinfo {pages} {121105} (\bibinfo {year}
  {2020})}\BibitemShut {NoStop}%
\bibitem [{\citenamefont {Ahmad}\ \emph {et~al.}(2023)\citenamefont {Ahmad},
  \citenamefont {Raman}, \citenamefont {Tewari},\ and\ \citenamefont
  {Sharma}}]{girish2023}%
  \BibitemOpen
  \bibfield  {author} {\bibinfo {author} {\bibfnamefont {A.}~\bibnamefont
  {Ahmad}}, \bibinfo {author} {\bibfnamefont {K.~V.}\ \bibnamefont {Raman}},
  \bibinfo {author} {\bibfnamefont {S.}~\bibnamefont {Tewari}},\ and\ \bibinfo
  {author} {\bibfnamefont {G.}~\bibnamefont {Sharma}},\ }\bibfield  {title}
  {\bibinfo {title} {{Longitudinal magnetoconductance and the planar Hall
  conductance in inhomogeneous Weyl semimetals}},\ }\href
  {https://doi.org/10.1103/PhysRevB.107.144206} {\bibfield  {journal} {\bibinfo
   {journal} {Phys. Rev. B}\ }\textbf {\bibinfo {volume} {107}},\ \bibinfo
  {pages} {144206} (\bibinfo {year} {2023})}\BibitemShut {NoStop}%
\bibitem [{\citenamefont {{Li}}\ \emph {et~al.}(2016)\citenamefont {{Li}},
  \citenamefont {{Kharzeev}}, \citenamefont {{Zhang}}, \citenamefont {{Huang}},
  \citenamefont {{Pletikosi{\'c}}}, \citenamefont {{Fedorov}}, \citenamefont
  {{Zhong}}, \citenamefont {{Schneeloch}}, \citenamefont {{Gu}},\ and\
  \citenamefont {{Valla}}}]{li_2016}%
  \BibitemOpen
  \bibfield  {author} {\bibinfo {author} {\bibfnamefont {Q.}~\bibnamefont
  {{Li}}}, \bibinfo {author} {\bibfnamefont {D.~E.}\ \bibnamefont
  {{Kharzeev}}}, \bibinfo {author} {\bibfnamefont {C.}~\bibnamefont {{Zhang}}},
  \bibinfo {author} {\bibfnamefont {Y.}~\bibnamefont {{Huang}}}, \bibinfo
  {author} {\bibfnamefont {I.}~\bibnamefont {{Pletikosi{\'c}}}}, \bibinfo
  {author} {\bibfnamefont {A.~V.}\ \bibnamefont {{Fedorov}}}, \bibinfo {author}
  {\bibfnamefont {R.~D.}\ \bibnamefont {{Zhong}}}, \bibinfo {author}
  {\bibfnamefont {J.~A.}\ \bibnamefont {{Schneeloch}}}, \bibinfo {author}
  {\bibfnamefont {G.~D.}\ \bibnamefont {{Gu}}},\ and\ \bibinfo {author}
  {\bibfnamefont {T.}~\bibnamefont {{Valla}}},\ }\bibfield  {title} {\bibinfo
  {title} {{Chiral magnetic effect in ZrTe$_{5}$}},\ }\href
  {https://doi.org/10.1038/nphys3648} {\bibfield  {journal} {\bibinfo
  {journal} {Nature Physics}\ }\textbf {\bibinfo {volume} {12}},\ \bibinfo
  {pages} {550} (\bibinfo {year} {2016})}\BibitemShut {NoStop}%
\bibitem [{\citenamefont {{Zhang}}\ \emph {et~al.}(2016)\citenamefont
  {{Zhang}}, \citenamefont {{Xu}}, \citenamefont {{Belopolski}}, \citenamefont
  {{Yuan}}, \citenamefont {{Lin}}, \citenamefont {{Tong}}, \citenamefont
  {{Bian}}, \citenamefont {{Alidoust}}, \citenamefont {{Lee}}, \citenamefont
  {{Huang}}, \citenamefont {{Chang}}, \citenamefont {{Chang}}, \citenamefont
  {{Hsu}}, \citenamefont {{Jeng}}, \citenamefont {{Neupane}}, \citenamefont
  {{Sanchez}}, \citenamefont {{Zheng}}, \citenamefont {{Wang}}, \citenamefont
  {{Lin}}, \citenamefont {{Zhang}}, \citenamefont {{Lu}}, \citenamefont
  {{Shen}}, \citenamefont {{Neupert}}, \citenamefont {{Zahid Hasan}},\ and\
  \citenamefont {{Jia}}}]{cheng-long}%
  \BibitemOpen
  \bibfield  {author} {\bibinfo {author} {\bibfnamefont {C.-L.}\ \bibnamefont
  {{Zhang}}}, \bibinfo {author} {\bibfnamefont {S.-Y.}\ \bibnamefont {{Xu}}},
  \bibinfo {author} {\bibfnamefont {I.}~\bibnamefont {{Belopolski}}}, \bibinfo
  {author} {\bibfnamefont {Z.}~\bibnamefont {{Yuan}}}, \bibinfo {author}
  {\bibfnamefont {Z.}~\bibnamefont {{Lin}}}, \bibinfo {author} {\bibfnamefont
  {B.}~\bibnamefont {{Tong}}}, \bibinfo {author} {\bibfnamefont
  {G.}~\bibnamefont {{Bian}}}, \bibinfo {author} {\bibfnamefont
  {N.}~\bibnamefont {{Alidoust}}}, \bibinfo {author} {\bibfnamefont {C.-C.}\
  \bibnamefont {{Lee}}}, \bibinfo {author} {\bibfnamefont {S.-M.}\ \bibnamefont
  {{Huang}}}, \bibinfo {author} {\bibfnamefont {T.-R.}\ \bibnamefont
  {{Chang}}}, \bibinfo {author} {\bibfnamefont {G.}~\bibnamefont {{Chang}}},
  \bibinfo {author} {\bibfnamefont {C.-H.}\ \bibnamefont {{Hsu}}}, \bibinfo
  {author} {\bibfnamefont {H.-T.}\ \bibnamefont {{Jeng}}}, \bibinfo {author}
  {\bibfnamefont {M.}~\bibnamefont {{Neupane}}}, \bibinfo {author}
  {\bibfnamefont {D.~S.}\ \bibnamefont {{Sanchez}}}, \bibinfo {author}
  {\bibfnamefont {H.}~\bibnamefont {{Zheng}}}, \bibinfo {author} {\bibfnamefont
  {J.}~\bibnamefont {{Wang}}}, \bibinfo {author} {\bibfnamefont
  {H.}~\bibnamefont {{Lin}}}, \bibinfo {author} {\bibfnamefont
  {C.}~\bibnamefont {{Zhang}}}, \bibinfo {author} {\bibfnamefont {H.-Z.}\
  \bibnamefont {{Lu}}}, \bibinfo {author} {\bibfnamefont {S.-Q.}\ \bibnamefont
  {{Shen}}}, \bibinfo {author} {\bibfnamefont {T.}~\bibnamefont {{Neupert}}},
  \bibinfo {author} {\bibfnamefont {M.}~\bibnamefont {{Zahid Hasan}}},\ and\
  \bibinfo {author} {\bibfnamefont {S.}~\bibnamefont {{Jia}}},\ }\bibfield
  {title} {\bibinfo {title} {{Signatures of the Adler-Bell-Jackiw chiral
  anomaly in a Weyl fermion semimetal}},\ }\href
  {https://doi.org/10.1038/ncomms10735} {\bibfield  {journal} {\bibinfo
  {journal} {Nature Communications}\ }\textbf {\bibinfo {volume} {7}},\
  \bibinfo {eid} {10735} (\bibinfo {year} {2016})}\BibitemShut {NoStop}%
\bibitem [{\citenamefont {Shama}\ \emph {et~al.}(2020)\citenamefont {Shama},
  \citenamefont {Gopal},\ and\ \citenamefont {Singh}}]{shama}%
  \BibitemOpen
  \bibfield  {author} {\bibinfo {author} {\bibnamefont {Shama}}, \bibinfo
  {author} {\bibfnamefont {R.}~\bibnamefont {Gopal}},\ and\ \bibinfo {author}
  {\bibfnamefont {Y.}~\bibnamefont {Singh}},\ }\bibfield  {title} {\bibinfo
  {title} {{Observation of planar Hall effect in the ferromagnetic Weyl
  semimetal Co$_3$Sn$_2$S$_2$}},\ }\href
  {https://doi.org/https://doi.org/10.1016/j.jmmm.2020.166547} {\bibfield
  {journal} {\bibinfo  {journal} {Journal of Magnetism and Magnetic Materials}\
  }\textbf {\bibinfo {volume} {502}},\ \bibinfo {pages} {166547} (\bibinfo
  {year} {2020})}\BibitemShut {NoStop}%
\bibitem [{\citenamefont {Tanwar}\ \emph {et~al.}(2023)\citenamefont {Tanwar},
  \citenamefont {Ahmad}, \citenamefont {Alam}, \citenamefont {Yao},
  \citenamefont {Tafti},\ and\ \citenamefont {Matusiak}}]{marcin}%
  \BibitemOpen
  \bibfield  {author} {\bibinfo {author} {\bibfnamefont {P.~K.}\ \bibnamefont
  {Tanwar}}, \bibinfo {author} {\bibfnamefont {M.}~\bibnamefont {Ahmad}},
  \bibinfo {author} {\bibfnamefont {M.~S.}\ \bibnamefont {Alam}}, \bibinfo
  {author} {\bibfnamefont {X.}~\bibnamefont {Yao}}, \bibinfo {author}
  {\bibfnamefont {F.}~\bibnamefont {Tafti}},\ and\ \bibinfo {author}
  {\bibfnamefont {M.}~\bibnamefont {Matusiak}},\ }\bibfield  {title} {\bibinfo
  {title} {{Gravitational anomaly in the ferrimagnetic topological Weyl
  semimetal NdAlSi}},\ }\href {https://doi.org/10.1103/PhysRevB.108.L161106}
  {\bibfield  {journal} {\bibinfo  {journal} {Phys. Rev. B}\ }\textbf {\bibinfo
  {volume} {108}},\ \bibinfo {pages} {L161106} (\bibinfo {year}
  {2023})}\BibitemShut {NoStop}%
\bibitem [{\citenamefont {{Zhang}}\ \emph {et~al.}(2017)\citenamefont
  {{Zhang}}, \citenamefont {{Zhang}}, \citenamefont {{Wang}}, \citenamefont
  {{Liu}}, \citenamefont {{Chen}}, \citenamefont {{Lu}}, \citenamefont
  {{Liang}}, \citenamefont {{Cao}}, \citenamefont {{Yuan}}, \citenamefont
  {{Tang}}, \citenamefont {{Li}}, \citenamefont {{Zhou}}, \citenamefont {{Gu}},
  \citenamefont {{Wu}}, \citenamefont {{Zou}},\ and\ \citenamefont
  {{Xiu}}}]{thete_dep}%
  \BibitemOpen
  \bibfield  {author} {\bibinfo {author} {\bibfnamefont {C.}~\bibnamefont
  {{Zhang}}}, \bibinfo {author} {\bibfnamefont {E.}~\bibnamefont {{Zhang}}},
  \bibinfo {author} {\bibfnamefont {W.}~\bibnamefont {{Wang}}}, \bibinfo
  {author} {\bibfnamefont {Y.}~\bibnamefont {{Liu}}}, \bibinfo {author}
  {\bibfnamefont {Z.-G.}\ \bibnamefont {{Chen}}}, \bibinfo {author}
  {\bibfnamefont {S.}~\bibnamefont {{Lu}}}, \bibinfo {author} {\bibfnamefont
  {S.}~\bibnamefont {{Liang}}}, \bibinfo {author} {\bibfnamefont
  {J.}~\bibnamefont {{Cao}}}, \bibinfo {author} {\bibfnamefont
  {X.}~\bibnamefont {{Yuan}}}, \bibinfo {author} {\bibfnamefont
  {L.}~\bibnamefont {{Tang}}}, \bibinfo {author} {\bibfnamefont
  {Q.}~\bibnamefont {{Li}}}, \bibinfo {author} {\bibfnamefont {C.}~\bibnamefont
  {{Zhou}}}, \bibinfo {author} {\bibfnamefont {T.}~\bibnamefont {{Gu}}},
  \bibinfo {author} {\bibfnamefont {Y.}~\bibnamefont {{Wu}}}, \bibinfo {author}
  {\bibfnamefont {J.}~\bibnamefont {{Zou}}},\ and\ \bibinfo {author}
  {\bibfnamefont {F.}~\bibnamefont {{Xiu}}},\ }\bibfield  {title} {\bibinfo
  {title} {{Room-temperature chiral charge pumping in Dirac semimetals}},\
  }\href {https://doi.org/10.1038/ncomms13741} {\bibfield  {journal} {\bibinfo
  {journal} {Nature Communications}\ }\textbf {\bibinfo {volume} {8}},\
  \bibinfo {eid} {13741} (\bibinfo {year} {2017})}\BibitemShut {NoStop}%
\bibitem [{\citenamefont {Zuev}\ \emph {et~al.}(2009)\citenamefont {Zuev},
  \citenamefont {Chang},\ and\ \citenamefont {Kim}}]{Philip2009}%
  \BibitemOpen
  \bibfield  {author} {\bibinfo {author} {\bibfnamefont {Y.~M.}\ \bibnamefont
  {Zuev}}, \bibinfo {author} {\bibfnamefont {W.}~\bibnamefont {Chang}},\ and\
  \bibinfo {author} {\bibfnamefont {P.}~\bibnamefont {Kim}},\ }\bibfield
  {title} {\bibinfo {title} {Thermoelectric and magnetothermoelectric transport
  measurements of graphene},\ }\href
  {https://doi.org/10.1103/PhysRevLett.102.096807} {\bibfield  {journal}
  {\bibinfo  {journal} {Phys. Rev. Lett.}\ }\textbf {\bibinfo {volume} {102}},\
  \bibinfo {pages} {096807} (\bibinfo {year} {2009})}\BibitemShut {NoStop}%
\bibitem [{\citenamefont {Nam}\ \emph {et~al.}(2010)\citenamefont {Nam},
  \citenamefont {Ki},\ and\ \citenamefont {Lee}}]{nam_PRB2010}%
  \BibitemOpen
  \bibfield  {author} {\bibinfo {author} {\bibfnamefont {S.-G.}\ \bibnamefont
  {Nam}}, \bibinfo {author} {\bibfnamefont {D.-K.}\ \bibnamefont {Ki}},\ and\
  \bibinfo {author} {\bibfnamefont {H.-J.}\ \bibnamefont {Lee}},\ }\bibfield
  {title} {\bibinfo {title} {Thermoelectric transport of massive {D}irac
  fermions in bilayer graphene},\ }\href
  {https://doi.org/10.1103/PhysRevB.82.245416} {\bibfield  {journal} {\bibinfo
  {journal} {Phys. Rev. B}\ }\textbf {\bibinfo {volume} {82}},\ \bibinfo
  {pages} {245416} (\bibinfo {year} {2010})}\BibitemShut {NoStop}%
\bibitem [{\citenamefont {Balduini}\ \emph {et~al.}(2024)\citenamefont
  {Balduini}, \citenamefont {Molinari}, \citenamefont {Rocchino}, \citenamefont
  {Hasse}, \citenamefont {Felser}, \citenamefont {Sousa}, \citenamefont {Zota},
  \citenamefont {Schmid}, \citenamefont {Grushin},\ and\ \citenamefont
  {Gotsmann}}]{claudia-multifold}%
  \BibitemOpen
  \bibfield  {author} {\bibinfo {author} {\bibfnamefont {F.}~\bibnamefont
  {Balduini}}, \bibinfo {author} {\bibfnamefont {A.}~\bibnamefont {Molinari}},
  \bibinfo {author} {\bibfnamefont {L.}~\bibnamefont {Rocchino}}, \bibinfo
  {author} {\bibfnamefont {V.}~\bibnamefont {Hasse}}, \bibinfo {author}
  {\bibfnamefont {C.}~\bibnamefont {Felser}}, \bibinfo {author} {\bibfnamefont
  {M.}~\bibnamefont {Sousa}}, \bibinfo {author} {\bibfnamefont
  {C.}~\bibnamefont {Zota}}, \bibinfo {author} {\bibfnamefont {H.}~\bibnamefont
  {Schmid}}, \bibinfo {author} {\bibfnamefont {A.~G.}\ \bibnamefont
  {Grushin}},\ and\ \bibinfo {author} {\bibfnamefont {B.}~\bibnamefont
  {Gotsmann}},\ }\bibfield  {title} {\bibinfo {title} {Intrinsic negative
  magnetoresistance from the chiral anomaly of multifold fermions},\ }\href
  {https://doi.org/10.1038/s41467-024-50451-5} {\bibfield  {journal} {\bibinfo
  {journal} {Nature Communications}\ }\textbf {\bibinfo {volume} {15}},\
  \bibinfo {pages} {6526} (\bibinfo {year} {2024})}\BibitemShut {NoStop}%
\bibitem [{\citenamefont {Abrikosov}\ and\ \citenamefont
  {Beneslavski\u{i}}()}]{abrikosov1996}%
  \BibitemOpen
  \bibfield  {author} {\bibinfo {author} {\bibfnamefont {A.~A.}\ \bibnamefont
  {Abrikosov}}\ and\ \bibinfo {author} {\bibfnamefont {S.~D.}\ \bibnamefont
  {Beneslavski\u{i}}},\ }\bibinfo {title} {Possible existence of substances
  intermediate between metals and dielectrics},\ in\ \href
  {https://doi.org/10.1142/9789814317344_0010} {\emph {\bibinfo {booktitle} {30
  Years of the Landau Institute — Selected Papers}}},\ pp.\ \bibinfo {pages}
  {64--73}\BibitemShut {NoStop}%
\bibitem [{\citenamefont {Abrikosov}(1974)}]{Abrikosov}%
  \BibitemOpen
  \bibfield  {author} {\bibinfo {author} {\bibfnamefont {A.~A.}\ \bibnamefont
  {Abrikosov}},\ }\bibfield  {title} {\bibinfo {title} {Calculation of critical
  indices for zero-gap semiconductors},\ }\href@noop {} {\bibfield  {journal}
  {\bibinfo  {journal} {Sov. Phys.-JETP}\ }\textbf {\bibinfo {volume} {39}},\
  \bibinfo {pages} {709} (\bibinfo {year} {1974})}\BibitemShut {NoStop}%
\bibitem [{\citenamefont {Moon}\ \emph {et~al.}(2013)\citenamefont {Moon},
  \citenamefont {Xu}, \citenamefont {Kim},\ and\ \citenamefont
  {Balents}}]{Moon2013PRL}%
  \BibitemOpen
  \bibfield  {author} {\bibinfo {author} {\bibfnamefont {E.-G.}\ \bibnamefont
  {Moon}}, \bibinfo {author} {\bibfnamefont {C.}~\bibnamefont {Xu}}, \bibinfo
  {author} {\bibfnamefont {Y.~B.}\ \bibnamefont {Kim}},\ and\ \bibinfo {author}
  {\bibfnamefont {L.}~\bibnamefont {Balents}},\ }\bibfield  {title} {\bibinfo
  {title} {Non-{F}ermi-liquid and topological states with strong spin-orbit
  coupling},\ }\href {https://doi.org/10.1103/PhysRevLett.111.206401}
  {\bibfield  {journal} {\bibinfo  {journal} {Phys. Rev. Lett.}\ }\textbf
  {\bibinfo {volume} {111}},\ \bibinfo {pages} {206401} (\bibinfo {year}
  {2013})}\BibitemShut {NoStop}%
\bibitem [{\citenamefont {Roy}\ \emph {et~al.}(2017)\citenamefont {Roy},
  \citenamefont {Goswami},\ and\ \citenamefont {Juri\ifmmode \check{c}\else
  \v{c}\fi{}i\ifmmode~\acute{c}\else \'{c}\fi{}}}]{broy}%
  \BibitemOpen
  \bibfield  {author} {\bibinfo {author} {\bibfnamefont {B.}~\bibnamefont
  {Roy}}, \bibinfo {author} {\bibfnamefont {P.}~\bibnamefont {Goswami}},\ and\
  \bibinfo {author} {\bibfnamefont {V.}~\bibnamefont {Juri\ifmmode
  \check{c}\else \v{c}\fi{}i\ifmmode~\acute{c}\else \'{c}\fi{}}},\ }\bibfield
  {title} {\bibinfo {title} {{Interacting Weyl fermions: Phases, phase
  transitions, and global phase diagram}},\ }\href
  {https://doi.org/10.1103/PhysRevB.95.201102} {\bibfield  {journal} {\bibinfo
  {journal} {Phys. Rev. B}\ }\textbf {\bibinfo {volume} {95}},\ \bibinfo
  {pages} {201102} (\bibinfo {year} {2017})}\BibitemShut {NoStop}%
\bibitem [{\citenamefont {{Mandal}}(2021)}]{ips-biref}%
  \BibitemOpen
  \bibfield  {author} {\bibinfo {author} {\bibfnamefont {I.}~\bibnamefont
  {{Mandal}}},\ }\bibfield  {title} {\bibinfo {title} {{Robust marginal Fermi
  liquid in birefringent semimetals}},\ }\href
  {https://doi.org/10.1016/j.physleta.2021.127707} {\bibfield  {journal}
  {\bibinfo  {journal} {Physics Letters A}\ }\textbf {\bibinfo {volume}
  {418}},\ \bibinfo {eid} {127707} (\bibinfo {year} {2021})}\BibitemShut
  {NoStop}%
\bibitem [{\citenamefont {Mandal}\ and\ \citenamefont
  {Freire}(2021)}]{ips-freire1}%
  \BibitemOpen
  \bibfield  {author} {\bibinfo {author} {\bibfnamefont {I.}~\bibnamefont
  {Mandal}}\ and\ \bibinfo {author} {\bibfnamefont {H.}~\bibnamefont
  {Freire}},\ }\bibfield  {title} {\bibinfo {title} {{Transport in the
  non-Fermi liquid phase of isotropic Luttinger semimetals}},\ }\href
  {https://doi.org/10.1103/PhysRevB.103.195116} {\bibfield  {journal} {\bibinfo
   {journal} {Phys. Rev. B}\ }\textbf {\bibinfo {volume} {103}},\ \bibinfo
  {pages} {195116} (\bibinfo {year} {2021})}\BibitemShut {NoStop}%
\bibitem [{\citenamefont {{Mandal}}\ and\ \citenamefont
  {{Freire}}(2022)}]{ips-freire-raman}%
  \BibitemOpen
  \bibfield  {author} {\bibinfo {author} {\bibfnamefont {I.}~\bibnamefont
  {{Mandal}}}\ and\ \bibinfo {author} {\bibfnamefont {H.}~\bibnamefont
  {{Freire}}},\ }\bibfield  {title} {\bibinfo {title} {{Raman response and
  shear viscosity in the non-Fermi liquid phase of Luttinger semimetals}},\
  }\href {https://doi.org/10.1088/1361-648X/ac6785} {\bibfield  {journal}
  {\bibinfo  {journal} {Journal of Physics Condensed Matter}\ }\textbf
  {\bibinfo {volume} {34}},\ \bibinfo {eid} {275604} (\bibinfo {year}
  {2022})}\BibitemShut {NoStop}%
\bibitem [{\citenamefont {Mandal}\ and\ \citenamefont
  {Freire}(2024)}]{ips-hermann-review}%
  \BibitemOpen
  \bibfield  {author} {\bibinfo {author} {\bibfnamefont {I.}~\bibnamefont
  {Mandal}}\ and\ \bibinfo {author} {\bibfnamefont {H.}~\bibnamefont
  {Freire}},\ }\bibfield  {title} {\bibinfo {title} {{Transport properties in
  non-Fermi liquid phases of nodal-point semimetals}},\ }\href
  {https://doi.org/10.1088/1361-648X/ad665e} {\bibfield  {journal} {\bibinfo
  {journal} {Journal of Physics: Condensed Matter}\ }\textbf {\bibinfo {volume}
  {36}},\ \bibinfo {pages} {443002} (\bibinfo {year} {2024})}\BibitemShut
  {NoStop}%
\bibitem [{\citenamefont {{Freire}}\ and\ \citenamefont
  {{Mandal}}(2021)}]{ips-freire-thermo}%
  \BibitemOpen
  \bibfield  {author} {\bibinfo {author} {\bibfnamefont {H.}~\bibnamefont
  {{Freire}}}\ and\ \bibinfo {author} {\bibfnamefont {I.}~\bibnamefont
  {{Mandal}}},\ }\bibfield  {title} {\bibinfo {title} {{Thermoelectric and
  thermal properties of the weakly disordered non-Fermi liquid phase of
  Luttinger semimetals}},\ }\href
  {https://doi.org/10.1016/j.physleta.2021.127470} {\bibfield  {journal}
  {\bibinfo  {journal} {Physics Letters A}\ }\textbf {\bibinfo {volume}
  {407}},\ \bibinfo {eid} {127470} (\bibinfo {year} {2021})}\BibitemShut
  {NoStop}%
\bibitem [{\citenamefont {{Kozii}}\ and\ \citenamefont
  {{Fu}}(2018)}]{kozii_plasmon}%
  \BibitemOpen
  \bibfield  {author} {\bibinfo {author} {\bibfnamefont {V.}~\bibnamefont
  {{Kozii}}}\ and\ \bibinfo {author} {\bibfnamefont {L.}~\bibnamefont {{Fu}}},\
  }\bibfield  {title} {\bibinfo {title} {{Thermal plasmon resonantly enhances
  electron scattering in Dirac/Weyl semimetals}},\ }\href
  {https://doi.org/10.1103/PhysRevB.98.041109} {\bibfield  {journal} {\bibinfo
  {journal} {\prb}\ }\textbf {\bibinfo {volume} {98}},\ \bibinfo {eid} {041109}
  (\bibinfo {year} {2018})}\BibitemShut {NoStop}%
\bibitem [{\citenamefont {{Mandal}}(2019)}]{ips-plasmon}%
  \BibitemOpen
  \bibfield  {author} {\bibinfo {author} {\bibfnamefont {I.}~\bibnamefont
  {{Mandal}}},\ }\bibfield  {title} {\bibinfo {title} {{Search for plasmons in
  isotropic Luttinger semimetals}},\ }\href
  {https://doi.org/10.1016/j.aop.2019.04.002} {\bibfield  {journal} {\bibinfo
  {journal} {Annals of Physics}\ }\textbf {\bibinfo {volume} {406}},\ \bibinfo
  {pages} {173} (\bibinfo {year} {2019})}\BibitemShut {NoStop}%
\bibitem [{\citenamefont {Wang}\ and\ \citenamefont {Mandal}(2023)}]{ips-jing}%
  \BibitemOpen
  \bibfield  {author} {\bibinfo {author} {\bibfnamefont {J.}~\bibnamefont
  {Wang}}\ and\ \bibinfo {author} {\bibfnamefont {I.}~\bibnamefont {Mandal}},\
  }\bibfield  {title} {\bibinfo {title} {Anatomy of plasmons in generic
  {L}uttinger semimetals},\ }\href
  {https://doi.org/10.1140/epjb/s10051-023-00596-x} {\bibfield  {journal}
  {\bibinfo  {journal} {The European Physical Journal B}\ }\textbf {\bibinfo
  {volume} {96}},\ \bibinfo {pages} {132} (\bibinfo {year} {2023})}\BibitemShut
  {NoStop}%
\bibitem [{\citenamefont {Mandal}\ and\ \citenamefont
  {Ziegler}(2021)}]{ips-klaus}%
  \BibitemOpen
  \bibfield  {author} {\bibinfo {author} {\bibfnamefont {I.}~\bibnamefont
  {Mandal}}\ and\ \bibinfo {author} {\bibfnamefont {K.}~\bibnamefont
  {Ziegler}},\ }\bibfield  {title} {\bibinfo {title} {Robust quantum transport
  at particle-hole symmetry},\ }\href
  {https://doi.org/10.1209/0295-5075/ac1a25} {\bibfield  {journal} {\bibinfo
  {journal} {EPL (Europhysics Letters)}\ }\textbf {\bibinfo {volume} {135}},\
  \bibinfo {pages} {17001} (\bibinfo {year} {2021})}\BibitemShut {NoStop}%
\bibitem [{\citenamefont {Nandkishore}\ and\ \citenamefont
  {Parameswaran}(2017)}]{rahul-sid}%
  \BibitemOpen
  \bibfield  {author} {\bibinfo {author} {\bibfnamefont {R.~M.}\ \bibnamefont
  {Nandkishore}}\ and\ \bibinfo {author} {\bibfnamefont {S.~A.}\ \bibnamefont
  {Parameswaran}},\ }\bibfield  {title} {\bibinfo {title} {Disorder-driven
  destruction of a non-{F}ermi liquid semimetal studied by renormalization
  group analysis},\ }\href {https://doi.org/10.1103/PhysRevB.95.205106}
  {\bibfield  {journal} {\bibinfo  {journal} {Phys. Rev. B}\ }\textbf {\bibinfo
  {volume} {95}},\ \bibinfo {pages} {205106} (\bibinfo {year}
  {2017})}\BibitemShut {NoStop}%
\bibitem [{\citenamefont {Mandal}\ and\ \citenamefont
  {Nandkishore}(2018)}]{ips-rahul}%
  \BibitemOpen
  \bibfield  {author} {\bibinfo {author} {\bibfnamefont {I.}~\bibnamefont
  {Mandal}}\ and\ \bibinfo {author} {\bibfnamefont {R.~M.}\ \bibnamefont
  {Nandkishore}},\ }\bibfield  {title} {\bibinfo {title} {Interplay of
  {C}oulomb interactions and disorder in three-dimensional quadratic band
  crossings without time-reversal symmetry and with unequal masses for
  conduction and valence bands},\ }\href
  {https://doi.org/10.1103/PhysRevB.97.125121} {\bibfield  {journal} {\bibinfo
  {journal} {Phys. Rev. B}\ }\textbf {\bibinfo {volume} {97}},\ \bibinfo
  {pages} {125121} (\bibinfo {year} {2018})}\BibitemShut {NoStop}%
\bibitem [{\citenamefont {Mandal}(2018)}]{ips-qbt-sc}%
  \BibitemOpen
  \bibfield  {author} {\bibinfo {author} {\bibfnamefont {I.}~\bibnamefont
  {Mandal}},\ }\bibfield  {title} {\bibinfo {title} {Fate of superconductivity
  in three-dimensional disordered {L}uttinger semimetals},\ }\href
  {https://doi.org/https://doi.org/10.1016/j.aop.2018.03.004} {\bibfield
  {journal} {\bibinfo  {journal} {Annals of Physics}\ }\textbf {\bibinfo
  {volume} {392}},\ \bibinfo {pages} {179 } (\bibinfo {year}
  {2018})}\BibitemShut {NoStop}%
\bibitem [{\citenamefont {Avdoshkin}\ \emph {et~al.}(2020)\citenamefont
  {Avdoshkin}, \citenamefont {Kozii},\ and\ \citenamefont
  {Moore}}]{kozii-cpge}%
  \BibitemOpen
  \bibfield  {author} {\bibinfo {author} {\bibfnamefont {A.}~\bibnamefont
  {Avdoshkin}}, \bibinfo {author} {\bibfnamefont {V.}~\bibnamefont {Kozii}},\
  and\ \bibinfo {author} {\bibfnamefont {J.~E.}\ \bibnamefont {Moore}},\
  }\bibfield  {title} {\bibinfo {title} {Interactions remove the quantization
  of the chiral photocurrent at {W}eyl points},\ }\href
  {https://doi.org/10.1103/PhysRevLett.124.196603} {\bibfield  {journal}
  {\bibinfo  {journal} {Phys. Rev. Lett.}\ }\textbf {\bibinfo {volume} {124}},\
  \bibinfo {pages} {196603} (\bibinfo {year} {2020})}\BibitemShut {NoStop}%
\bibitem [{\citenamefont {Mandal}(2020)}]{ips-cpge}%
  \BibitemOpen
  \bibfield  {author} {\bibinfo {author} {\bibfnamefont {I.}~\bibnamefont
  {Mandal}},\ }\bibfield  {title} {\bibinfo {title} {Effect of interactions on
  the quantization of the chiral photocurrent for double-weyl semimetals},\
  }\href {https://www.mdpi.com/2073-8994/12/6/919} {\bibfield  {journal}
  {\bibinfo  {journal} {Symmetry}\ }\textbf {\bibinfo {volume} {12}} (\bibinfo
  {year} {2020})}\BibitemShut {NoStop}%
\end{thebibliography}%

\end{document}